\documentclass[aps,prd,reprint,nofootinbib,superscriptaddress]{revtex4-1}
\usepackage{amsmath,amssymb,amsfonts}
\usepackage{mathrsfs}
\usepackage{bbm}
\usepackage{slashed}
\usepackage{graphicx}
\usepackage{verbatim}
\usepackage[T1]{fontenc}
\usepackage[utf8]{inputenc}
\usepackage{hyperref}
\usepackage{ifthen}
\usepackage{forloop}
\usepackage[english]{babel}
\usepackage{mathbbol}
\usepackage[usenames,dvipsnames]{xcolor}
\usepackage[normalem]{ulem}

\definecolor{bluecite}{HTML}{0875b7}

\hypersetup{
	colorlinks=true,
	linkcolor=bluecite,
	urlcolor=magenta,
	citecolor=bluecite
}

\newcommand{\UV}{{\small UV}}
\newcommand{\IR}{{\small IR}}
\newcommand{\FRG}{{\small FRG}}
\newcommand{\QFT}{{\small QFT}}
\newcommand{\RG}{{\small RG}}

\newcommand{\GFP}{{\small GFP}}
\newcommand{\SGFP}{{\small SGFP}}
\newcommand{\WGB}{{\small WGB}}
\newcommand{\eg}{{\textit{e.g.}}}
\newcommand{\ie}{{\textit{i.e.}}}

\newcommand{\Nmax}{\ensuremath{N_{\mathrm{max}}}{}}
\newcommand{\Omax}{\ensuremath{N_{\mathrm{max}}}{}}
\newcommand{\Pmax}{\ensuremath{N_{\mathrm{max}}}{}}

\newcommand{\scalarfielddimful}{\ensuremath{\Phi}}
\newcommand{\scalarfielddimless}{\ensuremath{\phi}}

\newcommand{\kineticoperatordimful}{\ensuremath{\mathcal X}}
\newcommand{\kineticoperatordimless}{\ensuremath{X}}
\newcommand{\kineticoperatordimlessGdimlessexpansion}{\ensuremath{\tilde{\kineticoperatordimless}}}
\newcommand{\kineticoperatordimlesscombinedexpansion}{\ensuremath{\hat{\kineticoperatordimless}}}

\newcommand{\GFPvariable}{\ensuremath{y}}

\newcommand{\kineticfunctiondimful}{\ensuremath{\mathcal K}_k}
\newcommand{\kineticfunctiondimless}{\ensuremath{K}}
\newcommand{\kineticfunctiondimlessFP}{\ensuremath{K_\ast}}
\newcommand{\scalarWFR}{\ensuremath{Z}}

\newcommand{\Kcoup}[1]{\ensuremath{K_{#1}}}
\newcommand{\KcoupFP}[1]{\ensuremath{K_{#1,\ast}}}

\newcommand{\kineticfunctionGdimlessexpansion}[1]{\ensuremath{L_{#1}}}
\newcommand{\kineticfunctioncombinedexpansion}[1]{\ensuremath{M_{#1}}}

\newcommand{\kineticfunctionGdimlessexpansionFP}[1]{\ensuremath{L_{#1,\ast}}}
\newcommand{\kineticfunctioncombinedexpansionFP}[1]{\ensuremath{M_{#1,\ast}}}

\newcommand{\scalaranomalousdimension}{\ensuremath{\eta_\scalarfielddimful}}
\newcommand{\scalaranomalousdimensionFP}{\ensuremath{\eta_{\scalarfielddimful,\ast}}}
\newcommand{\scalaranomalousdimensionhat}{\ensuremath{\hat\eta_\scalarfielddimful}}

\newcommand{\scalaranomalousdimensionGdimlessexpansion}[1]{\ensuremath{\eta_{#1}}}
\newcommand{\scalaranomalousdimensionGdimlessexpansionFP}[1]{\ensuremath{\eta_{#1,\ast}}}

\newcommand{\GFalpha}{\ensuremath{\alpha_h}}
\newcommand{\GFbeta}{\ensuremath{\beta_h}}

\newcommand{\Gdimful}{\ensuremath{G_k}}
\newcommand{\Gdimless}{\ensuremath{g}}
\newcommand{\Gdimlesscrit}{\ensuremath{g_{\text{crit}}}}
\newcommand{\Gdimlesscritnew}{\ensuremath{\tilde{g}_{\text{crit}}}}
\newcommand{\GEH}{\ensuremath{G_N}} 

\newcommand{\Lambdadimful}{\ensuremath{\Lambda_k}}
\newcommand{\Lambdadimfulcl}{\ensuremath{\Lambda}}
\newcommand{\Lambdadimless}{\ensuremath{\lambda}}

\newcommand{\thresholdBK}[1]{\ensuremath{Q_{#1}}}
\newcommand{\thresholdtildeBK}[1]{\ensuremath{\tilde Q_{#1}}}

\newcommand{\SLp}{\ensuremath{p}}
\newcommand{\SLmeasure}{\ensuremath{w}}
\newcommand{\SLEV}{\ensuremath{\lambda}}

\newcommand{\Int}[1]{\int \text{d}^{#1}x \, \sqrt{g}}
\newcommand{\Intb}[1]{\int \text{d}^{#1}x \, \sqrt{\bar{g}}}

\newcommand{\aref}[1]{\hyperref[#1]{appendix~\ref*{#1}}}

\allowdisplaybreaks

\begin{document}

	\title{On the weak-gravity bound for a shift-symmetric scalar field}
	\author{Gustavo P. de Brito\,\href{https://orcid.org/0000-0003-2240-528X}{\protect \includegraphics[scale=.07]{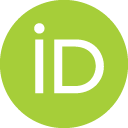}}\,}
	\email[]{gustavo@sdu.dk}
	\affiliation{
		CP3-Origins, University of Southern Denmark, Campusvej 55, DK-5230 Odense M, Denmark
	}
	\author{Benjamin Knorr\,\href{https://orcid.org/0000-0001-6700-6501}{\protect \includegraphics[scale=.07]{ORCIDiD_icon128x128.png}}\,}
	\email[]{benjamin.knorr@su.se}
	\affiliation{
		Nordita, Stockholm University and KTH Royal Institute of Technology, Hannes Alfv\'ens v\"ag 12, SE-106 91 Stockholm, Sweden
	}
	\author{Marc Schiffer\,\href{https://orcid.org/0000-0002-0778-4800}{\protect \includegraphics[scale=.07]{ORCIDiD_icon128x128.png}}\,}
	\email[]{mschiffer@perimeterinstitute.ca}
	\affiliation{
		Perimeter Institute for Theoretical Physics, 31 Caroline St. N., Waterloo, ON N2L 2Y5, Canada
	}

	\begin{abstract}
		The weak-gravity bound has been discovered in several asymptotically safe gravity-matter systems. It limits the strength of gravitational fluctuations that are compatible with an ultraviolet-complete matter sector, and results from the collision of two partial fixed points of the matter system as a function of the strength of the gravitational interactions. In this paper, we will investigate this mechanism in detail for a shift-symmetric scalar field. First, we will study the fixed point structure of the scalar system without gravity. We find indications that the Gaussian fixed point is the only viable fixed point, suggesting that a weak-gravity bound resulting from the collision of two partial fixed points is a truncation artefact. We will then couple the scalar system to gravity and perform different expansions to track the Gaussian fixed point as gravitational fluctuations become stronger. We also introduce a new notion of the weak-gravity bound that is based on the number of relevant operators.
	\end{abstract}

	\maketitle

	\tableofcontents
	\section{Motivation}\label{sec:intro}

	The consistent quantisation of the gravitational force is one of the major open problems in theoretical physics. General Relativity, which describes gravity ranging from sub-millimetre \cite{Murata:2014nra, Tan:2020vpf} through solar system \cite{Sanner:2018atx, MICROSCOPE:2022doy} all the way to cosmological scales \cite{Berti:2015itd, LIGOScientific:2016aoc, LIGOScientific:2016sjg, Planck:2018vyg,  EventHorizonTelescope:2019dse, EventHorizonTelescope:2019uob, EventHorizonTelescope:2019ggy, LIGOScientific:2020tif}, breaks down, \eg, in the centre of black holes. This breakdown occurs at the smallest distance scales, and indicates that new physics is required to describe the gravitational interaction in that regime. The most popular candidate for such new physics lies in quantum gravity, which encodes quantum fluctuations of spacetime itself. However, despite huge efforts, to date no fully consistent and phenomenologically viable theory of quantum gravity has been developed. One of the reasons for this situation is that perturbative quantisation, so successful for the Standard Model of Particle Physics, fails in the case of gravity due to the negative mass dimension of Newton's constant. This power counting argument was also confirmed by the explicit computations of the one-loop \cite{tHooft:1974toh, Deser:1974cy, Deser:1974zzd} and two-loop counterterms \cite{Goroff:1985sz, Goroff:1985th, vandeVen:1991gw}. The failure of perturbative quantisation requires that new concepts have to be considered to develop a quantum theory of gravity, for example by resorting to genuinely non-perturbative scenarios, imposing additional symmetries, or others. Crucially, since the characteristic energy scale of quantum gravity, the Planck scale, is extremely high, it is difficult to confront theories of quantum gravity with direct observational tests. In that light, internal consistency tests play an important role to exclude candidate theories.
	
	An important aspect in this conundrum is that a quantisation of gravity alone does not suffice: ultimately, to describe our universe, we have to find a consistent quantum theory of all fundamental forces and particles. This seemingly innocent and trivial statement has however dramatic consequences: since gravity is expected not to be a free theory at high energies, due to the power counting argument, it dictates that some specific pure matter couplings cannot vanish either. This expectation comes from the fact that all matter gravitates, such that the interacting nature of gravity directly percolates into the matter sector. Already the kinetic energy of a free particle is coupled to the metric, and thus quantum gravity fluctuations invariably induce higher order interaction terms consistent with the symmetries of the kinetic term \cite{Eichhorn:2011pc, Eichhorn:2012va, Eichhorn:2013ug, Meibohm:2016mkp, Christiansen:2017gtg, Eichhorn:2017sok, Eichhorn:2017eht, Eichhorn:2018nda, Ali:2020znq, deBrito:2021pyi, Laporte:2021kyp, Knorr:2022ilz, Eichhorn:2022gku}.
	
	This observation potentially creates another problem: if gravity is too strong, the induced interactions might prevent a description of the matter sector that holds at arbitrary high energies and is also consistent with low-energy data. The idea of such a limit on the strength of gravity was coined the weak-gravity bound (\WGB{})\footnote{This is not to be confused with the weak gravity conjecture \cite{Arkani-Hamed:2006emk, Harlow:2022gzl} that is generally not related to the \WGB{}.} in the literature \cite{Eichhorn:2017eht}.
	
	While such a bound can in principle appear in any approach to quantum gravity, in this work we will focus on the \WGB{} in asymptotically safe gravity \cite{Weinberg:1980gg}. The basic premise of the latter is that gravity can be quantised consistently as a quantum field theory (\QFT{}) in a \emph{non-perturbative} fashion. The physical mechanism behind this is quantum scale invariance at high energies, indicating a second-order phase transition in the language of condensed matter physics. This corresponds to a fixed point of the renormalisation group flow. Encouraging indications for the asymptotic safety scenario have been found in the past decades \cite{Reuter:2001ag, Codello:2006in, Codello:2007bd, Machado:2007ea, Codello:2008vh, Benedetti:2010nr, Manrique:2011jc, Falls:2013bv, Codello:2013fpa, Dona:2013qba, Christiansen:2014raa, Becker:2014qya, Christiansen:2015rva, Ohta:2015efa, Meibohm:2015twa, Gies:2016con, Biemans:2016rvp, Denz:2016qks, Gonzalez-Martin:2017gza, Biemans:2017zca,  Christiansen:2017cxa, Knorr:2017mhu, Christiansen:2017bsy, Alkofer:2018fxj, Eichhorn:2018akn, DeBrito:2018hur, Eichhorn:2018ydy, Knorr:2019atm, Kluth:2020bdv, Knorr:2020ckv, Bonanno:2021squ, Knorr:2021slg,  Knorr:2021niv, Baldazzi:2021orb, Mitchell:2021qjr, Sen:2021ffc, Fehre:2021eob, Kluth:2022vnq, Pastor-Gutierrez:2022nki}, see also \cite{Percacci:2017fkn, Reuter:2019byg, Pawlowski:2020qer, Bonanno:2020bil, Morris:2022btf, Wetterich:2022ncl, Martini:2022sll, Knorr:2022dsx,  Eichhorn:2022bgu, Eichhorn:2022gku, Platania:2023srt} for introductions and reviews, and \cite{Bonanno:2000ep, Shaposhnikov:2009pv, Harst:2011zx, Bonanno:2015fga, Oda:2015sma, Christiansen:2017gtg, Eichhorn:2017eht,  Eichhorn:2017ylw,  Eichhorn:2017lry, Bonanno:2017zen, Bonanno:2018gck, Gubitosi:2018gsl, Held:2019xde, Platania:2019kyx, Bosma:2019aiu, DeBrito:2019gdd, Reichert:2019car, Platania:2020lqb, Draper:2020bop, Draper:2020knh, Eichhorn:2020sbo, deBrito:2021akp, Kowalska:2022ypk, Borissova:2022mgd} for phenomenological implications. 
	
	Within the asymptotic safety scenario, one mechanism to observe the \WGB{} is a collision of partial fixed points of gravitationally induced interactions. For this, one starts with a pure matter \QFT{} without gravitational interactions, and then follows the fixed points of this system as one includes, and increases the strength of, gravitational interactions. If the partial fixed point that emanates from the free fixed point of the pure matter system collides with another partial fixed point at a critical value of the gravitational interaction, there is a \WGB{} in this system. This mechanism has been explored in a number of works, coupling gravity to scalars \cite{Eichhorn:2012va, deBrito:2021pyi, Knorr:2022ilz}, fermions \cite{Eichhorn:2011pc, Eichhorn:2017eht, deBrito:2020dta}, and Abelian gauge fields \cite{Christiansen:2017gtg, Eichhorn:2019yzm, Eichhorn:2021qet}.
	
	This particular mechanism clearly crucially relies on the fixed point structure of the pure matter \QFT{}, since generically fixed points are expected to only collide in pairs. There is also the opposite possibility: a pair of complex-conjugate partial fixed points can collide at a finite value of the gravitational coupling and turn real, but once again, this is generally expected to happen in pairs.
	
	In this paper, we will focus on a shift-symmetric scalar field coupled to gravity and extend previous studies of the \WGB{} in this system. In particular, we subsequently add more induced interactions and investigate the fate of the \WGB{} under these extensions. We have two main motivations to investigate this theory.
	
	First, it is believed that scalar field theories in four dimensions are trivial, \ie{}, they do not admit an ultraviolet (\UV{}) completion that results in an interacting infrared (\IR{}) theory \cite{Kleinert:2001ax, Shrock:2023xcu}. The arguments for this triviality rely on the study of non-shift-symmetric, $\phi^4$ theories. However, it is conceivable that the renormalisation group flow in a theory space defined by shift symmetry can feature different properties, possibly defining a new universality class for scalar theories in four dimensions \cite{Laporte:2022ziz}. Here, we extend previous work in \cite{Laporte:2022ziz} and explore this possibility, and we present several arguments that lead us to conclude that the existing candidates for non-trivial universality classes of shift-symmetric scalar theories are likely spurious.
	
	Second, shift-symmetric scalar theories coupled to gravity are good working examples to understand the \WGB{} in asymptotically safe gravity. The mechanism for the \WGB{} previously studied in the literature \cite{Eichhorn:2011pc,Eichhorn:2012va,Eichhorn:2017eht, Eichhorn:2019yzm, deBrito:2020dta, deBrito:2021pyi, Knorr:2022ilz,Christiansen:2017gtg, Eichhorn:2022gku, Eichhorn:2021qet} involves the collision of two fixed points that are already present at the pure matter level. Our arguments concerning the spurious nature of the existing candidates for non-trivial universality classes in shift-symmetric scalar theories imply that the \WGB{} (based on the mechanism of a partial fixed point collision) is not present. Still, we can see a partial fixed point collision as an indicator of the breakdown of certain expansion schemes in functional renormalisation group calculations. Thus, we also explore which expansion schemes automatically avoid this collision. We will also introduce a modified notion of the \WGB{} related to a different mechanism: instead of defining the strong gravity regime by the absence of a fixed point, we define it by the existence of additional relevant operators compared to the free theory. Our system possesses such a refined \WGB{}.
		
	This paper is structured as follows: in \autoref{sec:setup}, we introduce the method that our investigation relies on, the functional renormalisation group (\FRG{}). Furthermore, we use a simple toy model to illustrate how gravitational fluctuations induce matter self-interactions, and how they can spoil a \UV{} completion in the matter sector. In \autoref{sec:purematt}, we will discuss a single shift-symmetric scalar field whose action is described by a function of the kinetic term. We will investigate the fixed point structure of this system upon expansion of the function in powers of the kinetic term. In particular, we will search for viable interacting fixed points of this pure scalar system. In \autoref{sec:gravmatt}, we will couple the shift-symmetric scalar field to gravity, and investigate how the free fixed point of the pure matter system changes when turning on gravitational fluctuations. We will employ different expansions of the gravity-scalar system in terms of the kinetic term and the gravitational coupling, and discuss the absence of the partial fixed point collision, but the presence of the new notion of \WGB{}, in the system. Finally, in \autoref{sec:summary} we summarise our results and conclude.

	\section{Methodological Introduction}\label{sec:setup}
	In this section, we introduce the necessary ingredients for our study and the notation that we will use throughout the paper. First, we will introduce the \FRG{} and define the gravity-scalar system that we aim to study in \autoref{sec:FRGIntro}. Furthermore, we will review the mechanism with which gravitational fluctuations induce interactions in the matter sector in \autoref{sec:WGBIntro}. There, we will also review the emergence of the \WGB{} as it has been discussed in the literature.

	\subsection{Setup}\label{sec:FRGIntro}
	
	To explore the fixed point structure of the scalar system with and without the impact of gravitational fluctuations, we employ the \FRG{} \cite{Wetterich:1992yh, Morris:1993qb, Ellwanger:1993mw, Reuter:1996cp}. It is based on the flow equation for the scale-dependent effective action $\Gamma_k$, which reads
	\begin{equation}\label{eq:floweq}
	k\partial_k\,\Gamma_k=\frac{1}{2}\mathrm{Tr}\left[ \left(\Gamma_k^{(2)}+\mathfrak R_k\right)^{-1} k\partial_k\,\mathfrak R_k \right] \, ,
	\end{equation}
	where $\Gamma_k^{(2)}$ is the second functional derivative of $\Gamma_k$ with respect to all fields of the system, and where $\mathfrak R_k$ is the so-called regulator functional. The functional trace indicates a sum over discrete and an integral over continuous variables. The regulator functional $\mathfrak R_k$ acts like a scale-dependent mass term, and therefore ensures finiteness in the \IR. Together with its scale derivative $k\partial_k\,\mathfrak R_k$ that ensures \UV{} finiteness, it implements the Wilsonian idea of integrating out quantum fluctuations according to their momentum shell. Therefore, the scale-dependent action $\Gamma_k$ interpolates between $\Gamma_{k\to\infty}$, where no quantum fluctuations are integrated out, corresponding roughly to a bare action,\footnote{The precise relation between $\Gamma_{k\to\infty}$ and the bare action $S$ is known as the reconstruction problem, see, \eg{}, refs. \cite{Manrique:2009tj, Morris:2015oca, Fraaije:2022uhg}.} and the full quantum effective action $\Gamma_{k\to0}$, where all quantum fluctuations are integrated out. The scale dependence of couplings and operators can be extracted from the flow equation \eqref{eq:floweq} by projecting onto the corresponding tensor structure.
	
	The flow equation is not limited to the perturbative regime, but allows to extract the scale dependence of couplings in non-perturbative settings. Besides asymptotically safe gravity, it has been successfully employed, for example, to condensed matter physics and the strong nuclear force, see, \eg{}, \cite{Dupuis:2020fhh} for an up-to-date review.

	For a general theory, it is extremely difficult to solve \eqref{eq:floweq} exactly, and as a consequence one has to introduce approximations (so-called truncations). In the following, we describe the systematic approximation that we will employ to investigate the effect of gravitational fluctuations on a shift-symmetric scalar field.

    We approximate the dynamics of our system by the scale-dependent effective action
	\begin{equation}\label{eq:genaction}
	\Gamma_k=\Gamma^{\mathrm{grav}}_k+\Gamma^{\mathrm{scal}}_k\,.
	\end{equation}
	Since our focus lies on the scalar sector,  we employ the simplest approximation in the gravitational sector, the Einstein-Hilbert action,
	\begin{equation}
	\Gamma^{\mathrm{grav}}_k = \frac{1}{16\pi \Gdimful} \Int{4} \, \left[ -R + 2\Lambdadimful \right]+\Gamma_{\mathrm{gf}}\,,
	\end{equation}
	that contains the standard gauge-fixing term
	\begin{equation}
	\Gamma_{\mathrm{gf}} = \frac{1}{32\pi G_k \GFalpha{}} \Intb{4} \, \bar g^{\mu\nu} \mathcal F_\mu \, \mathcal F_\nu \, ,
	\end{equation}
	where
	\begin{equation}
	\mathcal F_\mu = \left( \delta_\mu^{(\alpha} \bar D^{\beta)} - \frac{1+\GFbeta{}}{4} \bar g^{\alpha\beta} \bar D_\mu \right) h_{\alpha\beta}
	\end{equation}
	is the gauge-fixing condition. Here, we have introduced the two gauge-fixing parameters \GFalpha{} and \GFbeta{}. In the following we will always employ the Landau limit $\GFalpha\to0$, since it is a fixed point of both gauge-fixing parameters \cite{Litim:2002ce, Knorr:2017fus}, and we keep \GFbeta{} arbitrary.
	The gauge-fixing term also gives rise to Faddeev-Popov ghosts. Since we will however neglect induced scalar-ghost interactions \cite{Eichhorn:2013ug}, the ghost sector will not contribute to the scale dependence of the scalar sector in our approximation. 
	
	In the following investigation, we will restrict ourselves to a linear parameterisation of gravitational fluctuations,
	\begin{equation}
	\label{eq:linsplit}
	g_{\mu\nu} = \bar g_{\mu\nu} + h_{\mu\nu} \, ,
	\end{equation}
	and we will choose a flat background, \ie{}, $\bar{g}_{\mu\nu}=\delta_{\mu\nu}$.\footnote{In \aref{app:gmSpuriousDep} we also discuss results obtained with the exponential parameterisation of metric fluctuations.}
		
	We will approximate the dynamics of the scalar sector in terms of the kinetic operator for the scalar field \scalarfielddimful{}, 
	\begin{equation}
	 \kineticoperatordimful =  \frac{1}{2} \left( D_\mu \scalarfielddimful \right) \left( D^\mu \scalarfielddimful \right) \, ,
	\end{equation}
	and consider a function of \kineticoperatordimful{}, \ie,
	\begin{equation}\label{eq:kinscal}
	\Gamma^{\mathrm{scal}}_k = \Int{4} \, \kineticfunctiondimful (\kineticoperatordimful) \, ,
	\end{equation}
	 where the dimensionful functional $\kineticfunctiondimful$ satisfies the boundary conditions
	\begin{equation}\label{eq:Kfunctconds}
	 \kineticfunctiondimful(0) = 0 \, , \quad \kineticfunctiondimful'(0) = \scalarWFR \, .
	\end{equation}
	The first equation ensures that $\kineticfunctiondimful$ does not contain \kineticoperatordimful{}-independent terms, that is a contribution to the cosmological constant, while the second equation ensures that $\kineticfunctiondimful$ contains the standard kinetic term of a scalar field. We have introduced the wavefunction renormalisation of the scalar field \scalarWFR, which gives rise to an anomalous dimension via
	\begin{equation}
	 \scalaranomalousdimension = -\frac{k\partial_k\,\scalarWFR}{\scalarWFR}\,.
	\end{equation}
	To uniquely project on the scale dependence of $\kineticfunctiondimful$, we choose a background for the scalar field where
	\begin{equation}
	 \kineticoperatordimful = \mathrm{const}\,.
	\end{equation}
	On this background, \kineticoperatordimful{} is the only shift-symmetric and $\mathbb{Z}_2$-invariant quantity that can be built with the scalar field \scalarfielddimful{} and covariant derivatives. On more general backgrounds for the scalar field, further invariants involving the scalar field and covariant derivatives can be constructed. 
	
	Finally, we will use the regulator
	\begin{equation}
	\mathfrak R_k(p^2) = \frac{k^2}{p^2} R_k\left(\frac{p^2}{k^2}\right) \,\Gamma^{(2)}_k \Big|_{\Lambdadimful=0,h_{\mu\nu}=0,\scalarfielddimful=0} \, ,
	\end{equation}
	which ensures that no mass-like contributions enter the regulator \cite{Gies:2002af, Pawlowski:2005xe, Benedetti:2010nr, Gies:2015tca}. In the following, where reasonable we will keep the regulator $R_k$ unspecified, and express results in terms of general threshold functions. When quoting numerical values such as fixed point values or critical exponents, we either employ a Litim  \cite{Litim:2001up} or an exponential \cite{Berges:2000ew} regulator,
	\begin{align}
	 &\text{Litim:} &R_k(z)=&\left(1-z\right)\Theta(1-z)\,,\label{eq:shapeLitim} \\
	 &\text{Exp.:} &R_k(z)=&\frac{z}{e^z-1}\,.\label{eq:shapeExp}
	\end{align}
	
	Since the condition of scale invariance is best expressed in terms of dimensionless quantities, we introduce the dimensionless counterparts of all dimensionful quantities. In particular, we introduce the dimensionless versions $\Gdimless$ and $\Lambdadimless$ of the Newton coupling and the cosmological constant, respectively, via
	\begin{equation}
	 \Gdimless = k^2 \, \Gdimful \, , \qquad \Lambdadimless = k^{-2} \, \Lambdadimful \, ,
	\end{equation}
	as well as dimensionless versions $\scalarfielddimless$, $\kineticoperatordimless$, and $\kineticfunctiondimless$ of the scalar field, the kinetic operator, and the function in the kinetic operator,
	\begin{equation}
	\scalarfielddimless = \sqrt{\scalarWFR} \, k^{-1}\,\scalarfielddimful\,,\quad \kineticoperatordimless = \scalarWFR \, k^{-4}\,\kineticoperatordimful\,,\quad \kineticfunctiondimless(\kineticoperatordimless) = k^{-4}\,\kineticfunctiondimful(\kineticoperatordimful)\,.
	\end{equation}
	A fixed point of the \RG{} flow is realised when the scale dependence of the dimensionless versions of all couplings vanishes.
	
	Critical exponents determine the universality class of a fixed point. They are defined by
		\begin{equation}
		\label{eq:critexp}
		\Theta_i=-\mathrm{eig}\left[\frac{\partial \beta_{g_i}}{\partial g_j}\right]\bigg|_{g_n=g_{n,*}}\,,
		\end{equation}
		where $g_i$ are all couplings of the system. With this definition
		$\Theta_i>0$ corresponds to a relevant (\IR{} repulsive) direction, while $\Theta_i<0$ corresponds to an irrelevant (\IR{} attractive) direction. 
	
	To compute the scale dependence of $\kineticfunctiondimless$, we used the Mathematica package suite \emph{xAct} \cite{2007CoPhC.177..640M, 2008CoPhC.179..586M, 2008CoPhC.179..597M, 2014CoPhC.185.1719N} and the \emph{FormTracer} \cite{Cyrol:2016zqb}.
	
	\subsection{Induced interactions and the weak-gravity bound}\label{sec:WGBIntro}
	
	Within asymptotically safe quantum gravity and absent implausible cancellations, it is clear that gravitational fluctuations necessarily induce specific matter self-interactions, see \cite{Eichhorn:2022gku} for an overview. On a diagrammatic level, this can be understood by considering the kinetic term of a scalar field, \ie, \eqref{eq:kinscal} with $\kineticfunctiondimless(\kineticoperatordimless)=\kineticoperatordimless$. 
	Expanding it in terms of metric fluctuations gives rise to an infinite series of interactions containing two scalar fields (contained in $\kineticoperatordimless$) and $n$-powers of the metric fluctuation $h_{\mu\nu}$. Focussing, \eg{}, on the second order term, we have a vertex with two scalar fields and two metric fluctuations $h_{\mu\nu}$. Using only graviton propagators, two of such vertices can then be combined to a diagram that contributes to the scale dependence of an interaction containing four scalar fields. Hence, once gravitational fluctuations are present, interactions of the form $\Kcoup{2}\,\kineticoperatordimless^2$ will be induced.
	
\begin{figure}
 		\includegraphics[width=\linewidth]{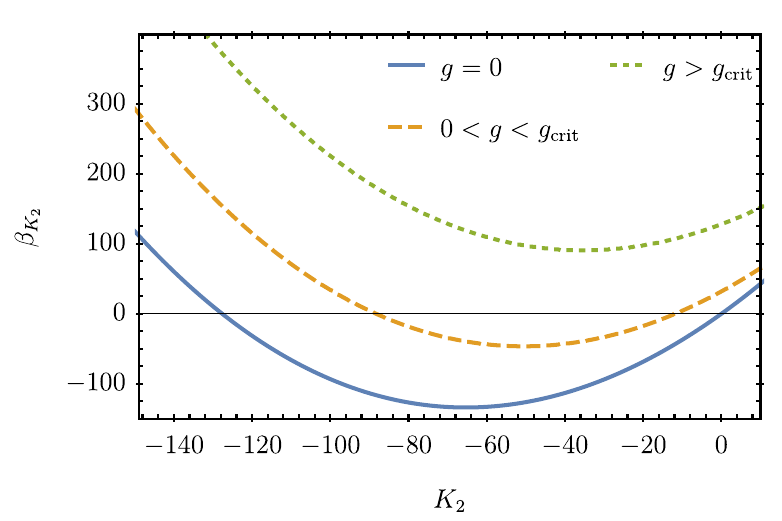}
 		\caption{Beta function of the induced four-scalar interaction $\Kcoup{2}$, cf.~\eqref{eq:betaschem}. In the absence of gravitational fluctuations (blue solid line), $\beta_{\Kcoup{2}}$ features two fixed points, the \GFP{} and an interacting fixed point. In the presence of weak gravitational fluctuations (orange dashed line), the \GFP{} has been shifted to the interacting \SGFP{}. Beyond the weak-gravity regime (green dotted line), $\beta_{\Kcoup{2}}$ does not feature any real zeroes. This indicates that the scalar sector does not admit a \UV{} completion via the \SGFP{}.}
 		\label{fig:BetaK2WGBIllust}
 	\end{figure}

	More specifically, the scale dependence of the induced coupling $\Kcoup{2}$ can be schematically written as
	\begin{equation}\label{eq:betaschem}
	\beta_{\Kcoup{2}}=C_0+C_1\,\Kcoup{2}+C_2\,\Kcoup{2}^2 \, ,
	\end{equation}
	where $C_0$ and $C_1$ are functions of the gravitational couplings $\Gdimless$ and $\Lambdadimless$, $C_2$ is a regulator-dependent number, and we neglected the anomalous dimension. In particular, $C_0(\Gdimless\to0)=0$. Hence, in the absence of gravitational fluctuations, and neglecting higher-order contributions, there are two fixed points, namely
	\begin{equation}
	\KcoupFP{2}=0\,,\quad\mathrm{and}\quad\KcoupFP{2}=-\frac{C_1}{C_2}\,,\quad\mathrm{for}\quad\Gdimless=0\,,
	\end{equation}
	see also the solid blue line in \autoref{fig:BetaK2WGBIllust}.
 	The first fixed point is the standard Gaussian fixed point (\GFP{}), where $\Kcoup{2}$ can be consistently set to zero in the absence of gravity. Since the coupling corresponds to an irrelevant direction at this fixed point, the theory will remain non-interacting at all scales. At the second fixed point, the pure scalar theory would be interacting, see also \cite{Laporte:2022ziz}.
	
 	When gravitational fluctuations are present, $C_0$ no longer vanishes. In this case, the two fixed point solutions of \eqref{eq:betaschem} are given by
 	\begin{equation}\label{eq:sgfpsol}
 	\KcoupFP{2}=-\frac{1}{2\,C_2}\left(C_1(\Gdimless)\pm\sqrt{C_1^2(\Gdimless)-4C_0(\Gdimless)C_2}\right) \, .
 	\end{equation}
	Assuming that the square root is real, both fixed point values for $\KcoupFP{2}$ are now non-zero. Hence, the scalar coupling $\Kcoup{2}$ cannot consistently be set to zero. Instead, the \GFP{} is shifted by gravity, becoming a so-called shifted \GFP{} (\SGFP), see also the dashed orange line in  \autoref{fig:BetaK2WGBIllust}. This indicates that the scalar sector cannot be non-interacting in the presence of gravitational fluctuations. Since $C_0$ is a continuous function in $\Gdimless$, the \SGFP{} is a continuous deformation of the \GFP{} for sufficiently small $\Gdimless$. This general idea has been confirmed by explicit computations in the context of asymptotically safe quantum gravity for different matter systems, see \cite{Eichhorn:2011pc, Eichhorn:2012va, Muneyuki:2013aba, Meibohm:2016mkp, Christiansen:2017gtg, Eichhorn:2017eht, Ali:2020znq, deBrito:2021pyi, Laporte:2021kyp,Eichhorn:2021qet, Knorr:2022ilz}.
	
	Depending on the exact form of the coefficients $C_i$ in \eqref{eq:betaschem}, the solutions given in \eqref{eq:sgfpsol} can either approach, or move away from each other as functions of the gravitational couplings $\Gdimless$ and $\Lambdadimless$. In the former case, the partial fixed points collide at a critical strength of the gravitational interaction, and move into the complex plane, see the dotted green line in  \autoref{fig:BetaK2WGBIllust}. In this region, the system does not admit a real partial fixed point anymore, such that the scalar sector does not admit a \UV{} completion via the \SGFP{}.
	
	The critical coupling at which the partial fixed points collide is what defines the \WGB{} \cite{Eichhorn:2017eht}. It separates the viable weak gravity regime from the excluded strong gravity regime in this toy model. The \WGB{} corresponds to the line in the $(\Lambdadimless,\Gdimless)$-plane where the two fixed point solutions \eqref{eq:sgfpsol} coincide. 
	
   The \WGB{} is typically found to be described by a line $\Gdimlesscrit(\Lambdadimless)$ in this plane. Therefore, in the following we will restrict our analysis to $\Lambdadimless=0$. Accordingly, the presence and location of the \WGB{} is determined by $\Gdimlesscrit(\Lambdadimless=0)$. Let us also emphasise that the general argument that we will present below is independent of the inclusion of a finite cosmological constant as it relies solely on the fixed point structure of the pure scalar system.

	\section{Pure scalar system}
	\label{sec:purematt}
	
	As we just saw, the analysis within a simple truncation involving a quartic shift-symmetric scalar self-interaction indicates the existence of an interacting fixed point \cite{Eichhorn:2012va, deBrito:2021pyi, Laporte:2021kyp, Knorr:2022ilz, Laporte:2022ziz}. In this section, we explore whether the pure scalar sector features a suitable interacting fixed point beyond the quadratic truncation in \kineticoperatordimless{} that was employed in \cite{Eichhorn:2012va, deBrito:2021pyi, Laporte:2021kyp, Knorr:2022ilz}. We extend recent studies \cite{Laporte:2022ziz} and search the pure scalar system for fixed points that converge under extensions of the truncation, and that feature desired regularity and normalisability properties.
	
	For the pure scalar system, the flow equation for the dimensionless functional $\kineticfunctiondimless$ reads 
	\begin{widetext}
	\begin{equation}\label{eq:flowKphi_purematter}
	\begin{aligned}
	k\partial_k \kineticfunctiondimless(\kineticoperatordimless) + 4 \kineticfunctiondimless(\kineticoperatordimless)-  \kineticoperatordimless(4+\scalaranomalousdimension)\kineticfunctiondimless'(\kineticoperatordimless) = \frac{1}{16\pi^3} \int_0^\infty \text{d}z \, z \int_{-1}^{1} \text{d}x\,\sqrt{1-x^2} \left( \tfrac{(2-\scalaranomalousdimension) R_k(z) - 2z R_k'(z)}{z + R_k(z) + z \, f(\kineticoperatordimless, x^2)} - (f\to0) \right) \, ,
	\end{aligned}
	\end{equation}
	\end{widetext}
	where we have defined $z=q^2/k^2$, $f(\kineticoperatordimless,x^2) = -1 + (\kineticfunctiondimless'(\kineticoperatordimless) + 2\,x^2\, \kineticoperatordimless \,\kineticfunctiondimless''(\kineticoperatordimless))$, and the variable $x$ as the cosine of the angle between $D_{\mu}\scalarfielddimless$ and the loop momentum $q_\mu$. The additional terms on the left-hand side in \eqref{eq:flowKphi_purematter} arise from switching to dimensionless quantities, and represent the explicit and implicit mass dimensions of $\kineticfunctiondimful$ and $\kineticoperatordimful$ . See also \cite{Laporte:2022ziz} for different representations of the flow equation. The flow equation \eqref{eq:flowKphi_purematter} is the starting point for the investigation of interacting scalar fixed points that are defined by $k\partial_k \kineticfunctiondimless(\kineticoperatordimless)=0$.
	
	It is clear that \eqref{eq:flowKphi_purematter} admits the scaling solution $\kineticfunctiondimlessFP(\kineticoperatordimless)=\kineticoperatordimless, \scalaranomalousdimension=0$ that corresponds to the \GFP{}, since only the kinetic term is non-vanishing.
	
	The situation is more complicated beyond the \GFP{}, since \eqref{eq:flowKphi_purematter} still has the radial and angular parts of the integration over the loop momentum. The extra dependence on the angle arises as a consequence of non-vanishing derivatives of $\kineticfunctiondimless$ that enter the scalar two-point function. While the angular integration can in principle be done analytically, in practice this is of little use due to the complexity of the primitive function. Due to the non-linear structure of the differential equation, it is difficult to find exact analytical solutions. It is thus more practical to employ an approximation first and then to carry out the integrations.

	\subsection{Expansion in \texorpdfstring{$\kineticoperatordimless$}{X}}
	\label{sec:puremattrho}

	To search analytically for interacting fixed points, and to make contact to previous work, we will first proceed by performing a polynomial expansion of $\kineticfunctiondimless(\kineticoperatordimless)$ in $\kineticoperatordimless$ around $\kineticoperatordimless=0$, \ie,
	\begin{align}\label{eq:polynomial_expansion}
	\kineticfunctiondimless(\kineticoperatordimless) \approx X + \sum_{n=2}^{\Nmax}	\Kcoup{n}\, \kineticoperatordimless^n \,,
	\end{align}
	with $\Nmax$ denoting the maximal order of our truncation, and where we have introduced the dimensionless and scale dependent couplings $\Kcoup{n}$. In general, we can compute the beta function of a coupling $\Kcoup{n}$ by taking $n$ derivatives with respect to $\kineticoperatordimless$ on both sides of the flow equation \eqref{eq:flowKphi_purematter} and projecting the result to $\kineticoperatordimless = 0$. Upon expansion, all angular integrations can be carried out analytically, and we obtain analytical flow equations for the couplings $\Kcoup{n}$ in terms of threshold functions. This expansion has been previously explored to order $\Nmax=14$ (and $\Nmax=20$ when the scalar anomalous dimension was neglected) in \cite{Laporte:2022ziz}.
	
	To extract the scale dependence of the couplings $\Kcoup{n}$, we employ Fa\`a di Bruno's formula \cite{diBRUNO:1855} for derivatives of the right-hand side of \eqref{eq:flowKphi_purematter}. This procedure is efficient since $f(0,x^2)=0$ by our normalisation conditions \eqref{eq:Kfunctconds}. Consequently, we find for $n>1$
	\begin{widetext}
	\begin{equation}
	 \beta_{\Kcoup{n}} = (4(n-1) + n \, \scalaranomalousdimension)\Kcoup{n} + \frac{1}{(2\pi)^3} \sum_{l=1}^n (-1)^l \frac{l!}{n!} \left( \thresholdBK{l+1} - \frac{\scalaranomalousdimension}{2} \thresholdtildeBK{l+1} \right) \int_{-1}^1 \text{d}x \,\sqrt{1-x^2} \, Y_{n,l}\left( \xi_1(x),\cdots , \xi_{n-l+1}(x) \right) \, ,
	\end{equation}
	\end{widetext}
	where $Y_{n,l}$ denotes the Bell polynomials, and where we introduced $\xi_j(x) = (j+1)!\, (1+2j x^2)\,\Kcoup{j+1}$ as well as the threshold functions
	\begin{align}
	 \thresholdBK{n} &= \int_0^{\infty} \mathrm{d}z \, \left( \frac{z}{z+R_k(z)} \right)^n \left( R_k(z) - z R_k'(z) \right) \, , \\
	 \thresholdtildeBK{n} &= \int_0^{\infty} \mathrm{d}z \, \left( \frac{z}{z+R_k(z)} \right)^n R_k(z) \, .
	\end{align}
	For the Litim regulator \eqref{eq:shapeLitim}, these integrals read
	\begin{equation}
	 \thresholdBK{n} = \frac{1}{n+1} \, , \qquad \thresholdtildeBK{n} = \frac{1}{(n+1)(n+2)} \, ,
	\end{equation}
	whereas for the exponential regulator \eqref{eq:shapeExp} we find
	\begin{equation}
	\begin{aligned}
	 \thresholdBK{n} &= \frac{1}{n-1} \left[ \frac{\pi^2}{6} + \left( \gamma+\psi(n) \right)^2 - \psi'(n) \right] \, , \\
	 \thresholdtildeBK{n} &= \frac{1}{n} H_n \, .
	\end{aligned}
	\end{equation}
	Here, $\psi(n)=\Gamma'(n)/\Gamma(n)$ is the polygamma function, $\gamma$ is the Euler-Mascheroni constant, and $H_n$ is the $n$-th harmonic number. Note that for the exponential regulator and $n=1$ we have to take a limit that evaluates to
	\begin{equation}
	 \thresholdBK{1} = 2\zeta(3) \, .
	\end{equation}
	
	The fixed point value of the anomalous dimension computed from the polynomial expansion \eqref{eq:polynomial_expansion} reads
	\begin{equation}\label{eq:etaphi}
	\scalaranomalousdimensionFP = \frac{6 \KcoupFP{2} \thresholdBK{2}}{32 \pi^2 + 3 \KcoupFP{2} \thresholdtildeBK{2}} \, .
	\end{equation}
	This expression is also valid beyond polynomial truncations with the replacement $\KcoupFP{2} \mapsto \kineticfunctiondimlessFP''(0)/2$.\\

	\subsubsection{Fixed point structure}
	Structurally, $\beta_{\Kcoup{n}}$ is linear in $\Kcoup{n+1}$. Specifically, the only Bell polynomial in $\beta_{\Kcoup{n}}$ that depends on $\Kcoup{n+1}$ is that with $l=1$, for which we have
	\begin{equation}
	 \begin{aligned}
	 	Y_{n,1}(z_1(x), \dots, z_n(x)) &= z_n(x) \\
	 	&= (n+1)! (1+2n x^2) \, \Kcoup{n+1} \, .
	 \end{aligned}
	\end{equation}
	We can thus solve $\beta_{\Kcoup{n}}=0$ explicitly for $\Kcoup{n+1}$ in terms of all lower $\Kcoup{i}$:
	\begin{widetext}
	\begin{equation}
	\begin{aligned}
	 \KcoupFP{n+1} &= \frac{32\pi^2}{(n+1)(n+2)} \frac{(-1)^n}{\thresholdBK{2} - \frac{\scalaranomalousdimensionFP}{2} \thresholdtildeBK{2}} \Bigg[ \left( 4(n-1) + n \, \scalaranomalousdimensionFP \right) \KcoupFP{n} \\
	 &\hspace{3cm} + \frac{1}{(2\pi)^3} \sum_{l=2}^n (-1)^l \frac{l!}{n!} \left( \thresholdBK{l+1} - \frac{\scalaranomalousdimensionFP}{2} \thresholdtildeBK{l+1} \right) \int_{-1}^1 \text{d}x \,\sqrt{1-x^2} \, Y_{n,l}\left( z_1(x),\cdots , z_{n-l+1}(x) \right) \Bigg] \, .
	\end{aligned}
	\end{equation}
	\end{widetext}
	
	 \begin{figure*}
	 	\includegraphics[width=.8\linewidth]{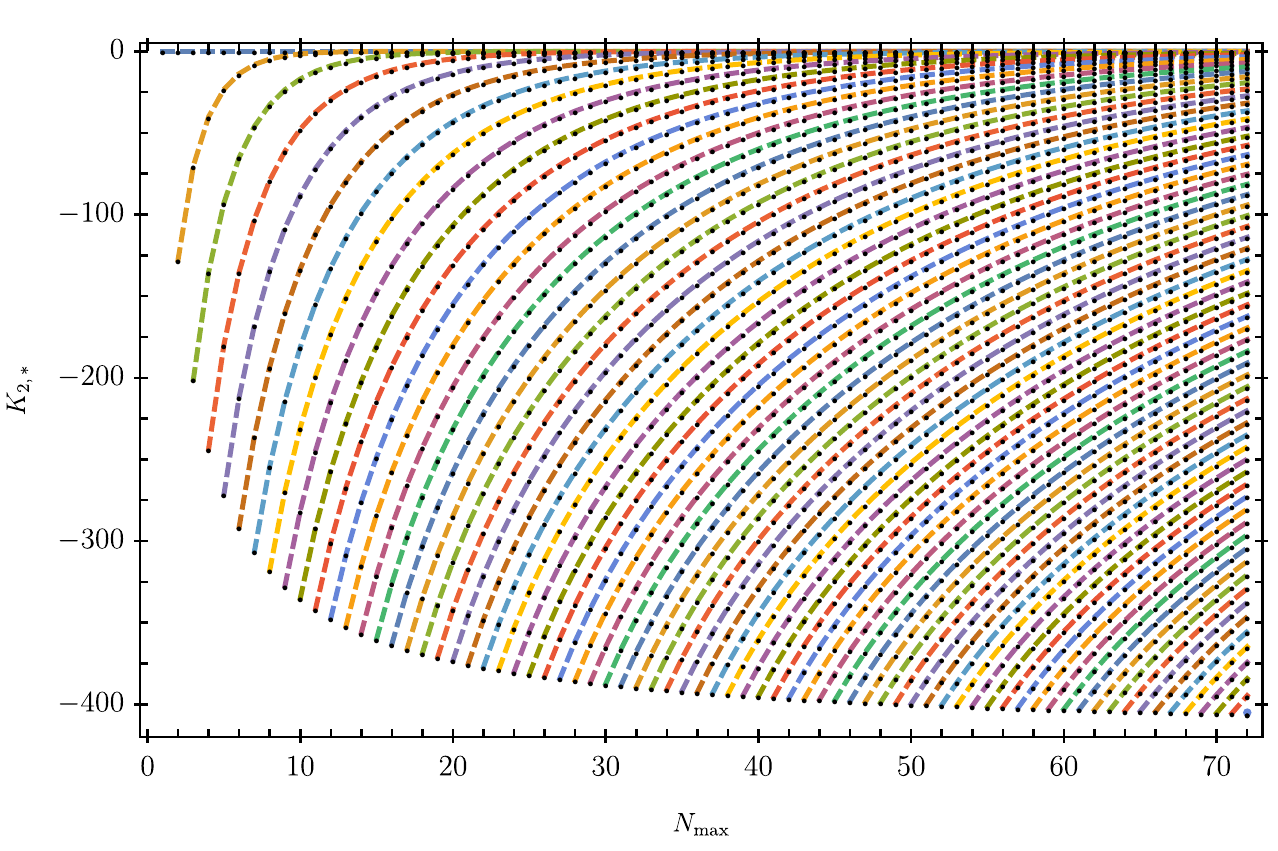}
	 	\caption{Fixed point structure of the pure scalar system for the Litim regulator as a function of the maximal order of the polynomial expansion $\Nmax$, see \eqref{eq:polynomial_expansion}, up to $\Nmax=72$. The black markers indicate the real fixed point values $\KcoupFP{2}$ for a given truncation. The dashed lines indicate the evolution of a given fixed point when increasing \Nmax{}, based on the number of relevant directions.}
	 	\label{fig:PMFPs}
	 \end{figure*}
	
	 Therefore, the set of beta functions $\beta_{\Kcoup{j}}$ for $j<\Nmax$ can be easily solved inductively for the couplings $\Kcoup{l}$ with $2<l\leq \Nmax$. Note that all $\KcoupFP{n}$, $n>2$ depend \emph{polynomially} on $\KcoupFP{2}$. To see this, first note the following relations:
	 \begin{equation}
	 \begin{aligned}
	  \frac{1}{\thresholdBK{2} - \frac{\scalaranomalousdimensionFP}{2} \thresholdtildeBK{2}} &= \frac{1 + \frac{3}{32\pi^2}\thresholdtildeBK{2} \, \KcoupFP{2}}{\thresholdBK{2}} \, , \\
	  \frac{\scalaranomalousdimensionFP}{\thresholdBK{2} - \frac{\scalaranomalousdimensionFP}{2} \thresholdtildeBK{2}} &= \frac{3}{16\pi^2} \KcoupFP{2} \, .
	 \end{aligned}
	 \end{equation}
	 From this it follows that $\KcoupFP{n+1}$ only depends polynomially on all $\KcoupFP{l}$, $2\leq l\leq n$. By induction, it follows then that all $\KcoupFP{n}$, $n>2$ depend polynomially on $\KcoupFP{2}$. For illustration, the first few couplings are
	\begin{align}
\label{eq:FPcondhigherorder}
	 \KcoupFP{3} &= \frac{1}{\thresholdBK{2}-\frac{\scalaranomalousdimensionFP}{2} \thresholdtildeBK{2}} \Bigg[ \frac{32\pi^2}{3} \left( 1 + \frac{\scalaranomalousdimensionFP}{2} \right) \KcoupFP{2} \nonumber \\
	 &\hspace{2.25cm} + \frac{5}{3} \left( \thresholdBK{3} - \frac{\scalaranomalousdimensionFP}{2} \thresholdtildeBK{3} \right) \KcoupFP{2}^2 \Bigg] \, , \\
	 \KcoupFP{4} &= \frac{1}{\thresholdBK{2}-\frac{\scalaranomalousdimensionFP}{2} \thresholdtildeBK{2}} \Bigg[ \frac{64\pi^2}{5} \left( 1 + \frac{3}{8} \scalaranomalousdimensionFP \right) \KcoupFP{3} \nonumber \\
	 &\hspace{2.25cm} + \frac{21}{5} \left( \thresholdBK{3} - \frac{\scalaranomalousdimensionFP}{2} \thresholdtildeBK{3} \right) \KcoupFP{2} \KcoupFP{3} \nonumber \\
	 &\hspace{2.25cm} - \frac{37}{10} \left( \thresholdBK{4} - \frac{\scalaranomalousdimensionFP}{2} \thresholdtildeBK{4} \right) \KcoupFP{2}^3 \Bigg] \, , \\
	 \KcoupFP{5} &=  \frac{1}{\thresholdBK{2}-\frac{\scalaranomalousdimensionFP}{2} \thresholdtildeBK{2}} \Bigg[ \frac{64\pi^2}{5} \left( 1 + \frac{1}{3} \scalaranomalousdimensionFP \right) \KcoupFP{4} \nonumber \\
	 &\hspace{2.25cm} + \frac{24}{5} \left( \thresholdBK{3} - \frac{\scalaranomalousdimensionFP}{2} \thresholdtildeBK{3} \right) \KcoupFP{2} \KcoupFP{4} \nonumber \\
	 &\hspace{2.25cm} + 3 \left( \thresholdBK{3} - \frac{\scalaranomalousdimensionFP}{2} \thresholdtildeBK{3} \right) \KcoupFP{3}^2 \nonumber \\
	 &\hspace{2.25cm} - \frac{81}{5} \left( \thresholdBK{4} - \frac{\scalaranomalousdimensionFP}{2} \thresholdtildeBK{4} \right) \KcoupFP{2}^2 \KcoupFP{3} \nonumber \\
	 &\hspace{2.25cm} + 10 \left( \thresholdBK{5} - \frac{\scalaranomalousdimensionFP}{2} \thresholdtildeBK{5} \right) \KcoupFP{2}^5 \Bigg] \, .
	\end{align}
	Here, we did not insert the expression for \scalaranomalousdimensionFP{} to make it more readable.

	 To close the system, the fixed point solutions for $\KcoupFP{2}$ are then found by setting $\KcoupFP{\Nmax+1}=0$.\footnote{We have also checked the boundary conditions $\KcoupFP{\Nmax+1} = \{\pm 1,\pm 10,\pm 10^2,\pm10^{3}\}$, and we found that the results for fixed points and critical exponents agree quantitatively. See \cite{Falls:2014tra,Bender:2022eze} for discussions on boundary conditions.} Hence, we can distinguish all fixed points of the full system by the fixed point value of $\KcoupFP{2}$. In the following, we will thus investigate the system by studying the different fixed point values $\KcoupFP{2}$ as a function of $\Nmax$. The functional dependence of the fixed point values $\KcoupFP{n\leq72}$ on $\KcoupFP{2}$ can be found in the ancillary notebook.
		
\begin{figure}
		\includegraphics[width=\linewidth]{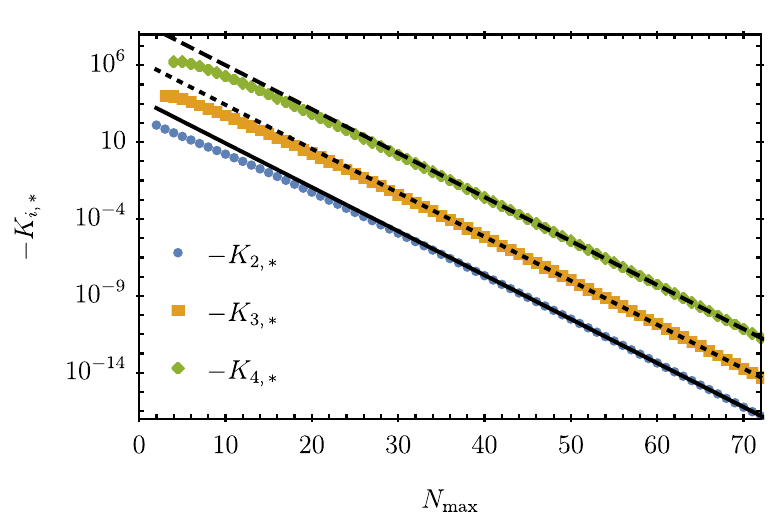}
		\caption{Fixed point values for the couplings $\Kcoup{2}$ (blue circles), $\Kcoup{3}$ (orange boxes), and $\Kcoup{4}$ (green diamonds) as a function of $\Nmax$ for the first interacting fixed point. The solid (dotted, dashed) black line indicates an exponential fit to the last 30 data points. The fitted exponential fall-off agrees for the three different fixed point values.}
		\label{fig:FPPMExp}
	\end{figure}

	In \autoref{fig:PMFPs} we show the fixed point structure of the pure scalar system as a function of $\Nmax$, up to $\Nmax=72$, and with the Litim regulator \eqref{eq:shapeLitim}. The black markers indicate the  fixed point values $\KcoupFP{2}$ at each order in the polynomial expansion. The coloured lines indicate how the fixed point values change when \Nmax{} is increased. For this, fixed points with a fixed number of relevant directions are connected. For example, the interacting fixed point for $\Nmax=2$ features one relevant direction and is connected by the dashed line to the fixed point at $\Nmax=3$ ($\Nmax=4$, $\Nmax=5$, $\dots$) that features one relevant and one (two, three, $\dots$) irrelevant direction(s). \autoref{fig:FPPMExp} shows $\KcoupFP{2,3,4}$ for the first interacting fixed point as a function of the truncation order. We find an exponential fall-off of $\KcoupFP{2}$, and as a consequence of the above analysis, also for all higher order couplings.
	
	Besides the fixed points shown in \autoref{fig:PMFPs}, there is one additional fixed point $\KcoupFP{2}$, as discussed in \cite{Laporte:2022ziz}, that only appears for even $\Nmax$. This fixed point is generated as a consequence of the analytic structure of the equation that determines the anomalous dimension, \eqref{eq:etaphi}. The fixed point is located at a negative value for $\KcoupFP{2}$, beyond the pole in the anomalous dimension \scalaranomalousdimension{}, see \eqref{eq:etaphi}. In particular, \scalaranomalousdimension{} is positive and large at this fixed point, such that some of the regulator properties might be violated \cite{Christiansen:2017cxa}. Furthermore, since it is separated from the other fixed points by a pole in the anomalous dimension, it cannot collide with one of those fixed points as a function of \Gdimless{}, or connect to a low-energy regime with small \Kcoup{2}. We will not discuss this fixed point further in the following.
	\begin{figure}
		\includegraphics[width=\linewidth]{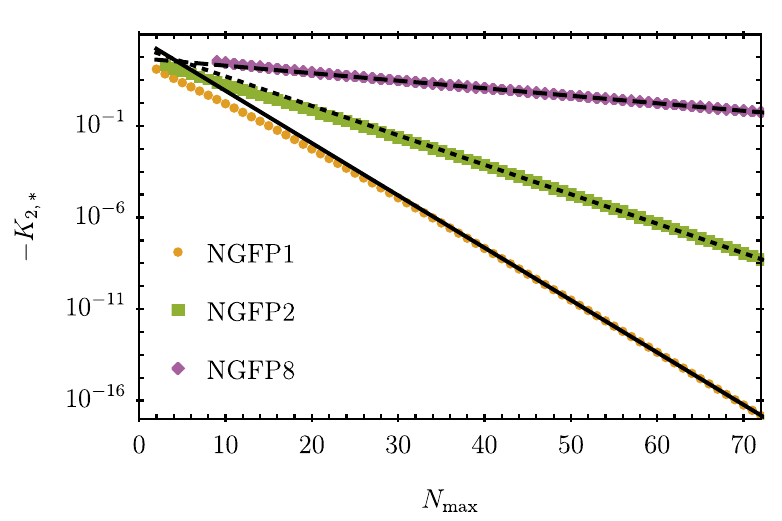}
		\caption{Fixed point values $\KcoupFP{2}$ for the first (orange circles), second (green boxes) and eighth (purple diamonds) interacting fixed point as a function of $\Nmax$. The solid (dotted, dashed) black line indicates an exponential fit to the last 30 data points. The fall-off remains exponential, but becomes slower for the more interacting fixed points.}
		\label{fig:FPPMExpv2}
	\end{figure}
	
	\begin{figure*}
		\includegraphics[width=.8\linewidth]{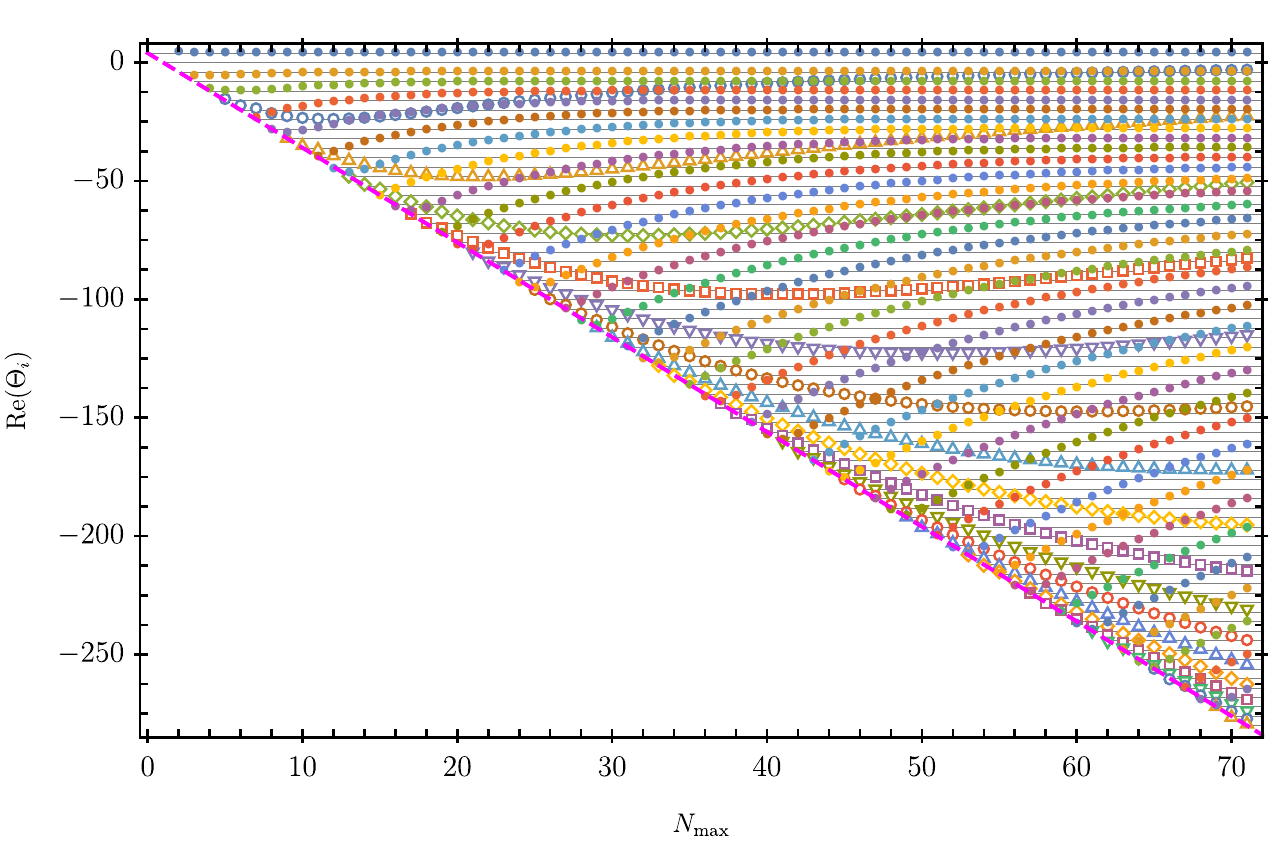}
		\caption{Critical exponents of the first interacting pure scalar fixed point as a function of the truncation order $\Nmax$. Open markers indicate a complex-conjugate pair of critical exponents. The horizontal lines indicate the spacing of $-4$ between critical exponents, which is expected at the \GFP{}. Furthermore, the magenta dashed line is given by $y=4-4\Nmax$, and indicates the canonical mass dimension of the canonically most irrelevant coupling added in a given truncation.}
		\label{fig:PMCEs}
	\end{figure*}

	As can be seen in \autoref{fig:FPPMExpv2}, the fixed point values $\KcoupFP{2}$ approach zero exponentially quickly as a function of $\Nmax$, see also \cite{Laporte:2022ziz}. This is not only true for the first, but also for the other interacting fixed points. As we discuss in \aref{app:pmSpuriousDep}, this fixed point structure of the system is also found with an exponential regulator. In particular, the exponential fall-off of fixed point values $\KcoupFP{2}(\Nmax)$ agrees on a quantitative level between both regulators.

	\subsubsection{Critical exponents and eigenvectors}
	Fixed point values are not universal quantities and can be changed by, \eg{}, rescalings of the couplings. By contrast, critical exponents are universal, and determine the universality class of a fixed point.  In the following, we focus on the least strongly interacting fixed point that only has one relevant direction, \ie{}, the fixed point that appears already at order $\Nmax=2$. We will discuss the structure of critical exponents first, and then comment on the eigenvectors corresponding to certain critical exponents.
	
	In \autoref{fig:PMCEs} we show the set of critical exponents for this interacting fixed point. There are three different sets of critical exponents: one positive, \ie{}, relevant critical exponent, a set of real-valued negative critical exponents, and a set of complex-conjugate pairs of critical exponents with negative real parts. 
	
The relevant direction is already present at the lowest order in the expansion, \ie, $\Nmax=2$, see \cite{Eichhorn:2012va, deBrito:2021pyi, Laporte:2021kyp, Knorr:2022ilz, Laporte:2022ziz}. At this level, it arises simply as a consequence of the quadratic form of $\beta_{\Kcoup{2}}$. When increasing $\Nmax$, the positive critical exponent converges quickly to $\Theta_1=4$, and agrees exactly with this value (up to 18 digits precision) at $\Nmax=71$.
	
	The negative critical exponents are approximately bounded from below by the canonical mass dimension of the highest order operator in a given truncation. Indeed, above $\Nmax=7$ the most irrelevant critical exponent does not deviate from the canonical mass dimension of the highest order operator by more than $3\%$, see the diagonal, dashed magenta line in \autoref{fig:PMCEs}. When increasing \Nmax{}, the negative critical exponents increase and seem to converge from below to integer values spaced by  $\delta\Theta=-4$, which corresponds to the spacing of canonical mass dimensions. In particular, at $\Nmax=71$, the first 10 real and negative critical exponents deviate less than one percent from the canonical mass dimension of the 10 lowest order couplings $\Kcoup{2}, \dots, \Kcoup{11}$, and the pattern continues with less precision to more negative critical exponents.

\begin{figure*}
	\begin{centering}	
		\includegraphics[width=.8\linewidth]{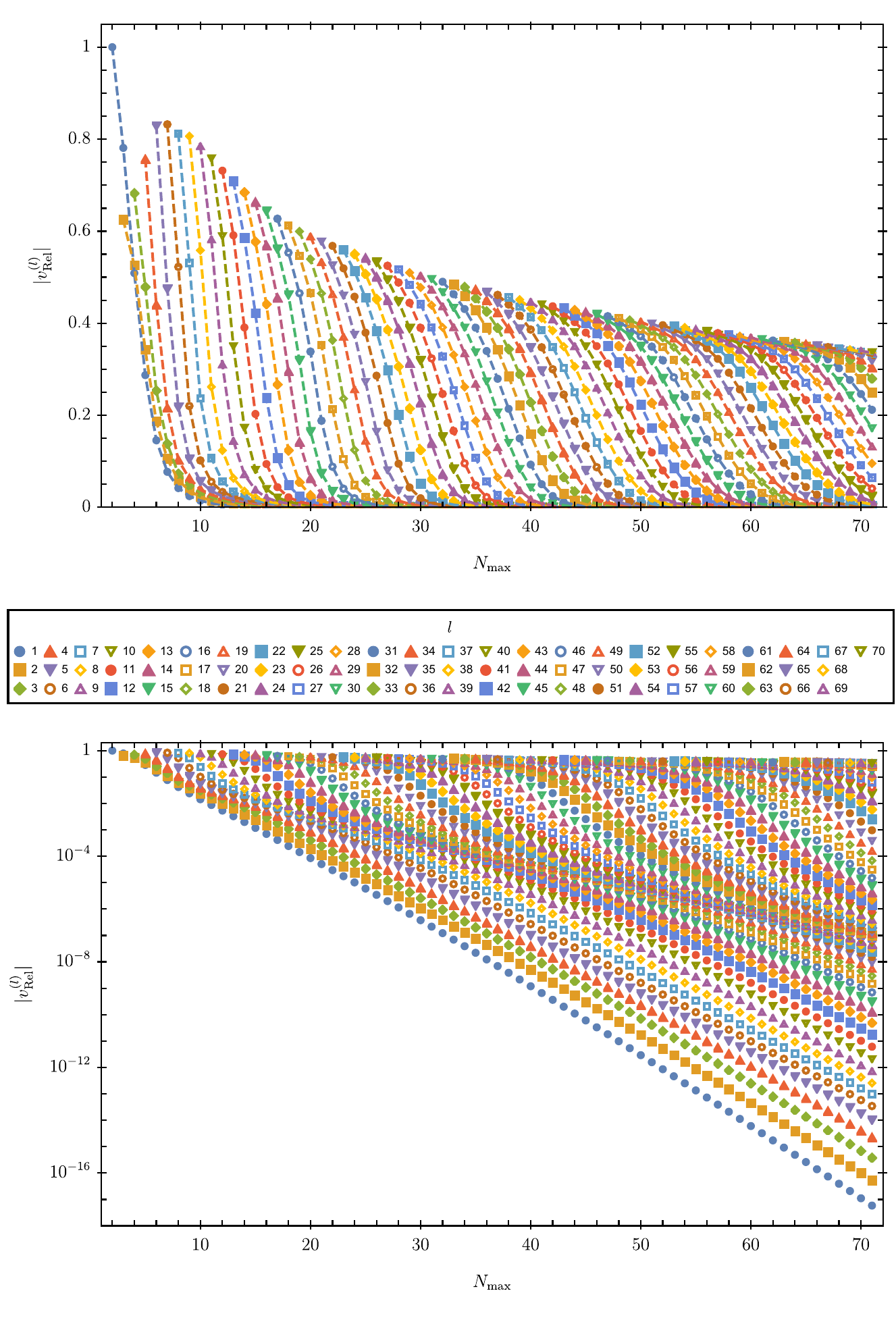}
	\end{centering}
	\caption{Components of the eigenvector corresponding to the relevant direction of the first interacting scalar fixed point for different truncations, on a linear (upper panel) and a logarithmic (lower panel) scale. The highest retained operator is almost always the most dominant one, but the contribution of all operators decreases when increasing $\Nmax$.}
	\label{fig:EigVectsRelLitim}
\end{figure*}
	
	The complex-conjugate pairs are denoted by open markers in \autoref{fig:PMCEs}. They do \emph{not} show convergence up to the explored order $\Nmax=71$. Indeed, the real part of the least irrelevant complex-conjugate pair still changes its value at this order in the truncation. Furthermore, it approaches $\text{Re}(\Theta)=0$, such that at larger \Nmax{}, new relevant directions might arise. We were however not able to increase \Nmax{} far enough to investigate whether the pair of critical exponents actually moves to positive real parts.
		
	We also investigated the critical exponents obtained with the exponential regulator \eqref{eq:shapeExp}. For $\Nmax=2$, the critical exponent obtained with the latter deviates by about $4\%$ from the value obtained with the Litim regulator. When increasing \Nmax{}, the deviation of all critical exponents between both regulators decreases. At $\Nmax=70$, all critical exponents deviate less than $5 \times 10^{-9}$ between the two regulators, including the most irrelevant critical exponent. In \aref{app:pmSpuriousDep} we present a more detailed comparison between the results obtained with the Litim and the exponential regulator.

\begin{figure*}
	\begin{centering}	
		\includegraphics[width=.8\linewidth]{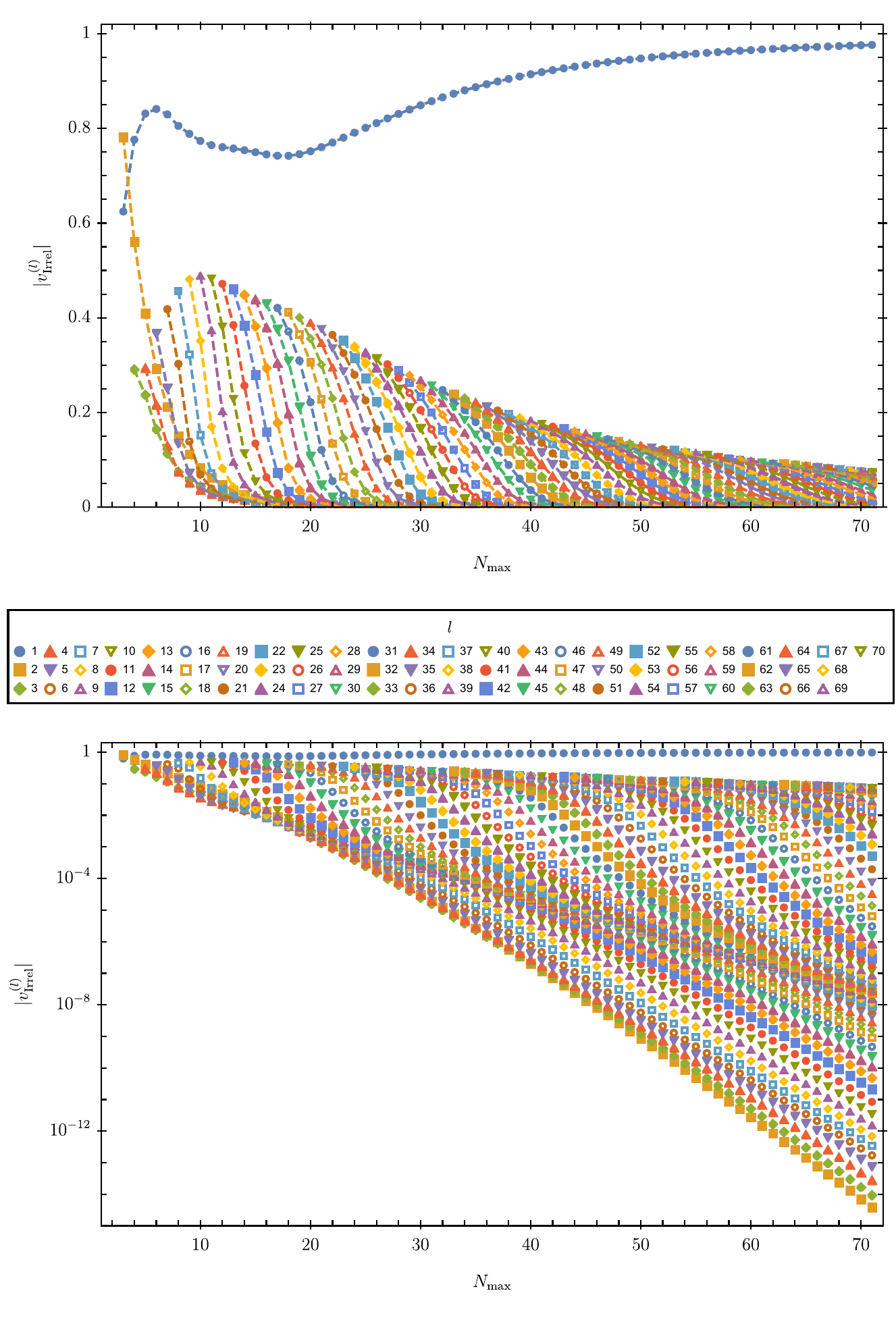}
	\end{centering}
	\caption{Components of the eigenvector corresponding to the first irrelevant direction of the first interacting scalar fixed point for different truncations, on a linear (upper panel) and a logarithmic (lower panel) scale. One can clearly see the dominant overlap with the operator $\kineticoperatordimless^2$, which converges to unity, whereas the overlap with all other operators goes to zero.}
	\label{fig:EigVectsIrrelLitim}
\end{figure*}

	We will now discuss the eigenvectors of the relevant direction, the first irrelevant direction and one complex-conjugate pair of critical exponents. Eigenvectors are non-universal quantities, and can therefore be changed by rescalings of the couplings. Nevertheless, they indicate the couplings  overlapping most with relevant or irrelevant directions.
	For a converged and controlled approximation scheme, we require that the critical exponents appearing at low orders in the truncation only overlap with some of the lower order couplings, and almost not at all with the higher order ones.
	
	When computing the eigenvectors of a system, it can happen that each eigenvector is dominated by the same component. To interpret the system of eigenvectors better, we employ the rescaling of couplings proposed in \cite{Kluth:2020bdv}. It ensures that each coupling contributes equally to the system of eigenvectors. Accordingly, both rows and columns of the matrix of eigenvectors are normalised to one. After this rescaling, we can directly compare the eigenvectors, and conclude which operator is the most important for a given critical exponent. We emphasise that the normalisation procedure is only a rescaling of the couplings and does not involve linear combinations of different couplings.
	
	In \autoref{fig:EigVectsRelLitim}, we show the absolute values of all components of the eigenvector $v_{\mathrm{Rel}}$ corresponding to the relevant direction, after the rescaling procedure. The component $v^{(l)}_{\mathrm{Rel}}$ points in the direction of $\Kcoup{l+1}$. We can see that, except for small \Nmax{}, $v_{\mathrm{Rel}}$ always points most dominantly in the direction of $\Kcoup{\Nmax}$. Furthermore, the overlap with couplings $\Kcoup{l}$ for $l<\Nmax$ decreases rapidly when increasing \Nmax{}. Hence, the canonically most irrelevant coupling has the largest overlap with the only relevant direction. This indicates that the fixed point, if it exists, is highly non-perturbative, despite the exponentially decreasing fixed point values. Furthermore, since in every truncation, the canonically most irrelevant coupling dominates the relevant eigendirection, it is questionable how reliably the polynomial expansion of $\kineticfunctiondimless(\kineticoperatordimless)$ can describe such a fixed point.
	
	In \autoref{fig:EigVectsIrrelLitim} we show the absolute values of all components of the eigenvector $v_{\mathrm{Irrel}}$ corresponding to the first irrelevant direction, \ie, the eigenvector corresponding to the critical exponent converging to $\Theta_{\mathrm{Irrel}}=-4$. We can see that for $\Nmax\geq3$, the component $l=1$, which points in the direction of $\Kcoup{2}$, dominates $v_{\mathrm{Irrel}}$. Furthermore, this contribution increases when increasing \Nmax{}, while the total contribution of all components with $l\neq1$ to $v_{\mathrm{Irrel}}$ decreases. In particular, all contributions to higher $l$ decrease rapidly when further increasing $\Nmax$. This indicates that the irrelevant direction with $\Theta_{\mathrm{Irrel}}\approx-4$ exactly corresponds to the direction of $\Kcoup{2}$ in theory space. This is not surprising, since this critical exponent corresponds to the canonical mass dimension of $\Kcoup{2}$, together with the fact that $\KcoupFP{2}$ is falling off exponentially as a function of $\Nmax$. A similar behaviour is present for the eigendirections corresponding to the critical exponent that approach $\Theta_{\mathrm{Irrel}}=-4n$, which at large \Nmax{} are dominated by the component pointing into the direction of $\Kcoup{n+1}$.
	
	For the eigenvector corresponding to the least irrelevant complex-conjugate pair of critical exponents, we see a similar behaviour as for the relevant direction: the overlap with the canonically most relevant couplings decreases quickly when increasing $\Nmax{}$. Furthermore, the eigenvector is dominated by the component pointing in the direction of the canonically most irrelevant coupling. Hence, this complex-conjugate pair of eigenvectors constitutes another example for an eigendirection that appears for relatively small $\Nmax$, but seems to be dominated by the canonically most irrelevant operator of the system.

	\subsubsection{Expansion around \texorpdfstring{$\Kcoup{2}=0$}{K2=0}}

	We have seen that the value of $\KcoupFP{2}$ at the interacting fixed points decreases exponentially when increasing $\Nmax$, and that all higher order couplings are determined in terms of $\KcoupFP{2}$. As a consequence, for sufficiently large \Nmax{}, it is enough to keep the term linear in $\KcoupFP{2}$ in these expressions to accurately represent the full fixed point function \kineticfunctiondimlessFP{}. For example, at $\Nmax=72$, for the first interacting fixed point and with the Litim regulator, the linearised fixed point solutions of $\KcoupFP{n}$ up to $n=66$ deviate less than $1\%$ from the full result.
	
	It turns out that we can actually find a closed form expression for the linear part with the help of the Mathematica function \emph{FindSequenceFunction}. This procedure leads to the relation
	\begin{equation}\label{eq:pochhammer}
	 \KcoupFP{n} = 3\,\KcoupFP{2}\frac{2^{7 n -12}\, \pi ^{2 n-4}}{n (n-1) \Gamma(n+2) (\thresholdBK{2})^{n-2}} + \mathcal O(\KcoupFP{2}^2) \, .
	\end{equation}
	We can then resum the Taylor expansion \eqref{eq:polynomial_expansion} to linear order in \KcoupFP{2}. We find
	\begin{widetext}
	\begin{equation}
	\label{eq:fullkresumlin}
	\begin{aligned}
	 \kineticfunctiondimlessFP(\kineticoperatordimless) &= \kineticoperatordimless - \frac{3\KcoupFP{2}}{\kineticoperatordimless} \left( \frac{\thresholdBK{2}}{128\pi^2} \right)^3 \Bigg[ 2 - 2 e^{\frac{128\pi^2}{\thresholdBK{2}}\kineticoperatordimless{}} \left( 1 - \frac{128\pi^2}{\thresholdBK{2}}\kineticoperatordimless{} \right) - 5 \left( \frac{128\pi^2}{\thresholdBK{2}}\kineticoperatordimless{} \right)^2 \\
	 &\hspace{4cm} + 2 \frac{128\pi^2}{\thresholdBK{2}}\kineticoperatordimless{} \left( 2 - \frac{128\pi^2}{\thresholdBK{2}}\kineticoperatordimless{} \right) \left( \text{Ei}\left( \frac{128\pi^2}{\thresholdBK{2}}\kineticoperatordimless{} \right) - \ln \left( \frac{128\pi^2}{\thresholdBK{2}}\kineticoperatordimless{} \right) - \gamma \right) \Bigg] + \mathcal O(\KcoupFP{2}^2) \, .
	\end{aligned}
	\end{equation}
	\end{widetext}
	Here Ei is the exponential integral function. For large \kineticoperatordimless{}, this function grows exponentially,
	\begin{equation}
	 \kineticfunctiondimlessFP(\kineticoperatordimless) \sim \frac{12\KcoupFP{2}}{\kineticoperatordimless^3} \left( \frac{\thresholdBK{2}}{128\pi^2} \right)^5 e^{\frac{128\pi^2}{\thresholdBK{2}}\kineticoperatordimless{}} + \mathcal O(\KcoupFP{2}^2) \, .
	\end{equation}
	This behaviour is rather dubious, since it is fundamentally inconsistent with the flow equation \eqref{eq:flowKphi_purematter}: if the solution were to grow exponentially, the right-hand side would fall off exponentially quickly, but the left-hand side would still grow exponentially. On the other hand, $\KcoupFP{2}$ tends to zero, so there are competing effects for large \kineticoperatordimless{}. For this reason, we shall investigate this limit subsequently in more depth.

	\subsection{Perturbations about the GFP}
	
	We have observed that for all interacting fixed points, $\KcoupFP{2}\to0$ exponentially quickly when increasing \Nmax{}. 
	This motivates investigating the linearised flow about the \GFP{} in more detail to gain analytical insights into its critical exponents and eigenvectors. To this end, we start from the flow equation \eqref{eq:flowKphi_purematter} for $\kineticfunctiondimless(\kineticoperatordimless)$ and expand it around the \GFP, \ie,
	\begin{equation}
	\label{eq:linpertPM}
	\begin{aligned}
	\kineticfunctiondimless(\kineticoperatordimless) &= X + \epsilon\, e^{-\Theta\,t} \,\delta \kineticfunctiondimless(\kineticoperatordimless)\,,\\
	\scalaranomalousdimension &= 0 + \epsilon\,e^{-\Theta\,t}\,\delta\scalaranomalousdimension\,,
	\end{aligned}
	\end{equation}
	where $t=\ln k/k_0$, $k_0$ is some reference scale, and $\Theta$ is the critical exponent corresponding to an eigenperturbation $\left\{\delta \kineticfunctiondimless(\kineticoperatordimless), \delta\scalaranomalousdimension \right\}$. Expanding the resulting perturbed flow equation up to linear order in $\epsilon$, the zeroth order vanishes by construction. To order $\epsilon$, the angular integration can be performed analytically, and the flow equation \eqref{eq:flowKphi_purematter} with perturbations \eqref{eq:linpertPM} results in 
	\begin{equation}\label{eq:linpertPMpertvanilla}
	\begin{aligned}
	 (4-\Theta)\,\delta \kineticfunctiondimless(\kineticoperatordimless) - &4\kineticoperatordimless\,\delta \kineticfunctiondimless'(\kineticoperatordimless) - \kineticoperatordimless \, \delta\scalaranomalousdimension \\
	 &= -\frac{2\delta \kineticfunctiondimless'(\kineticoperatordimless) + \kineticoperatordimless \, \delta \kineticfunctiondimless''(\kineticoperatordimless)}{32\pi^2} \, \thresholdBK{2} \, .
	\end{aligned}
	\end{equation}
	Note that $\thresholdBK{2}$ is always positive. To simplify the analysis, it is convenient to rescale the variable \kineticoperatordimless{} and the perturbations such that
	\begin{equation}
	 \kineticoperatordimless = \frac{\thresholdBK{2}}{128\pi^2} \GFPvariable \, , \, \delta \kineticfunctiondimless(\kineticoperatordimless) = \delta \hat\kineticfunctiondimless(\GFPvariable) \, , \,\delta\scalaranomalousdimension = \frac{128\pi^2}{\thresholdBK{2}} \delta\scalaranomalousdimensionhat \, .
	\end{equation}
	With these redefinitions, \eqref{eq:linpertPMpertvanilla} reads
	\begin{equation}\label{eq:linpertPMpert}
	\begin{aligned}
	 (4-\Theta) \delta\hat\kineticfunctiondimless(\GFPvariable) - &4\GFPvariable \, \delta\hat\kineticfunctiondimless'(\GFPvariable) - \GFPvariable \, \delta\scalaranomalousdimensionhat \\
	 &= -4 \left( 2\delta\hat\kineticfunctiondimless'(\GFPvariable) + \GFPvariable \, \delta \hat\kineticfunctiondimless''(\GFPvariable) \right) \, .
	\end{aligned}
	\end{equation}
	We can see that all regulator dependence is absorbed in the rescaling of the field, hence the critical exponents will be regulator-independent. 
	
	Mathematically, \eqref{eq:linpertPMpert} is a linear second-order differential equation for the perturbation $\delta \hat\kineticfunctiondimless$, where the term including $\delta\scalaranomalousdimensionhat$ is the inhomogeneous part. We can eliminate this inhomogeneous part by shifting
	\begin{equation}
	\label{eq:eigpertshift}
	 \delta \hat\kineticfunctiondimless(\GFPvariable) = \delta \tilde\kineticfunctiondimless(\GFPvariable) - \frac{\delta\scalaranomalousdimensionhat}{\Theta} \left( \frac{8}{\Theta-4} + \GFPvariable \right) \, .
	\end{equation}
	For this to be well-defined, we have to assume that $\Theta\notin\{0,4\}$. These two cases will be discussed separately below. With this shift, \eqref{eq:linpertPMpert} reads
	\begin{equation}
	 -4\GFPvariable \, \delta\tilde \kineticfunctiondimless''(\GFPvariable) + 4(\GFPvariable-2) \delta \tilde \kineticfunctiondimless'(\GFPvariable) + (\Theta-4) \delta\tilde \kineticfunctiondimless(\GFPvariable) = 0 \, .
	\end{equation}
	This equation can be brought into Sturm-Liouville form, which makes it easy to formulate conditions so that the resulting spectrum of critical exponents is discrete. We can write this equation as
	\begin{equation}\label{eq:PMpertrescaled}
	 \partial_\GFPvariable \left[ \SLp(\GFPvariable) \, \delta\tilde \kineticfunctiondimless'(\GFPvariable) \right] = -\SLEV \, \SLmeasure(\GFPvariable) \, \delta\tilde \kineticfunctiondimless(\GFPvariable) \, ,
	\end{equation}
	where
	\begin{equation}
	 \SLp(\GFPvariable) = \GFPvariable^2 e^{-\GFPvariable} \geq 0 \, , \quad \SLmeasure(\GFPvariable) = \GFPvariable \, e^{-\GFPvariable} \geq 0 \, , \quad \SLEV = 1 - \frac{\Theta}{4} \, .
	\end{equation}
	By standard Sturm-Liouville theory, we can thus expect a discrete spectrum for \SLEV{} that is bounded from below, each with a unique normalisable eigenfunction with $n$ zeroes. In particular, the set of eigenfunctions forms an orthonormal basis for the Hilbert space $\mathcal L = L^2([0,\infty),\SLmeasure(\GFPvariable)\text{d}\GFPvariable)$.
	Incidentally, we can write down the exact general solution to \eqref{eq:PMpertrescaled}. It reads
	\begin{equation}
	 \delta \tilde \kineticfunctiondimless(\GFPvariable) = c_1 \, {}_1F_1 \left( \frac{\Theta}{4}-1 ; 2 \bigg| \GFPvariable \right) + c_2 \, G_{1,2}^{2,0}\left( \GFPvariable \bigg| \begin{array}{c}
	2-\frac{\Theta}{4} \\
	-1,0 \\
	\end{array} \right) \, ,
	\end{equation}
	where $c_{1,2}$ are constants, ${}_1F_1$ is a hypergeometric function, and $G$ is the Meijer G function. To achieve regularity at $\GFPvariable=0$, we have to set $c_2=0$. Moreover, to achieve normalisability with respect to the measure $\SLmeasure(\GFPvariable)$, we have to investigate the asymptotic behaviour of the hypergeometric function. Generically,  the leading term reads
	\begin{equation}
	 {}_1F_1 \left( \frac{\Theta}{4}-1 ; 2 \bigg| \GFPvariable \right) \sim \frac{1}{\Gamma\left(\frac{\Theta}{4}-1\right)} \GFPvariable^{\frac{\Theta}{4}-3} \, e^\GFPvariable \, , \qquad \GFPvariable \to \infty \, .
	\end{equation}
	This indicates that generally these functions are not normalisable on $\mathcal L$, except if the critical exponent is quantised as
	\begin{equation}
	 \Theta = 4-4n \, , \qquad n \in \mathbbm N \,\,\mathrm{and}\,\,n>1\, .
	\end{equation}
	In this case, the hypergeometric function is just a polynomial, however recall that for the moment we have to exclude $n=0,1$ since in these cases the shift in \eqref{eq:eigpertshift} is singular.
	Let us stress again that it is the physical condition of having a discrete spectrum of critical exponents that selects the correct set of eigenfunctions, and transforms the equation into a regular Sturm-Liouville problem. This automatically excludes any exponentially growing perturbations.
	To completely fix the perturbations $\delta \hat\kineticfunctiondimless$, we still have to impose our boundary conditions, namely
	\begin{equation}\label{eq:PMBC}
	 \delta \hat \kineticfunctiondimless(0) = \delta \hat \kineticfunctiondimless'(0) = 0 \, .
	\end{equation}
	This uniquely and consistently fixes $\delta\scalaranomalousdimensionhat$ to be
	\begin{equation}
	 \delta\scalaranomalousdimensionhat = \frac{\Theta(\Theta-4)}{8} \, c_1 \, .
	\end{equation}
	As it must be, we are still left with the overall multiplying constant $c_1$ that can be used to normalise the eigenfunctions. Note also that this specific form cancels the divergencies for $\Theta=0,4$ in the rescaling \eqref{eq:eigpertshift}, however the corresponding eigenfunctions then vanish identically for these two cases, so we still have to treat these two cases more carefully. We will do this next.

	Let us first discuss the case $\Theta=4$. For this, we implement the shift
	\begin{equation}
	 \delta \hat\kineticfunctiondimless(\GFPvariable) = \delta \tilde\kineticfunctiondimless(\GFPvariable) - \frac{\delta\scalaranomalousdimensionhat}{4} \left( -\frac{2}{\GFPvariable} + \GFPvariable + 2\ln \GFPvariable \right) \, .
	\end{equation}
	This yields the differential equation
	\begin{equation}
	 \GFPvariable \, \delta \tilde \kineticfunctiondimless''(\GFPvariable) + (2-\GFPvariable) \delta \tilde \kineticfunctiondimless'(\GFPvariable) = 0 \, .
	\end{equation}
	This has the simple general solution
	\begin{equation}
	 \delta \tilde \kineticfunctiondimless(\GFPvariable) = c_2 + c_1 \left( -\frac{1}{\GFPvariable} e^\GFPvariable + \text{Ei}(\GFPvariable) \right) \, .
	\end{equation}
	Imposing regularity and boundary conditions, we can fix
	\begin{equation}
	 \delta\scalaranomalousdimensionhat = 2 c_1 \, , \qquad c_2 = \left( 1 - \gamma \right) c_1 \, .
	\end{equation}
	This perturbation is not normalisable in $\mathcal L$ due to the asymptotic behaviour
	\begin{equation}
	 \delta \hat\kineticfunctiondimless(\GFPvariable) \sim \frac{c_1}{y^2} \,e^y \, , \qquad \GFPvariable \to \infty \, .
	\end{equation}
	
	We can finally discuss the last case, $\Theta=0$. In this case we have to introduce the shift
	\begin{equation}
	 \delta \hat\kineticfunctiondimless(\GFPvariable) = \delta \tilde\kineticfunctiondimless(\GFPvariable) - \frac{\delta\scalaranomalousdimensionhat}{4} \left( \frac{1-5\GFPvariable}{\GFPvariable} - (2-\GFPvariable) \ln \GFPvariable \right) \, .
	\end{equation}
	With this shift, we arrive at
	\begin{equation}
	 \GFPvariable \, \delta \tilde \kineticfunctiondimless''(\GFPvariable) + (2-\GFPvariable) \delta \tilde \kineticfunctiondimless'(\GFPvariable) + \delta \tilde \kineticfunctiondimless(\GFPvariable) = 0 \, .
	\end{equation}
	The general solution for this equation reads
	\begin{equation}
	 \delta \tilde \kineticfunctiondimless(\GFPvariable) = c_1 (\GFPvariable-2) + \frac{c_2}{2\GFPvariable} \left( (1-\GFPvariable)e^\GFPvariable + \GFPvariable(\GFPvariable-2) \text{Ei}(\GFPvariable) \right) \, .
	\end{equation}
	Once again we have to impose regularity and boundary conditions at vanishing field. This fixes
	\begin{equation}
	 \delta\scalaranomalousdimensionhat = 2c_2 = \frac{8}{5-2\gamma} c_1 \, .
	\end{equation}
	Imposing these conditions, the resulting perturbation is not normalisable in $\mathcal L$ since it grows exponentially,
	\begin{equation}
	 \delta \hat\kineticfunctiondimless(\GFPvariable) \sim \frac{4c_1}{(5-2\gamma) \GFPvariable^3} e^\GFPvariable \, , \qquad \GFPvariable \to \infty \, .
	\end{equation}
	It thus has to be discarded.

	From this analysis we conclude that, if we restrict perturbations to lie in the Hilbert space $\mathcal L$ while also imposing the boundary conditions \eqref{eq:PMBC}, we find the expected spectrum
	\begin{equation}
	 \Theta_n = -4n \, , \qquad n \in \mathbbm N \, , \qquad n\geq1 \, .
	\end{equation}

Summarising the two special cases, we conclude that the two eigenperturbations corresponding to $\Theta=0,4$ are not polynomials, and they grow exponentially for large $\GFPvariable$. Hence, these eigenperturbations are not part of the Hilbert space $\mathcal L$.

\subsection{Summary and discussion of the pure scalar system}

Before coupling the system to gravity, let us briefly summarise and discuss our findings in the pure scalar system.

As a first step towards computing the scale dependence of the function $\kineticfunctiondimless(\kineticoperatordimless)$, we employed the polynomial expansion  \eqref{eq:polynomial_expansion} to different orders $\Nmax$ in $\kineticoperatordimless$. Due to the structure of the scale dependence of the expansion coefficients $\Kcoup{n}$, we were able to characterise all fixed point solutions of the full system in terms of the fixed point value of $\KcoupFP{2}$. The fixed point values of all other couplings are then polynomials in $\KcoupFP{2}$, see, \eg, \eqref{eq:FPcondhigherorder}. To order $\Nmax$, we find that the system admits $\Nmax$ fixed points, see \autoref{fig:PMFPs}. The fixed point values $\KcoupFP{2}$, and therefore also those of the higher-order couplings, of all fixed points decreases exponentially fast when increasing $\Nmax$, see \autoref{fig:FPPMExp} and \autoref{fig:FPPMExpv2}.

Focussing on the first interacting fixed point, we find that it features one relevant direction, whose critical exponent approaches $\Theta_1=4$. This eigenvalue has the largest overlap with the operator corresponding to the coupling $\Kcoup{\Nmax}$, \ie{}, the canonically most irrelevant operator of the system. The negative critical exponents of the first interacting fixed point approach $\Theta_2\approx-4$, $\Theta_3\approx-8$, $\dots$, at least for the first 10 negative critical exponents. These critical exponents overlap most dominantly with the operators corresponding to the couplings $\Kcoup{2}$,  $\Kcoup{3}$,$\dots$, and they approximate their respective canonical mass dimension. Furthermore, the fixed point features complex-conjugate pairs of critical exponents whose real part does not show convergence up to $\Nmax=71$. A similar pattern is found for the other interacting fixed points with more relevant directions.

Motivated by the finding that $\KcoupFP{2}$, and therefore  $\kineticfunctiondimlessFP(\kineticoperatordimless)-\kineticoperatordimless$, approaches zero exponentially fast, we linearised the fixed point equations for all $\Kcoup{n}$ with $n\geq3$ in $\KcoupFP{2}$. In this approximation, we can find a closed form of the fixed point value for any $\KcoupFP{n}$, see \eqref{eq:pochhammer}. Crucially, this allows us to resum the expansion in $\kineticoperatordimless$, which gives \eqref{eq:fullkresumlin}. However, the resulting approximate fixed point solution $\kineticfunctiondimlessFP(\kineticoperatordimless)$ grows exponentially in $\kineticoperatordimless$, which leads to inconsistencies in the flow equation.

To investigate this behaviour further, we explored the eigenperturbations of the system about the \GFP{} by expanding the flow equation to linear order in perturbations. For critical exponents $\Theta\notin\{0,4\}$, the differential equation for the perturbations can be brought into Sturm-Liouville form, see \eqref{eq:PMpertrescaled}, which immediately defines an integral measure with respect to which viable eigenperturbations have to be square-integrable. This leads to a quantisation of critical exponents as $\Theta = 4-4n$ with $ n \in \mathbbm N$. Furthermore, an explicit analysis of the cases $\Theta\in\{0,4\}$ shows that both corresponding eigenperturbations are not square-integrable with respect to the Sturm-Liouville measure. They hence need to be discarded.

At this point, we can finally make the connection between the linear analysis about the \GFP{} and the fixed points found in the previous subsection. All of the latter approach the \GFP{} for increasing \Nmax{}, and the leading order Taylor series coefficients are \eqref{eq:pochhammer}. On the other hand, it is clear that these linearised coefficients sum up to the eigenfunction of the \GFP{} corresponding to $\Theta=0$. From the above analysis, we thus conclude that these fixed points do not lie in the space spanned by the perturbations of the \GFP{} that are in the Hilbert space $\mathcal L$. As a matter of fact, the situation is very similar to that of the well-known Halpern-Huang interactions \cite{Halpern:1994vw, Halpern:1995vf, Halpern:1996dh}. If one were to include such interactions, many of the well-known properties of the \RG{} flow would not hold anymore. For example, as we discussed above, critical exponents might not be quantised anymore. Subsequently, we thus follow \cite{Morris:1994ki, Morris:1996nx, Morris:2022rvd} and discard these fixed points.

This leads us to one of the main results of our investigation: at least in the truncation that we investigated, there cannot be a \WGB{} in the gravity-scalar system induced by a collision of partial fixed points. The reason is that the potential collision partners for the \SGFP{} seem to be spurious, that is they are artefacts of finite order truncations. Of course, this does not exclude the general existence of a \WGB{} due to other mechanisms, or the non-existence of a combined gravity-scalar \UV{} completion. As a matter of fact, in the next section we will introduce a new notion of a \WGB{} that separates the weak from the strong gravity regime in terms of the number of relevant operators. Within this notion, in the strong gravity regime there is still a \UV{} completion via the \SGFP{}, but it has more relevant directions than the free theory.

\section{Gravity-scalar system}\label{sec:gravmatt}

We will now move on to discuss the full gravity-scalar system described by the action \eqref{eq:genaction}, extending previous work on shift-symmetric scalar fields by investigating the fixed point structure for the function $\kineticfunctiondimless(\kineticoperatordimless)$. In this analysis, following the \WGB{} literature \cite{Eichhorn:2017eht,deBrito:2021pyi,Eichhorn:2021qet}, we will keep \Gdimless{} as a free parameter. As mentioned in the last section, any collision of the \SGFP{} with one of the interacting pure scalar fixed points at finite \Gdimless{} is likely fiducial, and should be treated as a truncation artefact, at least within our truncation of the action. In this sense, we argue that the system in this truncation does not display a \WGB{} via a fixed point collision. In the following, we will thus mainly focus on the \SGFP{}. A key aspect of our analysis will correspondingly be how one can avoid such spurious collisions in different truncations.

The full fixed point equation for the gravity-scalar system has the form 
\begin{equation}\label{eq:fullFPeq_GravityMatter}
	4 \kineticfunctiondimlessFP(\kineticoperatordimless) - X (4+\scalaranomalousdimensionFP) \kineticfunctiondimlessFP'(\kineticoperatordimless) = \mathcal{F}[\kineticoperatordimless,\scalaranomalousdimensionFP,\kineticfunctiondimlessFP,\kineticfunctiondimlessFP',\kineticfunctiondimlessFP'';\Gdimless{}] \,,
\end{equation}
where
\begin{equation}\label{eq:fullFPeq_GravityMatter_RHS}
	\mathcal{F} =
	\frac{1}{16\pi^3} \int_0^\infty dz \,z \int_{-1}^{1}dx \, \sqrt{1-x^2} \, F \,.
\end{equation}
The function $F = F(z,x;\kineticoperatordimless,\scalaranomalousdimension,\kineticfunctiondimless,\kineticfunctiondimless',\kineticfunctiondimless'';\Gdimless{})$ is a generalisation of the integrand appearing on the right-hand side of \eqref{eq:flowKphi_purematter}. We include the full expression for $F$ in the ancillary notebook.

Similarly to the pure scalar case, it is not possible to analytically solve the flow equation for the function $\kineticfunctiondimless(\kineticoperatordimless)$. Hence, also for the gravity-scalar system, we will resort to different systematic expansions of $\kineticfunctiondimless(\kineticoperatordimless)$, and then analyse the partial fixed point structure within these expansions as a function of the external parameter \Gdimless{}.

For the gravity-scalar system there are two obvious expansions: an expansion in $\kineticoperatordimless$, and an expansion in the Newton coupling $\Gdimless$. We will hence first employ the expansion of $\kineticfunctiondimless(\kineticoperatordimless)$ in $\kineticoperatordimless$ and explore the impact of gravitational fluctuations on the expansion we have studied in \autoref{sec:purematt}. This allows us to investigate the precise pattern of spurious collisions as a function of \Gdimless{}. Furthermore, this is the expansion that has been studied in the literature before \cite{Eichhorn:2012va, deBrito:2021pyi, Laporte:2021kyp, Knorr:2022ilz}, so that we can easily connect our results to previous studies. 

The expansion in $\Gdimless$ allows us to track the \SGFP{} as a function of \Gdimless{} while taking the full dependence on $\kineticoperatordimless$ into account at each order. This expansion therefore is, at least in principle, an orthogonal expansion to study the fate of the \SGFP{} for increasing strength of the gravitational interaction. From this analysis, we will finally find a more refined, partially resummed expansion, that will turn out to indicate the limitations of the expansion about an off-shell background.

\subsection{Expansion in \texorpdfstring{\kineticoperatordimless}{X}}\label{sec:rhoexp}
Our starting point is again the polynomial expansion \eqref{eq:polynomial_expansion} of the  function $\kineticfunctiondimless(\kineticoperatordimless)$ in powers of the kinetic term  of the scalar field $\kineticoperatordimless$. The only difference to the setup in \autoref{sec:purematt} is the fact that the scale dependence, and therefore also the fixed point values of the expansion coefficients $\Kcoup{n}$, additionally depend on \Gdimless{}. 
In the following we will mostly focus on the \SGFP{}, \ie, the extension of the \GFP{} to $\Gdimless>0$, which is then shifted to become interacting in the presence of gravitational fluctuations. If not stated otherwise, the quoted results were obtained using the linear parameterisation, a Litim regulator \eqref{eq:shapeLitim} and with the gauge choice $\GFbeta=1$. 

\begin{figure*}
\includegraphics[width=\linewidth]{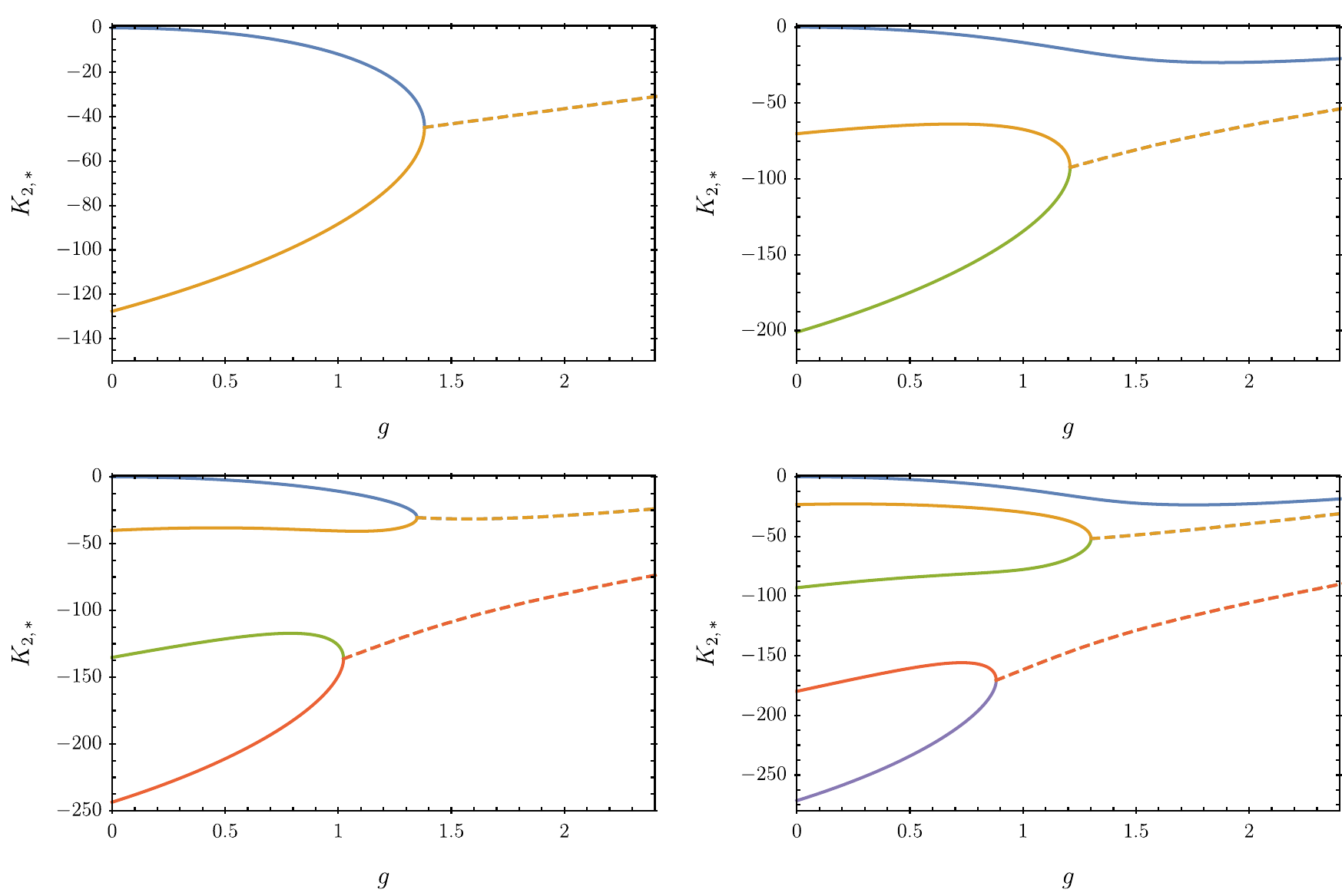}
	\caption{Partial fixed point structure of the gravity-scalar system as a function of $\Gdimless$, and for different orders $\Nmax=2,3,4,5$ (from left to right and top to bottom) in the polynomial expansion \eqref{eq:polynomial_expansion} of the function $\kineticfunctiondimless(\kineticoperatordimless)$, with $\GFbeta=1$ and the Litim regulator. Solid lines represent real-valued partial fixed points, while dashed lines indicate the real part of a complex-conjugate pair of partial fixed points. We only show the continuation for finite $\Gdimless$ of those partial fixed points that are real at $\Gdimless=0$. Furthermore, we do not show one additional partial fixed point, since it lies beyond a pole in the anomalous dimension and therefore cannot collide with any of the shown partial fixed points. We can see that in all cases the most interacting partial fixed point that is shown collides with the second-most interacting partial fixed point when increasing $\Gdimless$. Therefore, in truncations with an even $\Nmax$, where an even number of real-valued partial fixed points can collide with each other, the \SGFP{} eventually vanishes into the complex plane. This does not happen in truncations with odd $\Nmax$, since the \SGFP{} remains as the single real-valued partial fixed point.}
	\label{fig:FPCol}
\end{figure*}

As a first step towards understanding the impact of gravitational fluctuations on the pure scalar system, we compute all real-valued (spurious) fixed points for fixed $\Nmax$, and explore their fate when turning on \Gdimless{}. In \autoref{fig:FPCol}, we show the fixed point structure for $\Nmax\in\{2, 3, 4, 5\}$, where we only show those fixed points that are real-valued for $\Gdimless=0$. Furthermore, we continue to neglect the additional fixed point that is located beyond a pole in the anomalous dimension $\scalaranomalousdimension$. It is therefore disconnected from the other fixed points and cannot collide with any of them. 

We can see that for all displayed orders $\Nmax$, the most interacting fixed point collides with the second-most interacting fixed point when increasing $\Gdimless$. Since for even $\Nmax$ there is an even number of fixed points, also the \SGFP{} collides with another fixed point, and vanishes into the complex plane at some finite value of $\Gdimless$. For odd $\Nmax$ however, the \SGFP{} remains as the only real-valued fixed point. Therefore, the \WGB{} that has been discussed in the literature \cite{Eichhorn:2012va, deBrito:2021pyi, Laporte:2021kyp, Knorr:2022ilz} and that results from the collision of the \SGFP{} with another fixed point, is only present in truncations with an even $\Nmax$. This is another indication for the spurious nature of these collisions of partial fixed points. We can further observe that the first interacting fixed point begins to approach the \SGFP{} as a function of $\Gdimless$ when increasing \Nmax{}. 

Since even and odd values of $\Nmax$ show a different qualitative behaviour, we will study the \SGFP{} for both cases separately in the following. 
\begin{figure*}
	\includegraphics[width=\linewidth]{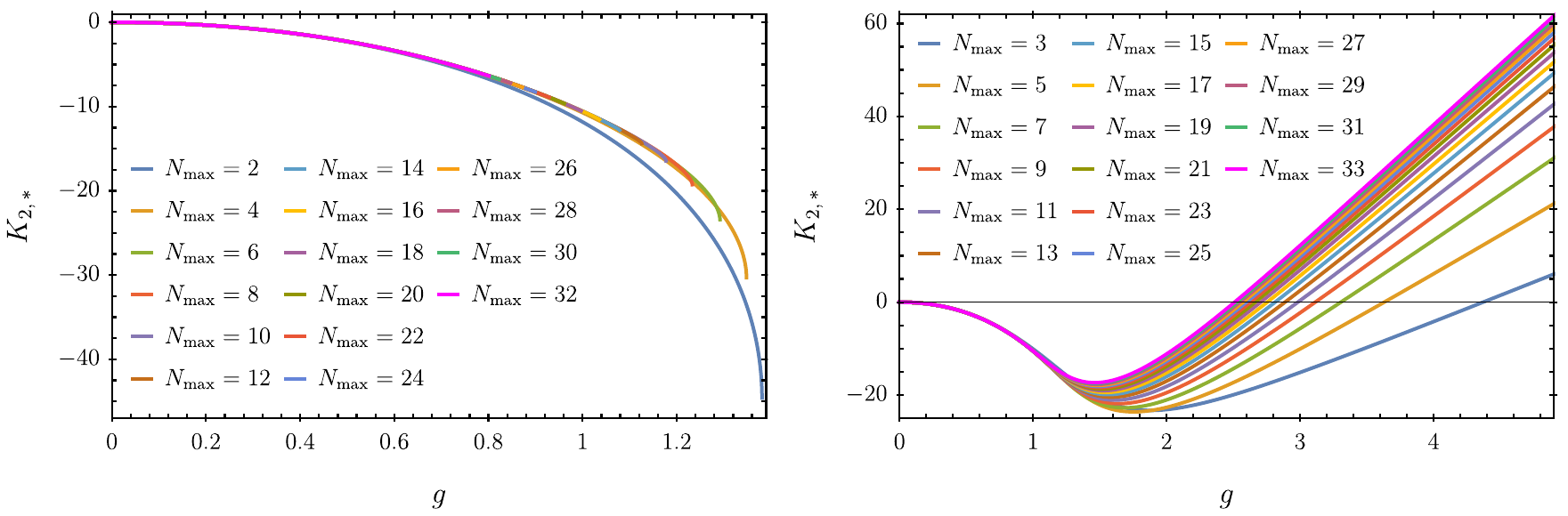}
	\caption{\SGFP{} for even (left panel) and odd (right panel) values of $\Nmax$. We can see that for even $\Nmax$ the \SGFP{} is numerically stable until the fixed point collision. For odd \Nmax{} the \SGFP{} is numerically very stable up to about $\Gdimless\approx1$, from where on convergence is slower.}
	\label{fig:SGFP}
\end{figure*}
\begin{figure*}
	\includegraphics[width=\linewidth]{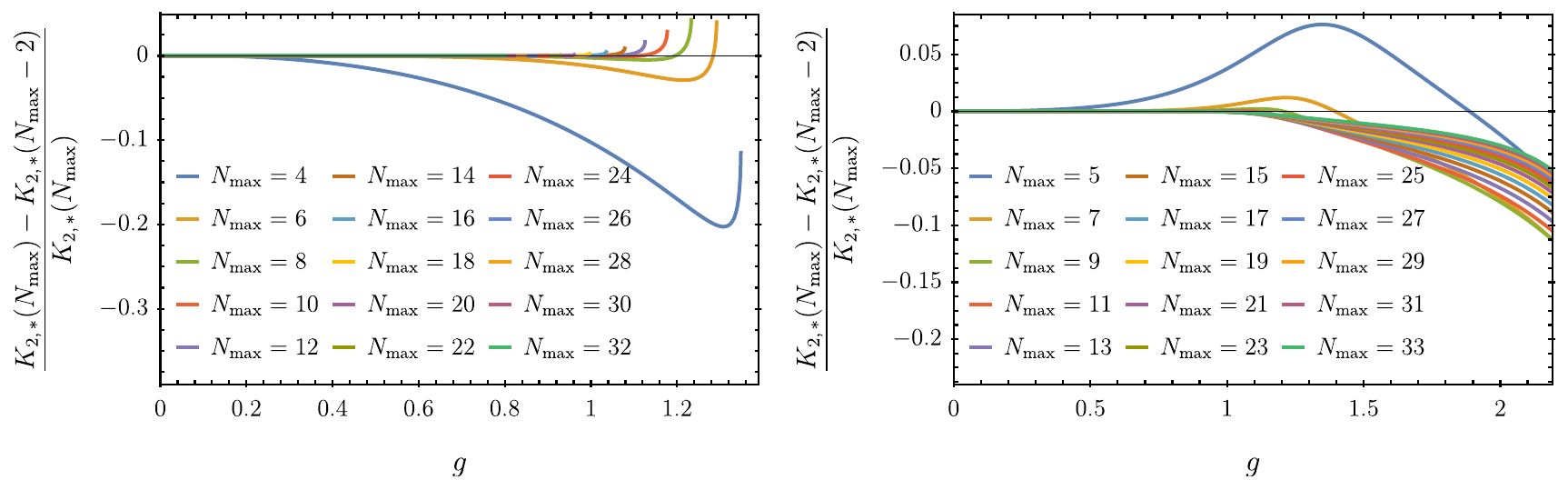}
	\caption{Relative deviations of the \SGFP{} as a function of $\Gdimless$, when increasing the truncation from a given even (odd) $\Nmax$ to the next even (odd) $\Nmax$. We see that generically the relative deviation increases when increasing $\Gdimless$, and that it overall decreases when increasing $\Nmax$.}
	\label{fig:SGFPRelDiff}
\end{figure*}
\begin{figure*}
	\includegraphics[width=\linewidth]{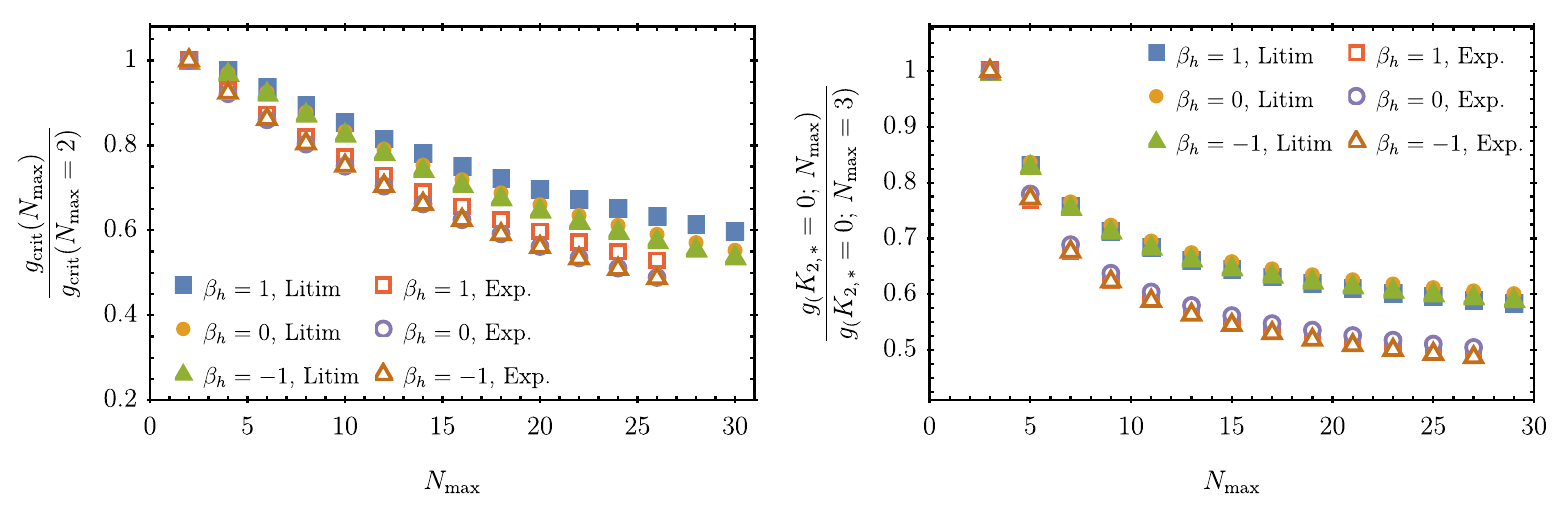}
	\caption{Aspects of gauge and regulator dependences of the \SGFP{}. Left panel:  (normalised) critical value of the Newton coupling $\Gdimlesscrit$ as a function of $\Nmax$ for different choices of the gauge parameter $\GFbeta$ and the regulator. For all displayed choices, a fixed point collision of the \SGFP{} is only present for even $\Nmax$. Furthermore, the behaviour of $\Gdimlesscrit$ as a function of $\Nmax$ agrees for all displayed choices on a quantitative level. Right panel: value of the Newton coupling where the partial fixed point value $\KcoupFP{2}$ at the \SGFP{} crosses zero, normalised to the case $\Nmax=3$. We see that the qualitative behaviour agrees for all displayed choices of gauge fixing and regulator.}
	\label{fig:GcritGK2Zero}
\end{figure*}

The left panel of \autoref{fig:SGFP} shows the \SGFP{} for even $\Nmax$ as a function of $\Gdimless$ and up to the point where it collides with another partial fixed point and vanishes into the complex plane. We can see that the critical value of the Newton coupling where the partial fixed point collision occurs, shifts to lower values. Furthermore, for $\Nmax\geq4$, the \SGFP{} is numerically stable when increasing \Nmax{}, up to the critical value for $\Gdimless$. To quantify this numerical stability further, in the left panel of \autoref{fig:SGFPRelDiff} we show the relative difference of \KcoupFP{2} at a given even \Nmax{} and at $\Nmax-2$. We see that for $\Nmax>4$ the \SGFP{} is stable, and deviates by less than $5\%$ from the previous even \Nmax{}, for $\Gdimless\leq1$, or until the point where it vanishes into the complex plane.

The right panel of \autoref{fig:SGFP} shows the \SGFP{} for odd $\Nmax$ as a function of $\Gdimless$. We can observe that the \SGFP{} follows a similar trajectory as for even $\Nmax$. However, instead of colliding with another partial fixed point, the value of $\KcoupFP{2}$ at the \SGFP{} eventually increases and becomes positive once \Gdimless{} is larger than some value around $2.5$. We can also see that for large $\Gdimless$, \KcoupFP{2} at the \SGFP{} seems to grow linearly in $\Gdimless$, and the slope approaches a fixed finite value when increasing $\Nmax$. The right panel of \autoref{fig:SGFPRelDiff} shows the relative difference between \KcoupFP{2} at a given odd \Nmax{} and at $\Nmax-2$. As in the case for even $\Nmax$, the \SGFP{} is stable, with deviations well below $5\%$ up to $\Gdimless=1$, see the right panel of \autoref{fig:SGFPRelDiff}. Furthermore, for $\Nmax>6$, the deviation up to $\Gdimless=1$ is below the per-mille level. 

In summary, the \SGFP{} is stable up to the point of collision for even $\Nmax$, or up to at least $\Gdimless=1$ for odd $\Nmax$. Therefore, we find further indications that the shift-symmetric scalar sector is genuinely interacting in the presence of gravitational fluctuations.

To further explore the stability of this picture, we will study the behaviour under changes of the gauge choice $\GFbeta$ and the regulator, again treating even and odd orders of the truncation separately. For even $\Nmax$, we study the critical value of the Newton coupling $\Gdimlesscrit$, where the \SGFP{} vanishes into the complex plane. The left panel of \autoref{fig:GcritGK2Zero} shows \Gdimlesscrit{} as a function of $\Nmax$ for different choices of the gauge fixing parameter $\GFbeta$ and both  regulator functions \eqref{eq:shapeLitim} and \eqref{eq:shapeExp}. The boxes (circles, triangles) indicate how \Gdimlesscrit{} changes when increasing \Nmax{} for the most common gauge choices $\GFbeta=1$ ($\GFbeta=0$, $\GFbeta=-1$) for the Litim regulator (full markers) and the exponential regulator (open markers), respectively.

For odd $\Nmax$, we study the value for $\Gdimless$ where the fixed point value $\KcoupFP{2}$ of the \SGFP{} is zero. This value is not expected to be a universal quantity, but its behaviour as a function of $\Nmax$ entails qualitative features of the system that nevertheless are expected to be robust. The right panel of \autoref{fig:GcritGK2Zero} shows this quantity for the three most common gauge choices. It clearly indicates that the system only mildly depends on the gauge for this range of gauge parameters for each of the two regulators. In fact, the relative evolution of the quantities in both panels of \autoref{fig:GcritGK2Zero} is in quantitative agreement across the shown values for $\GFbeta$ and the two regulators. We will discuss the dependence on unphysical choices such as the gauge and regulator in more detail in \aref{app:gmSpuriousDep}.

Applying Aitken's delta squared method \cite{aitken_1927} to estimate the limit of the sequence shown in the right panel of \autoref{fig:GcritGK2Zero} gives a convergent result for all shown choices of $\GFbeta$ and both regulators. For the example of $\GFbeta=1$ and the Litim  regulator, we estimate
\begin{equation}
\frac{\Gdimless(\KcoupFP{2}=0\,,\Nmax\to\infty)}{\Gdimless(\KcoupFP{2}=0\,,\Nmax=3)}\approx0.5\,.
\end{equation}
This indeed indicates that the \SGFP{} is convergent at least to the point where $\KcoupFP{2}$ crosses zero. Applying the same method to the left panel of \autoref{fig:GcritGK2Zero} however does not indicate a convergent behaviour; the estimate for the limit value when applying the method once, twice or three times, varies strongly. This indicates that while the \SGFP{} itself is convergent for a range of values for \Gdimless{}, the value for \Gdimlesscrit{} is not converged within our truncation. This is yet another piece of evidence for the fixed point collision being spurious.

We will now propose a new notion of the \WGB{} that separates the weak from the strong gravity regime: instead of a partial fixed point collision and a related absence of a \UV{} completion, we define the \WGB{} as the gravitational interaction strength where the \SGFP{} receives more relevant operators. Below (above) this critical strength, we are in the weak (strong) gravity regime. Since we treat $\Gdimless$ as an external parameter in our study, we do not have access to the full set of critical exponents of the gravity-scalar system. Instead, we define the \textit{pseudo critical exponents}
\begin{equation}
\label{eq:pseudocritexp}
\vartheta_i(\Gdimless)=-\mathrm{eig}\left[\frac{\partial \beta_{\Kcoup{i}}}{\partial \Kcoup{j}}\right]\bigg|_{\Kcoup{n}=\KcoupFP{n}}\,,
\end{equation}
and assume that they capture this information approximately. We furthermore define $\vartheta_1(\Gdimless)$ as the pseudo critical exponent that is continuously connected (as a function of \Gdimless{}) to the most relevant one at $\Gdimless=0$,  \ie, $\vartheta_1(0)=-4$. 

We thus propose to define the \WGB{} as the value of \Gdimless{} where the scalar (or in general any matter) system acquires an additional relevant direction. In our system this can  happen for odd values of $\Nmax$, since $\vartheta_1$ and $\vartheta_2$ form a complex-conjugate pair for some value of $\Gdimless$. This complex-conjugate pair can then become relevant for larger values of $\Gdimless$ without a partial fixed point collision, as the imaginary part stays finite through this transition. The \WGB{} defined by $\Gdimlesscritnew=\Gdimless(\mathrm{Re}(\vartheta_1)=0)$ therefore separates the regime where the \UV{}-complete scalar sector does not feature additional relevant directions ($\Gdimless<\Gdimlesscritnew$), \ie{}, a weak gravity regime, from the regime where at least two more relevant directions are present ($\Gdimless>\Gdimlesscritnew$), \ie{} a strong gravity regime. 

We focus on $\vartheta_1$, since it is the pseudo critical exponent that is numerically most controlled, and that is likely to be most insensitive to the higher order operators that were neglected in our truncation. 
We add that the more irrelevant pseudo critical exponents, at high enough \Nmax{}, become more relevant than $\vartheta_1$ already at $\Gdimless < \Gdimlesscritnew$, but they are expected to receive larger corrections from operators that we have neglected. This is due to the approximately triangular structure of the \FRG{} equation, where the flow of the $n$-point function is driven by correlation functions of order $\leq (n+2)$ only. We leave a more complete investigation of these aspects for future work.

\begin{figure}
	\includegraphics[width=\linewidth]{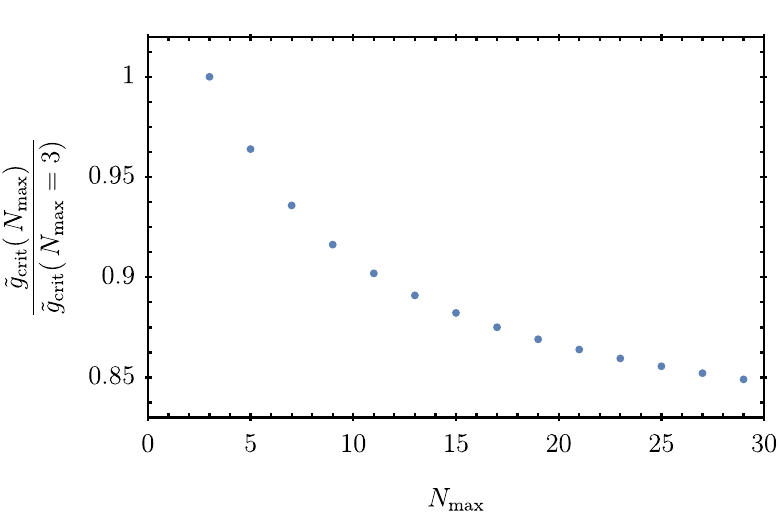} 
	\caption{Value of $\Gdimless$ where the pseudo critical exponent $\vartheta_1$, see \eqref{eq:pseudocritexp}, becomes relevant, as a function of $\Nmax$, representing our new definition of the \WGB{} that does not rely on the collision of partial fixed points: for $\Gdimless>\Gdimlesscritnew$, the \UV{} completion of the scalar sector requires the presence of additional relevant directions.}
	\label{fig:RelTheta}
\end{figure}

In \autoref{fig:RelTheta} we show $\Gdimlesscritnew(\Nmax)$ for $\GFbeta=1$ and the Litim regulator. We see that it features a qualitatively similar behaviour as $\Gdimless(\KcoupFP{2}=0;\,\Nmax)$: it decreases quickly for small $\Nmax$ and flattens out for larger $\Nmax$. Applying Aitken's delta squared method to $\Gdimlesscritnew(\Nmax)$ gives a convergent result, and we estimate 
\begin{equation}
\frac{\Gdimlesscritnew(\Nmax\to\infty)}{\Gdimlesscritnew(\Nmax=3)}\approx 0.81\,,
\end{equation}
corresponding to 
\begin{equation}
\Gdimlesscritnew(\Nmax\to\infty)\approx 3.5\,.
\end{equation} 
This indicates that our new definition of a \WGB{} via $\Gdimlesscritnew{}$ is indeed stable and convergent.

In summary we find a qualitatively different behaviour for even and odd orders in the truncation. For truncations with an odd $\Nmax$, the value of $\Gdimless>0$ where $\KcoupFP{2}=0$ converges to a finite value for each displayed choice of $\GFbeta$ and the regulator. In this case, the \SGFP{} shows a very stable behaviour up to relatively large values of \Gdimless{}, where additional relevant directions in the scalar sector appear, giving rise to a new notion of the \WGB{}. For even values $\Nmax$, the location of the partial fixed point collision seems not to converge within the truncations that we have studied. This is in line with our observations in \autoref{sec:purematt} where we have shown that the interacting partial fixed point in the pure scalar system that gives rise to the collision is only a truncation artefact, and not part of the desired function space. Hence, we expect the \SGFP{} to be controlled and convergent, while the collision with a truncation artefact is not expected to show any convergent behaviour.

\subsection{Expansion in \texorpdfstring{\Gdimless}{g}}\label{sec:gexp}

We will now investigate the \SGFP{} in a systematic expansion in powers of the (dimensionless) Newton's constant $\Gdimless$. This means that we expand the kinetic function as
\begin{equation}\label{eq:gexpansion}
 \kineticfunctiondimless(\kineticoperatordimless) \approx \kineticoperatordimless + \left(\frac{1}{16\pi}\right)^{\!2} \, \sum_{n=1}^{\Omax} \left( \frac{\Gdimless}{16\pi} \right)^n \, \kineticfunctionGdimlessexpansion{n}(\kineticoperatordimlessGdimlessexpansion) \, ,
\end{equation}
and the scalar anomalous dimension as
\begin{equation}
 \scalaranomalousdimension = \sum_{n=1}^{\Omax} \left( \frac{\Gdimless}{16\pi} \right)^n \, \scalaranomalousdimensionGdimlessexpansion{n} \, .
\end{equation}
Here, we introduced factors of $16\pi$ in convenient places, as well as
\begin{equation}
 \kineticoperatordimlessGdimlessexpansion = \frac{3}{2} \left(16\pi \right)^2\kineticoperatordimless \, .
\end{equation}
This expansion is, in principle, orthogonal to the polynomial expansion \eqref{eq:polynomial_expansion} in $\kineticoperatordimless$, and allows keeping the full dependence on $\kineticoperatordimless$ at each order in $\Gdimless$. From the boundary conditions \eqref{eq:Kfunctconds}, we get $\kineticfunctionGdimlessexpansion{n}(0) = \kineticfunctionGdimlessexpansion{n}'(0) = 0$ for all $n$.
We will again use $\GFbeta=1$ and the Litim regulator \eqref{eq:shapeLitim} to illustrate the resulting structure.

We will now systematically expand the fixed point equation in powers of $\Gdimless$. Generically, this expansion leads to the following structure
\begin{equation}
	\begin{aligned}\label{eq:expansionG_order-n}
		& 4 \kineticfunctionGdimlessexpansionFP{n}(\kineticoperatordimlessGdimlessexpansion) - 4 \kineticoperatordimlessGdimlessexpansion \kineticfunctionGdimlessexpansionFP{n}'(\kineticoperatordimlessGdimlessexpansion) - \frac{2}{3} \scalaranomalousdimensionGdimlessexpansionFP{n} \kineticoperatordimlessGdimlessexpansion = \\
		&\qquad \mathcal{F}_n \big(\kineticoperatordimlessGdimlessexpansion,\{ \scalaranomalousdimensionGdimlessexpansionFP{j} \}_{j\leq n-1},\{\kineticfunctionGdimlessexpansionFP{j},\kineticfunctionGdimlessexpansionFP{j}',\kineticfunctionGdimlessexpansionFP{j}''\}_{j\leq n} \big) \,,
	\end{aligned}	
\end{equation}
where $ \mathcal{F}_n$ has a polynomial dependence on all of its arguments. This structure allows us to find analytical solutions for $\kineticfunctionGdimlessexpansionFP{n}(\kineticoperatordimlessGdimlessexpansion)$ in an iterative way.

For $n=1$, we find the differential equation
\begin{equation}\label{eq:expansionG_order-1}
\begin{aligned}
 4 \kineticfunctionGdimlessexpansionFP{1}(\kineticoperatordimlessGdimlessexpansion) &- 4 \kineticoperatordimlessGdimlessexpansion \kineticfunctionGdimlessexpansionFP{1}'(\kineticoperatordimlessGdimlessexpansion) - \frac{2}{3} \scalaranomalousdimensionGdimlessexpansionFP{1} \kineticoperatordimlessGdimlessexpansion \\
 &= - 4 \left( 2 \kineticfunctionGdimlessexpansionFP{1}'(\kineticoperatordimlessGdimlessexpansion) + \kineticoperatordimlessGdimlessexpansion \, \kineticfunctionGdimlessexpansionFP{1}''(\kineticoperatordimlessGdimlessexpansion) \right) \, .
\end{aligned}
\end{equation}
Since this is a second order differential equation, the general solution has two free parameters that we call $c_1$ and $c_2$. This general solution reads
\begin{equation}\label{eq:expansionG_order-1_sol1}
	\begin{aligned}
		\kineticfunctionGdimlessexpansionFP{1}(\kineticoperatordimlessGdimlessexpansion) &= \left( \kineticoperatordimlessGdimlessexpansion - 2 \right) c_1 \\
		& + \frac{(1-\kineticoperatordimlessGdimlessexpansion) e^{\kineticoperatordimlessGdimlessexpansion} + \kineticoperatordimlessGdimlessexpansion (\kineticoperatordimlessGdimlessexpansion - 2) \text{Ei}(\kineticoperatordimlessGdimlessexpansion) }{2\kineticoperatordimlessGdimlessexpansion} \,c_2 \\
		&- \frac{2-5\kineticoperatordimlessGdimlessexpansion^2 + 2\kineticoperatordimlessGdimlessexpansion ( \kineticoperatordimlessGdimlessexpansion - 2 ) \ln \kineticoperatordimlessGdimlessexpansion}{12\kineticoperatordimlessGdimlessexpansion}\, \scalaranomalousdimensionGdimlessexpansionFP{1}\,.
	\end{aligned}
\end{equation}
To fix the parameters $c_{1,2}$, we look at the behaviour about $\kineticoperatordimless =0$.
Demanding that no divergencies occur at $\kineticoperatordimless = 0$ as well as imposing \eqref{eq:Kfunctconds}, we find
\begin{equation}
 c_1 = -\frac{\gamma}{6} \scalaranomalousdimensionGdimlessexpansionFP{1} \, , \qquad c_2 = \frac{1}{3} \scalaranomalousdimensionGdimlessexpansionFP{1} \, .
\end{equation}
To fix $\scalaranomalousdimensionGdimlessexpansionFP{1}$, we have to consider the behaviour of $\kineticfunctionGdimlessexpansionFP{1}(\kineticoperatordimlessGdimlessexpansion)$ at large \kineticoperatordimlessGdimlessexpansion{}. We note that $\kineticfunctionGdimlessexpansionFP{1}(\kineticoperatordimlessGdimlessexpansion)$ grows exponentially for large \kineticoperatordimlessGdimlessexpansion{}. Such behaviour is undesirable as discussed previously. To remove the exponentially growing terms, we thus have to set $\scalaranomalousdimensionGdimlessexpansionFP{1}=0$. Therefore, to first order in $\Gdimless$, we find
\begin{align}
	 \kineticfunctionGdimlessexpansionFP{1}(\kineticoperatordimlessGdimlessexpansion) = 0 \, , \qquad \scalaranomalousdimensionGdimlessexpansionFP{1} = 0 \, .
\end{align}
Note that for other gauge choices, we instead find a non-vanishing \scalaranomalousdimensionGdimlessexpansionFP{1}, but \kineticfunctionGdimlessexpansionFP{1} still vanishes.

Moving on to the second order in $\Gdimless$, we find the differential equation
\begin{equation}
	\begin{aligned}
		 &4 \kineticfunctionGdimlessexpansionFP{2}(\kineticoperatordimlessGdimlessexpansion) - 4 \kineticoperatordimlessGdimlessexpansion \kineticfunctionGdimlessexpansionFP{2}'(\kineticoperatordimlessGdimlessexpansion) - \frac{2}{3} \scalaranomalousdimensionGdimlessexpansionFP{2} \kineticoperatordimlessGdimlessexpansion = \\
		 &\qquad\qquad \frac{128}{9} \kineticoperatordimlessGdimlessexpansion^2 - 4 \left(2 \kineticfunctionGdimlessexpansionFP{2}'(\kineticoperatordimlessGdimlessexpansion) + \kineticoperatordimlessGdimlessexpansion \, \kineticfunctionGdimlessexpansionFP{2}''(\kineticoperatordimlessGdimlessexpansion) \right) \, .
	\end{aligned}
\end{equation}
Notably, there is now an inhomogeneous term quadratic in $\kineticoperatordimlessGdimlessexpansion$ that will allow for a non-trivial solution. Once again, we fix one integration constant by removing divergent terms at $\kineticoperatordimless=0$ and imposing the boundary conditions \eqref{eq:Kfunctconds}. Furthermore, we can fix $\scalaranomalousdimensionGdimlessexpansionFP{2}$ by demanding that the solution does not grow exponentially for large $\kineticoperatordimlessGdimlessexpansion$. This gives the unique solution
\begin{equation}
 \kineticfunctionGdimlessexpansionFP{2}(\kineticoperatordimlessGdimlessexpansion) = -\frac{32}{9} \kineticoperatordimlessGdimlessexpansion^2 \, , \qquad \scalaranomalousdimensionGdimlessexpansionFP{2} = -128 \, .
\end{equation}
This procedure can be continued order by order in $\Gdimless$. Generally, we find that $\kineticfunctionGdimlessexpansionFP{n}$ is a polynomial in \kineticoperatordimlessGdimlessexpansion{} of order $n$,
\begin{equation}
 \kineticfunctionGdimlessexpansionFP{n}(\kineticoperatordimlessGdimlessexpansion) = \sum_{i=2}^n \ell_{n,i} \kineticoperatordimlessGdimlessexpansion^i \, ,
\end{equation}
and we include the coefficients $\ell_{n,i}$ and $\scalaranomalousdimensionGdimlessexpansionFP{n}$ up to $\Omax=28$ in the ancillary notebook.

This result is remarkable, for the following reasons. First, by requiring that the solution is in the Hilbert space $\mathcal L$, we eliminated \emph{all} integration constants. This entails that the free parameter $\KcoupFP{2}$ that we had to choose in the polynomial expansion of the last section by an additional condition, can be computed order by order in $\Gdimless$.
Second, the expansion coefficient of the term $\kineticoperatordimlessGdimlessexpansion^n$ starts at $\Gdimless^n$. This means, on the one hand, that we have found another way to compute $\KcoupFP{2}$ order by order in $\Gdimless$ from the general polynomial expansion of \autoref{sec:rhoexp}.
On the other hand, this suggests a different, partially resummed expansion of the form
\begin{equation}
\label{eq:combinedexp}
 \kineticfunctiondimless(\kineticoperatordimless) \approx \kineticoperatordimless + \left( \frac{1}{16\pi} \right)^2\, \sum_{n=0}^{\Pmax} \kineticfunctioncombinedexpansion{n}( \kineticoperatordimlesscombinedexpansion ) \left( \frac{\Gdimless}{16\pi} \right)^n \, ,
\end{equation}
where
\begin{equation}
 \kineticoperatordimlesscombinedexpansion = 16\pi \Gdimless \kineticoperatordimless \, .
\end{equation}
This is an expansion that retains \emph{global} information in $\kineticoperatordimless$. It works by combining first the terms in the $\kineticfunctionGdimlessexpansion{n}$ with the largest exponent into $\kineticfunctioncombinedexpansion{0}$, and then subsequently combines all the smaller exponent terms. It also suggests that the natural variable is actually the combination $\Gdimless \kineticoperatordimless$. As a matter of fact, this combination precisely appears in the non-trivial interaction between gravity and the scalar field in the two-point function. We will consider this expansion in the next subsection.

Before moving on, let us remark that, due to the relatively large value of \Gdimlesscritnew{} indicating the new \WGB{} as obtained in the \kineticoperatordimless{}-expansion, we refrain from a similar analysis within the \Gdimless{}-expansion.

\subsection{Combined expansion}\label{sec:combexp}
In the combined expansion we expand the function $\kineticfunctiondimless(\kineticoperatordimless)$ according to \eqref{eq:combinedexp}. The anomalous dimension is independent of \kineticoperatordimless{} by definition, and therefore it is expanded in powers of $\Gdimless$ as in the previous section. In fact, we can use the coefficients $\scalaranomalousdimensionGdimlessexpansionFP{i}$ that we computed previously. Structurally, the combined expansion leads to first order differential equations for the coefficients $\kineticfunctioncombinedexpansion{n}$, namely
\begin{equation}\label{eq:combinedexp_fixpteq_1}
	\begin{aligned}
		&\kineticoperatordimlesscombinedexpansion \kineticfunctioncombinedexpansionFP{n}'( \kineticoperatordimlesscombinedexpansion) - \kineticfunctioncombinedexpansionFP{n}( \kineticoperatordimlesscombinedexpansion) =\\
		&\qquad \mathcal{F}_n^{\textmd{comb}}\big(\kineticoperatordimlesscombinedexpansion,\{ \scalaranomalousdimensionGdimlessexpansionFP{j} \}_{j\leq n+1},\{\kineticfunctioncombinedexpansionFP{j}, \kineticfunctioncombinedexpansionFP{j}', \kineticfunctioncombinedexpansionFP{j}''\}_{j< n} \big) \,,
	\end{aligned}
\end{equation}
This is a consequence of the fact that the right-hand side of the flow equation is one order higher in \Gdimless{}, such that at each order in the expansion it only provides an inhomogeneous part for the differential equation.

Formally, the fixed point solutions to all orders in the combined expansion read
\begin{align}
\label{eq:combexpintegrl}
	\kineticfunctioncombinedexpansionFP{n}(\kineticoperatordimlesscombinedexpansion)=\kineticoperatordimlesscombinedexpansion \left( \int^{\kineticoperatordimlesscombinedexpansion{}} \!\! d \kineticoperatordimlesscombinedexpansion{}' \, \frac{ \mathcal{F}_n^{\textmd{comb}}}{\big(\kineticoperatordimlesscombinedexpansion{}'\big)^2} + C_n\right).
\end{align} 
where $C_n$ is an integration constant that we can fix by demanding $\kineticfunctioncombinedexpansionFP{n}'(0)=0$. However, due to the involved form of $\mathcal{F}_n^{\textmd{comb}}$, we were only able to analytically perform the integral for $n=0$ and $n=1$. For this reason, we will again not investigate our new notion for the \WGB{}.

\begin{figure}
	\includegraphics[width=\linewidth]{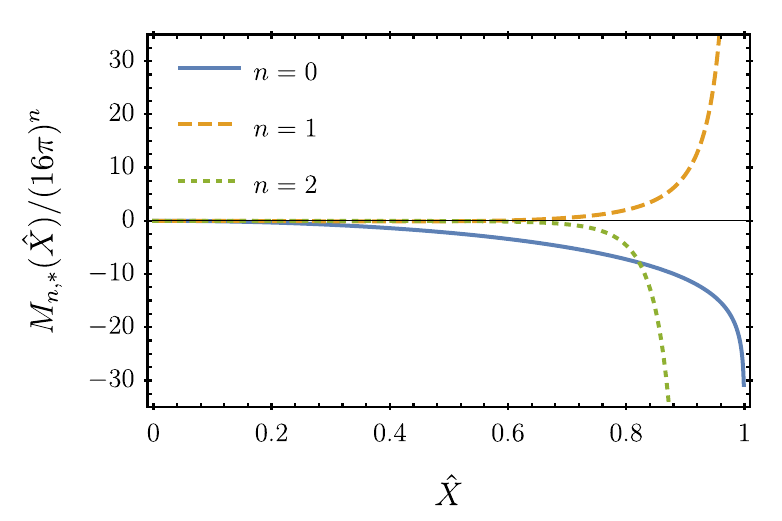} 
	\caption{Functions $\kineticfunctioncombinedexpansionFP{n}(\kineticoperatordimlesscombinedexpansion)$ for $n=0,1,2$. For convenience of the plot, we divide the $\kineticfunctioncombinedexpansionFP{n}(\kineticoperatordimlesscombinedexpansion)$ by a factor $(16\pi)^n$. As we see, the curves vary slowly for $\kineticoperatordimlesscombinedexpansion < 0.5$, but they approach a singularity at $\kineticoperatordimlesscombinedexpansion=1$. Such a singularity sets an upper bound on the validity of the combined expansion.}
	\label{fig:SolutionsCombined}
\end{figure}

To zeroth order in \Gdimless{} and with $\scalaranomalousdimensionGdimlessexpansion{1}=0$, we find
\begin{equation}
	\begin{aligned}
		\mathcal{F}_0^{\textmd{comb}} =- \frac{4}{\kineticoperatordimlesscombinedexpansion} 
		\left( \frac{2-\kineticoperatordimlesscombinedexpansion}{1-\kineticoperatordimlesscombinedexpansion} - 3 \kineticoperatordimlesscombinedexpansion - 2 \sqrt{\frac{1-\kineticoperatordimlesscombinedexpansion}{1+\kineticoperatordimlesscombinedexpansion}} \right) \, ,
	\end{aligned}
\end{equation}
which leads to
\begin{equation}\label{eq:SolutionCombinedOrder0}
\begin{aligned}
\kineticfunctioncombinedexpansionFP{0}(\kineticoperatordimlesscombinedexpansion) &= \frac{4}{\kineticoperatordimlesscombinedexpansion} \left( 1 - 2 \kineticoperatordimlesscombinedexpansion \right) \left( 1 - \sqrt{1-\kineticoperatordimlesscombinedexpansion^2} \right) \\
&\,+ 2 \kineticoperatordimlesscombinedexpansion \left( 2 \ln \left[ 2 \frac{1-\kineticoperatordimlesscombinedexpansion}{1+\sqrt{1-\kineticoperatordimlesscombinedexpansion^2}} \right] - 1 \right) \, .
\end{aligned}
\end{equation}
Here, we have already fixed the constant of integration. 
In \autoref{fig:SolutionsCombined} we plot $\kineticfunctioncombinedexpansionFP{0}(\kineticoperatordimlesscombinedexpansion)$. In addition, we also show $\kineticfunctioncombinedexpansionFP{1}(\kineticoperatordimlesscombinedexpansion)$ that we obtained analytically, and $\kineticfunctioncombinedexpansionFP{2}(\kineticoperatordimlesscombinedexpansion)$ obtained from numerical integration.

As we can see in \autoref{fig:SolutionsCombined}, the functions $\kineticfunctioncombinedexpansionFP{0}(\kineticoperatordimlesscombinedexpansion)$, $\kineticfunctioncombinedexpansionFP{1}(\kineticoperatordimlesscombinedexpansion)$, and $\kineticfunctioncombinedexpansionFP{2}(\kineticoperatordimlesscombinedexpansion)$ diverge at $\kineticoperatordimlesscombinedexpansion=1$, putting a hard limit on the validity of this expansion. From \eqref{eq:SolutionCombinedOrder0}, we see that $\kineticfunctioncombinedexpansionFP{0}(\kineticoperatordimlesscombinedexpansion)$ diverges logarithmically for $\kineticoperatordimlesscombinedexpansion = 1$.
The divergence of $\kineticfunctioncombinedexpansionFP{1}(\kineticoperatordimlesscombinedexpansion)$ at $\kineticoperatordimlesscombinedexpansion = 1$ is a pole of order $3/2$. Beyond that order, we were not able to solve the differential equations \eqref{eq:combinedexp_fixpteq_1} in a closed form. However, we can still extract the leading pole by analysing the fixed point equation about $\kineticoperatordimlesscombinedexpansion = 1$. In general, the leading order contribution to $\mathcal{F}_n^{\textmd{comb}}$ about $\kineticoperatordimlesscombinedexpansion = 1$ comes from terms proportional to $\big[\kineticfunctioncombinedexpansion{0}''(\kineticoperatordimlesscombinedexpansion)\big]^{n}$. Keeping only the relevant terms to extract the leading order divergence, we can integrate \eqref{eq:combinedexp_fixpteq_1} order by order in $n$, which we have done explicitly up to $n=48$. With the help of \emph{FindSequenceFunction}, we find (for $n\geq1$)\begin{equation}\label{eq:combexppole}
	\kineticfunctioncombinedexpansionFP{n}^\textmd{pole}(\kineticoperatordimlesscombinedexpansion) \sim \frac{\sqrt{2}\,(-6)^{n+1} \mathcal{B}_n}{(1-\kineticoperatordimlesscombinedexpansion)^{3(n-1)+3/2}}
\end{equation}
with coefficients $\mathcal{B}_n$ satisfying the recursive relation
\begin{equation}\label{eq:B_recursion}
	\begin{aligned}
		&\mathcal{B}_n = -\frac{16 (n-2) (2 n-5) (2 n-3) (4 n+5)}{3 n (4 n+1)(1-2 n)^2} \mathcal{B}_{n-2} \\ 
		&\,\, +\frac{(3-2 n)^2 \left(448 n^3-976 n^2-412 n+805\right)}{12  n (2 n-5) (4 n+1)(1-2 n)^2} \mathcal{B}_{n-1},
	\end{aligned}
\end{equation}
with initial conditions $\mathcal{B}_1=1/6$ and  $\mathcal{B}_2 = 113/3888$.

Unfortunately, we were not able to solve the recursion \eqref{eq:B_recursion} to obtain more information on the pole structure of the combined expansion. One might entertain the hope that all the poles of the different orders sum up to yield a finite result at $\kineticoperatordimlesscombinedexpansion=1$, so that a global fixed point solution could be found. While we can neither confirm nor deny this idea, in \autoref{sec:understanding_the_pole} we present some arguments why we find this scenario unlikely.

\subsection{Off-shell origin of the pole at \texorpdfstring{$\kineticoperatordimlesscombinedexpansion=1$}{Xhat=1}}\label{sec:understanding_the_pole}

As we have seen in \autoref{sec:combexp}, the functions \kineticfunctioncombinedexpansionFP{n} in the combined expansion feature a pole at $\kineticoperatordimlesscombinedexpansion=1$. While we were able to extract the general structure of the leading-order divergent behaviour for each expansion coefficient $\kineticfunctioncombinedexpansionFP{n}$, see \eqref{eq:combexppole}, we could not determine whether or not the poles cancel under resummation of the $\kineticfunctioncombinedexpansionFP{n}$.

To understand the origin of this pole, we study the right-hand side of the fixed point equation for the gravity-scalar system \eqref{eq:fullFPeq_GravityMatter} before integrating over loop momenta and before performing any expansion, \ie{} we focus on the function $F$ in \eqref{eq:fullFPeq_GravityMatter_RHS}. 
In particular, we will investigate the integrand for large \kineticoperatordimless{}. Assuming that $\scalaranomalousdimensionFP{}>-4$, a scaling analysis fixes $\kineticfunctiondimlessFP(\kineticoperatordimless)\sim A \kineticoperatordimless^{\frac{4}{4+\scalaranomalousdimensionFP{}}}$. To have an action bounded from below, we further need $A>0$. We will now investigate $F$ in this regime, and find that parts of its denominator change sign, indicating a pole.

Generally speaking, the function $F$ of the gravity-scalar system consists of two parts: one contribution involving the propagator of the scalar field, and one from the propagator of metric fluctuations. We will focus on the scalar contribution in the following for simplicity, and denote the denominator of this contribution as $\textmd{Den}(F_\scalarfielddimful)$.

Inserting the asymptotic scaling, let us first consider $z=0$ and $x=0$. With a Litim regulator, we find
\begin{equation}
	\begin{aligned}
		&\textmd{Den}(F_\scalarfielddimful)\big|_{z=x=0}
		\propto\\
		 &\qquad\bigg(1+\frac{256 \pi ^2 A^2\, \Gdimless^2\, \left(\scalaranomalousdimensionFP^3+64\right) \kineticoperatordimless^{\frac{8}{\scalaranomalousdimensionFP +4}}}{(\scalaranomalousdimensionFP +4)^3}
		\\
		&\hspace{1cm} -\frac{32 \pi  A \,\Gdimless\, (\scalaranomalousdimensionFP  (\scalaranomalousdimensionFP +2)-16) \kineticoperatordimless^{\frac{4}{\scalaranomalousdimensionFP +4}}}{(\scalaranomalousdimensionFP +4)^2}\bigg)\, .
	\end{aligned}
\end{equation}
For $\scalaranomalousdimensionFP>-4$, the second term is the dominant contribution for large $\kineticoperatordimless$. Hence, for $z=0$ and $x=0$, the denominator is strictly positive in this limit. By contrast, let us now consider $z=0$ and $x=1/\sqrt{2}$. At this point, the denominator reads
\begin{equation}
	\begin{aligned}
		&\textmd{Den}(F_\scalarfielddimful)\big|_{z=0,x=1/\sqrt{2}} \propto \\
		&\qquad \bigg(1+\frac{256 \pi ^2 A^2 \,\Gdimless^2\, \scalaranomalousdimension  \left(\scalaranomalousdimension ^2+16\right) \kineticoperatordimless^{\frac{8}{\scalaranomalousdimension +4}}}{(\scalaranomalousdimension +4)^3}\\
		&\hspace{2.5cm} -\frac{32 \pi  A \, \Gdimless \, \scalaranomalousdimension  (\scalaranomalousdimension +2) \kineticoperatordimless^{\frac{4}{\scalaranomalousdimension +4}}}{(\scalaranomalousdimension +4)^2}\bigg)\,,
	\end{aligned}
\end{equation}
where again the second term is dominant for large $\kineticoperatordimless$. For $\Gdimless \leq 2.5$, we found in \autoref{sec:rhoexp} that $\scalaranomalousdimensionFP<0$, see also \autoref{fig:SGFP}. In that regime, the dominant term in the denominator of $F$ for $z=0$ and $x=1/\sqrt{2}$ is thus negative, and hence $\textmd{Den}(F_\scalarfielddimful)$ will be negative for large enough $\kineticoperatordimless$ for these values.

The change of sign of $\textmd{Den}(F_\scalarfielddimful)$ as a function of $x$ indicates that the angular integration over $x$ gives rise to a pole. Such a pole is typically related to the fact that the chosen background does not satisfy the equations of motion. In fact, the standard pole of the graviton propagator about a flat background at $\Lambdadimless=1/2$ has a similar origin: an integral over the flat, regularised propagator in this case reads
\begin{equation}
	\int_0^\infty \text{d}z \, \frac{\mathcal P(z)}{z+R(z)-2\lambda} \, ,
\end{equation}
for some integral kernel $\mathcal P$. For non-compact regulators that are normalised to unity at $z=0$, and for $\lambda>1/2$, we find that for small $z$, the denominator is negative, while for large enough $z$, it is positive.

Therefore, the pole at $\kineticoperatordimlesscombinedexpansion=1$ that we discovered in the combined expansion is most likely an off-shell pole. To corroborate this conclusion, we will analyse the classical equations of motion of the system. Its action is
\begin{equation}
	S=-\frac{1}{16\pi \GEH}\Int{4}\, \,[R-2\Lambdadimfulcl]+\Int{4}{}\,\,\kineticoperatordimful\,,
\end{equation}
giving rise to the classical equations of motion
\begin{align}
	R^{\mu\nu}- \frac{1}{4} R\,g^{\mu\nu} =& 16\pi\, \GEH\left(\frac{1}{2}D^\mu \scalarfielddimful\,D^\nu \scalarfielddimful-\frac{1}{4}g^{\mu\nu} \kineticoperatordimful\right) \, , \label{eq:tracelessEOM} \\
	R-4\Lambdadimfulcl = &16 \pi\, \GEH\, \kineticoperatordimful \, , \\
	D^\mu D_\mu\scalarfielddimful=&0 \, .
\end{align}
The first and second lines are the traceless and trace part of the gravitational field equations, respectively, and the third line is the equation of motion for the scalar field. Contracting \eqref{eq:tracelessEOM} with two factors of $D \scalarfielddimful$ leads to
\begin{equation}
	R^{\mu\nu}\,D_\mu \scalarfielddimful\,D_\nu \scalarfielddimful-\frac{1}{2}R \,\kineticoperatordimful=24\pi\,\GEH\, \kineticoperatordimful^2\,.
\end{equation}
We see that a metric with vanishing curvature is only on-shell whenever $\kineticoperatordimful=0$ (or $\GEH=0$). Conversely, any configuration with non-vanishing $\kineticoperatordimful$, and hence also non-vanishing $\kineticoperatordimless$, will be off-shell in a flat spacetime.

In this spirit, the expansions of $\kineticfunctiondimless$ in $\kineticoperatordimless$ or $\Gdimless$, as we employed in \autoref{sec:rhoexp} and \autoref{sec:gexp}, are expansions about an on-shell background. By contrast, the combined expansion of \autoref{sec:combexp} beyond $\Nmax=0$ is clearly an off-shell expansion. Divergences in this expansion are thus likely to be caused by the choice of an off-shell background. From this reasoning, we do not expect the divergences at $\Gdimless\,\kineticoperatordimless=1/16\pi$ in the combined expansion to cancel when resumming the coefficients $\kineticfunctioncombinedexpansionFP{n}$.

\subsection{Analysis of the truncation error}\label{sec:NumericalCheckSolution}

\begin{figure*}
	\includegraphics[width=\linewidth]{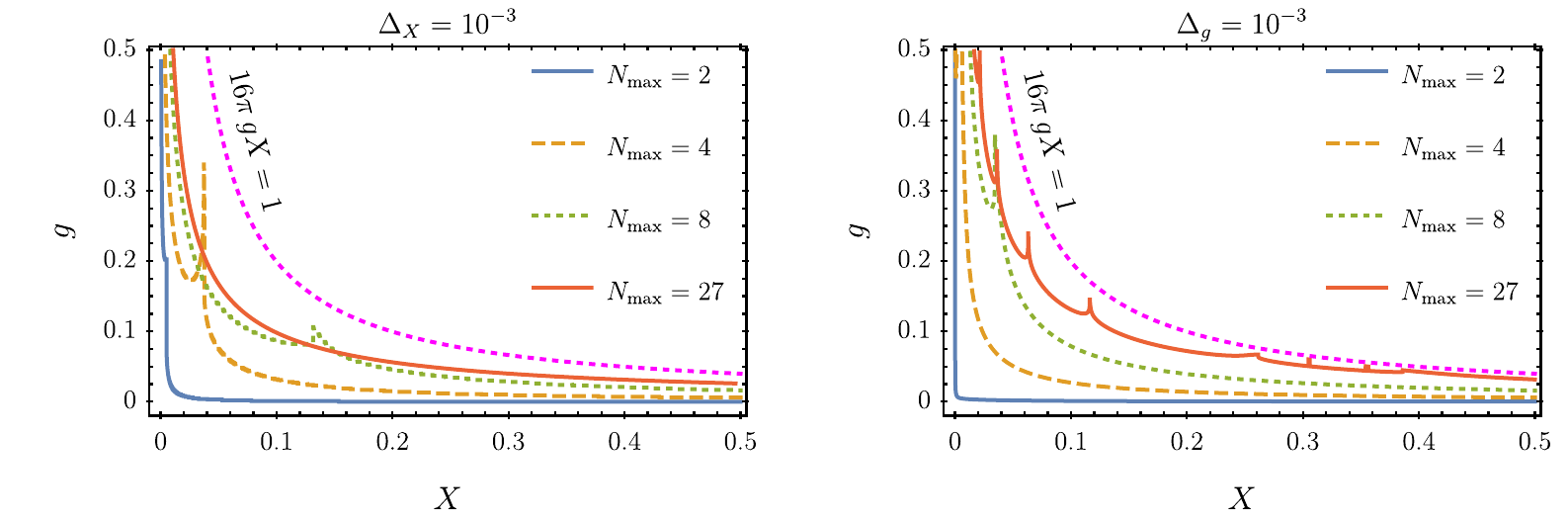} 
	\caption{Contour lines in the $(\Gdimless,\kineticoperatordimless)$-plane where the relative error associated with the approximate fixed point solutions obtained via an expansion in $\kineticoperatordimless$ (left panel) and $\Gdimless$ (right panel) is $0.1\%$ (\textit{i.e.}, $\Delta = 10^{-3}$). In both panels we plot contour lines associated with various values of $\Nmax$. Below each of the contour lines, the value of the relative error for the corresponding truncation decreases. The contour lines exhibit some spikes that we attribute to features of the truncation, and not numerical instabilities. 
		For visual reference, we also plot a dashed, magenta line corresponding to $\kineticoperatordimlesscombinedexpansion = 16\pi \Gdimless \kineticoperatordimless = 1$. For all values of $\Nmax$, we have verified that the relative error exceeds $\Delta = 1$ above the line $16\pi \Gdimless \kineticoperatordimless = 1$.}
	\label{fig:RelativeError_XandG}
\end{figure*}
\begin{figure*}
	\includegraphics[width=1\linewidth]{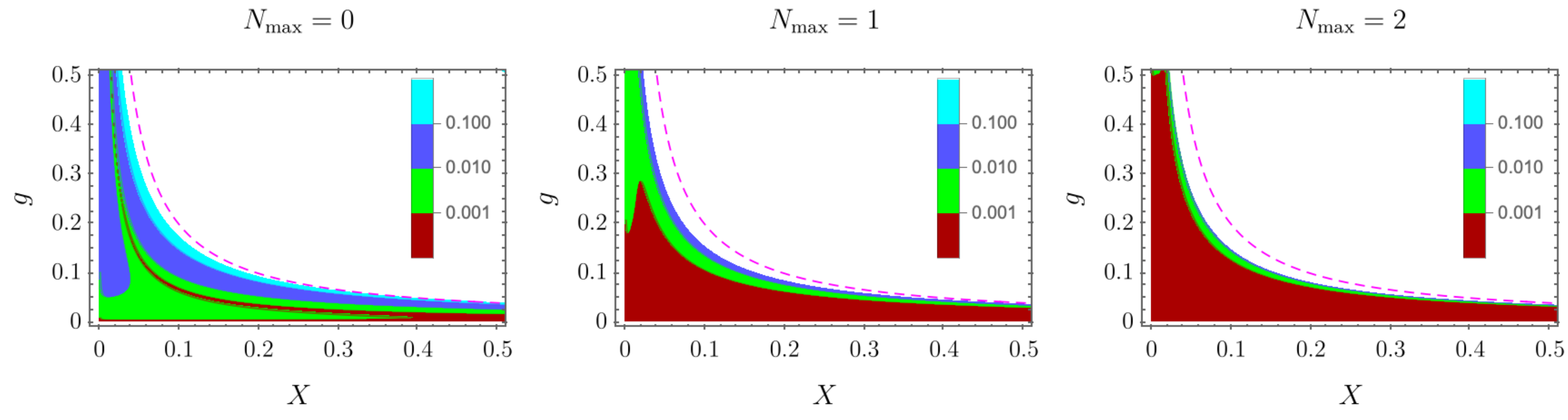} 
	\caption{Regions corresponding to different upper bounds on the relative error $\Delta_{\Gdimless \kineticoperatordimless}$ for the combined expansion. From left to right, we show the results for $\Nmax = 0,1,2$. The dashed, purple line indicates the position of the pole $\kineticoperatordimlesscombinedexpansion =16\pi \Gdimless \kineticoperatordimless = 1$.}
	\label{fig:RelativeError_Combined}
\end{figure*}

We have presented approximations to the \SGFP{} for the gravity-scalar system based on three different expansion schemes. A natural question to ask is about the quality of the different approximation schemes and the respective truncation error in lack of a global solution.
In this section, we study the relative error between the left-hand and the right-hand side of the fixed point equation \eqref{eq:fullFPeq_GravityMatter} obtained in the three expansion schemes. 

We define the relative error as
\begin{equation}\label{eq:relative_error}
	\hspace*{-.15cm}
	\Delta(\kineticoperatordimless,\Gdimless) =
	\Bigg|\frac{4 \kineticfunctiondimless(\kineticoperatordimless) \!-\! X (4+\scalaranomalousdimension) \kineticfunctiondimless'(\kineticoperatordimless) \!-\! \mathcal{F}}{4 \kineticfunctiondimless(\kineticoperatordimless) - X (4+\scalaranomalousdimension) \kineticfunctiondimless'(\kineticoperatordimless)} \Bigg| \, ,
\end{equation}
where we insert a given approximation for \kineticfunctiondimless{} and \scalaranomalousdimension{} at the \SGFP{}.
For an exact fixed point solution satisfying \eqref{eq:fullFPeq_GravityMatter} the relative error $\Delta(\kineticoperatordimless,\Gdimless)$  vanishes for all values of $\kineticoperatordimless$ and $\Gdimless$. For approximate fixed point solutions obtained by one of the expansion schemes discussed above, the relative error $\Delta(\kineticoperatordimless,\Gdimless)$ will be non-vanishing. Thus, we can probe the quality of a given approximation by evaluating $\Delta(\kineticoperatordimless,\Gdimless)$ in the $(\Gdimless,\kineticoperatordimless)$-plane. In the following, we use the notation $\Delta_\kineticoperatordimless(\kineticoperatordimless,\Gdimless)$, $\Delta_\Gdimless(\kineticoperatordimless,\Gdimless)$ and $\Delta_{\Gdimless\kineticoperatordimless}(\kineticoperatordimless,\Gdimless)$ to denote the relative error associated to the expansion in $\kineticoperatordimless$, $\Gdimless$ and the combined expansion, respectively.

Before commenting on the results, let us briefly discuss the evaluation of the relative error $\Delta(\kineticoperatordimless,\Gdimless)$. From the definition of $\Delta(\kineticoperatordimless,\Gdimless)$, we need to evaluate $\mathcal{F}$ on the approximate fixed point solution obtained by the different expansion schemes. From \eqref{eq:fullFPeq_GravityMatter_RHS}, we can see that the evaluation of $\mathcal{F}$ involves two integrals. With the Litim regulator, we can perform the radial integral analytically. However, we can only perform the angular integral numerically. Thus, to evaluate the relative error $\Delta(\kineticoperatordimless,\Gdimless)$ in the $(\Gdimless,\kineticoperatordimless)$-plane, we first need to fix numerical values for $\kineticoperatordimless$ and $\Gdimless$, and then compute $\mathcal{F}$ for a given approximate partial fixed point solution. We repeated this procedure for values of $\kineticoperatordimless$ and $\Gdimless$ between $0$ and $1/2$, with grid size $\delta \kineticoperatordimless = \delta \Gdimless = 10^{-3}$.

In \autoref{fig:RelativeError_XandG}, we plot contour lines where $\Delta(\kineticoperatordimless,\Gdimless) = 10^{-3}$ (0.1\% error) for the $\kineticoperatordimless$-expansion (left panel) and the $\Gdimless$-expansion (right panel), for different values of $\Nmax$. Each contour line defines an upper bound to the region where $\Delta(\kineticoperatordimless,\Gdimless) \leq 10^{-3}$. 
Comparing the two panels in \autoref{fig:RelativeError_XandG}, we can see that the overall qualitative picture is relatively similar for both expansion schemes. As we see, some of the contour lines exhibit a few spikes. By increasing the numerical precision, we have checked that these are not numerical instabilities, but rather features of the given truncation. In particular, they are caused by near cancellations in the numerator of \eqref{eq:relative_error}. 
In both expansion schemes, the main result is that the quality of the truncation gets better when increasing $\Nmax$ and when reducing the value of the product $\Gdimless \kineticoperatordimless$.
Apart from the spikes, each contour line can be approximated by a hyperbola of the type $\Gdimless \propto \kineticoperatordimless^{-1}$, where the proportionality coefficient depends on $\Nmax$ and on the values of the relative error.

Furthermore, we note that all contour lines shown in \autoref{fig:RelativeError_XandG} satisfy the inequality $16 \pi \Gdimless \kineticoperatordimless < 1$. For values of $\kineticoperatordimless$ and $\Gdimless$ such that $16 \pi \,\Gdimless \kineticoperatordimless > 1$, our numerical checks lead to $\Delta(\kineticoperatordimless,\Gdimless) \geq 1$, indicating that the approximations obtained in the $\kineticoperatordimless$- and $\Gdimless$-expansions are not reliable in that region. This result is in line with the combined expansion studied in \autoref{sec:combexp}, where the functions \kineticfunctioncombinedexpansionFP{n} display a pole at $\kineticoperatordimlesscombinedexpansion = 16 \pi \Gdimless \kineticoperatordimless = 1$, and the general on-shell assessment in \autoref{sec:understanding_the_pole}.

Now, we evaluate the quality of the combined expansion discussed in \autoref{sec:combexp}. In \autoref{fig:RelativeError_Combined}, we show contour plots representing the relative error $\Delta_{\Gdimless\kineticoperatordimless}(\kineticoperatordimless,\Gdimless)$ for $\Nmax \in \{ 0,1,2 \}$. The different colours correspond to different upper bounds on $\Delta_{\Gdimless\kineticoperatordimless}(\kineticoperatordimless,\Gdimless)$. We focus on the region defined by $16 \pi \Gdimless \kineticoperatordimless < 1$. 

From \autoref{fig:RelativeError_Combined}, we note that the region in red where the relative error is smaller than $0.1 \%$, gets larger when we increase $\Nmax$. This is an indication that the quality of the combined expansion gets better with increasing $\Nmax$. This result is in accordance with the previous expansion schemes. However, comparing \autoref{fig:RelativeError_XandG} and \autoref{fig:RelativeError_Combined} we see a different pattern in the way that the boundary of the regions with a relative error smaller than $0.1 \%$ evolves with respect to $\Nmax$.

To have a quantitative measure on how the relative error changes as a function of $\Nmax$, we introduce the following quantity:
\begin{align}
	\mathcal{C}_{i}(\Nmax,\Delta) =  16\pi \,  \Gdimless_{\Delta}^{2}\,,
\end{align}
where we define $\Gdimless_\Delta$ as the smallest value for which $\Delta(\kineticoperatordimless=\Gdimless_\Delta,\Gdimless_\Delta)=\Delta$, and where the subscript $i$ indicates the expansion scheme of $\kineticfunctiondimless$.

\begin{figure}
	\includegraphics[width=\linewidth]{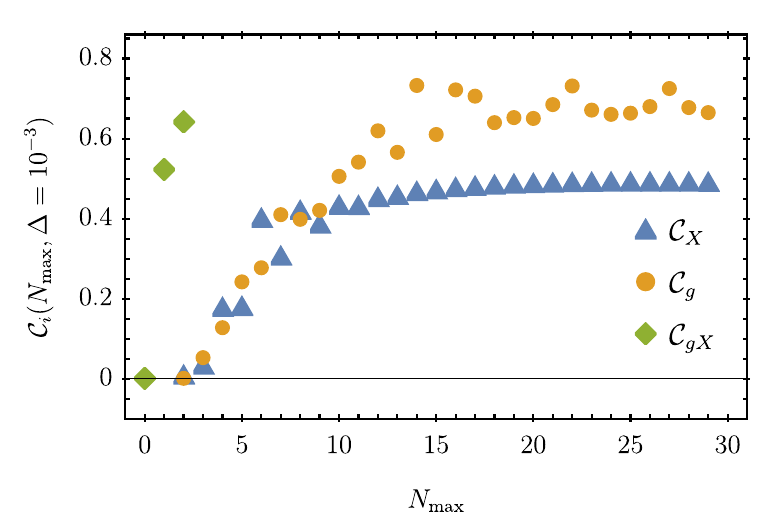} 
	\caption{$\mathcal{C}_i(\Nmax,\Delta=10^{-3})$ as function of $\Nmax$, for the three expansion schemes discussed in \autoref{sec:gravmatt}. In both the \kineticoperatordimless{}-expansion and the \Gdimless{}-expansion, $\mathcal{C}_i$ saturates at a value lower than unity. The combined expansion shows overall a much smaller error, and correspondingly a much quicker convergence.}
	\label{fig:NumCheckCoeffs}
\end{figure}

In \autoref{fig:NumCheckCoeffs}, we plot $\mathcal{C}_{i}(\Nmax,\Delta=10^{-3})$ as a function of $\Nmax$, for the three expansion schemes used in this work. 
Comparing the data corresponding to the expansions in $\kineticoperatordimless$ and $\Gdimless$, we note that both truncation schemes share the same qualitative behaviour: the distance to the line where the error exceeds $\Delta$ increases, but eventually saturates with $\mathcal{C}_i<1$. On the quantitative level, the most important difference is that $\mathcal{C}_\Gdimless$ grows larger than $\mathcal{C}_\kineticoperatordimless$. This indicates that the $\Gdimless$-expansion has a larger radius of convergence than the $\kineticoperatordimless$-expansion.

Concerning the oscillation of the points in \autoref{fig:NumCheckCoeffs}, we attribute this behaviour as a consequence of the same type of truncation fluctuations that generate the spikes in \autoref{fig:RelativeError_XandG}. For example, for $\Nmax = 27$ (in the $\Gdimless$-expansion) the point $\kineticoperatordimless_\Delta=\Gdimless_\Delta$ lies partially within one of the spikes in the corresponding contour line shown in \autoref{fig:RelativeError_XandG}. This sometimes leads to larger values for $\mathcal{C}_\Gdimless(\Nmax,\Delta=10^{-3})$.

Finally, in \autoref{fig:NumCheckCoeffs} we also included points corresponding to the combined expansion for $\Nmax\in\{0,1,2\}$. In this case, we note that for $\Nmax=2$, $\mathcal{C}_{\Gdimless\kineticoperatordimless}$ already exceeds $\mathcal{C}_\kineticoperatordimless$ for all $\Nmax \leq 29$, and reaches approximately the same level as the $\mathcal{C}_\Gdimless$ with the highest value of $\Nmax$. This remarkable feature indicates the superiority of the combined expansion compared with the other expansion schemes explored in this paper.

\subsection{Summary and discussion of the gravity-scalar system}

In this section, we studied the shift-symmetric scalar system when minimally coupled to gravity. As a first step, we explored the same expansion of $\kineticfunctiondimless$ in powers of $\kineticoperatordimless$ as in \autoref{sec:purematt}, which is the only expansion that has been studied in the literature so far. We discovered a qualitatively different behaviour of the \SGFP{} for even and odd $\Nmax$: while the \SGFP{} vanishes into the complex plane due to the collision with a spurious partial fixed point for even $\Nmax$, the \SGFP{} remains real and numerically stable for values of \Gdimless{} of at least $\Gdimless \approx 2$ for odd $\Nmax$, see \autoref{fig:SGFP}. Therefore, within our truncation, the presence of a partial fixed point collision can be easily circumvented by restricting to an odd $\Nmax$. We have also introduced a new notion of the \WGB{} related to the number of relevant operators at the \SGFP{}.

Investigating different choices for the gauge parameter $\GFbeta$ and two different regulators revealed that the location of the partial fixed point collision for even $\Nmax$ is not numerically stable, see the left panel of \autoref{fig:GcritGK2Zero}. This finding is not surprising in light of the results of \autoref{sec:purematt}: there, we already concluded that only the \GFP{} of the pure scalar system is a stable fixed point. Accordingly, a \WGB{} that results from a collision of the \SGFP{} with another fixed point can only be a truncation artefact, since all other fixed points of the scalar system are truncation artefacts.

Conversely, for odd \Nmax{} the \SGFP{} is numerically stable, also under changes of the gauge parameter $\GFbeta$ and the two choices for the regulator, see the right panel of \autoref{fig:GcritGK2Zero}.

We then studied the \SGFP{} within an expansion in powers of \Gdimless{}. After fixing the constants of integration by demanding regularity at $\kineticoperatordimless=0$ and normalisability for $\kineticoperatordimless\to\infty$, the fixed point solution for the expansion coefficients in this expansion are polynomials in $\kineticoperatordimless$. This means that the expansion of $\kineticfunctiondimless$ in $\Gdimless$ and in $\kineticoperatordimless$ are not independent expansions. The expansion also fixes all integration constants order by order in \Gdimless{}, complementing the expansion in \kineticoperatordimless{}.

Furthermore, the expansion in \Gdimless{} shows that a combined expansion of the form \eqref{eq:combinedexp} is better suited to retain global information in $\kineticoperatordimless$. Indeed, the expansion coefficients in this combined expansion at the fixed point are non-polynomial in $\Gdimless\,\kineticoperatordimless$, see \eqref{eq:SolutionCombinedOrder0}. However, this expansion also features a divergence at $\Gdimless\,\kineticoperatordimless=1/16\pi$, which puts a strict limit on the region of validity of this expansion.

In \autoref{sec:understanding_the_pole} we have studied the origin of the pole at $\Gdimless\,\kineticoperatordimless=1/16\pi$. We found two independent indications that this pole is an off-shell pole similar to the well-known one at $\Lambdadimless=1/2$ in the gravitational propagator on a flat background: first, by studying the scalar contribution to the flow equation, we were able to show that its denominator changes sign as a function of the angular variable $x$. Upon integration over $x$ this will lead to a pole. Second, by studying the classical equations of motion of the gravity-scalar system, we concluded that only a vanishing $\kineticoperatordimless$  or $\Gdimless$ can be on-shell on a flat background. Therefore, the expansions in $\kineticoperatordimless$ and $\Gdimless$ employed in this section are on-shell. Conversely, the combined expansion, which exhibits the pole at $\Gdimless\,\kineticoperatordimless=1/16\pi$, is an expansion about an off-shell background.

Finally, in \autoref{sec:NumericalCheckSolution} we studied in which region in the $(\Gdimless,\kineticoperatordimless)$-plane our truncation is reliable. For this purpose we  evaluated the left- and right-hand side of the flow equation at a truncated fixed point, and investigated for which values of $\kineticoperatordimless$ and $\Gdimless$ the relative difference between both sides is small. For all employed expansions, we find that the viable region generally grows when increasing $\Nmax$, see \autoref{fig:RelativeError_XandG} and \autoref{fig:RelativeError_Combined}. Comparing the different expansions, we furthermore find that the combined expansion (to order $\Nmax=2$) outperforms the expansion in $\kineticoperatordimless$ (to order $\Nmax=29$), and performs similarly well as the expansion in $\Gdimless$ (to order $\Nmax=29$), see \autoref{fig:NumCheckCoeffs}.

\section{Summary and Conclusions}\label{sec:summary}

In this paper, we have investigated a system of a shift-symmetric scalar field with and without the inclusion of gravitational fluctuations. In particular, we considered the flow equation for a function of the kinetic term of the scalar field. The goal of our study was to investigate the \WGB{} that separates the theory space into a weak and a strong gravity regime.

The \WGB{}, as discussed in the literature \cite{Eichhorn:2011pc,Eichhorn:2012va,Eichhorn:2017eht, deBrito:2020dta, deBrito:2021pyi, Knorr:2022ilz,Christiansen:2017gtg, Eichhorn:2022gku}, results from the collision of two partial fixed points that are already present in the absence of gravity. Thus, to better understand this version of the \WGB{}, in \autoref{sec:purematt} we performed a broad analysis of the fixed point structure of a shift-symmetric scalar theory, extending previous work in \cite{Laporte:2022ziz}. We found robust indications that the only reliable fixed point in a shift-symmetric scalar theory is the \GFP{}. We interpret the interacting fixed point candidates that appear in polynomial truncations as being truncation artefacts. This conclusion is based on the findings discussed in \autoref{sec:purematt}:
\begin{itemize}
 \item the kinetic function of the potential interacting fixed points approaches $\kineticfunctiondimlessFP(\kineticoperatordimless)=\kineticoperatordimless$ exponentially quickly when increasing the truncation order $\Nmax$, and
 \item the analysis of perturbations of the \GFP{} shows that the interacting fixed point candidates are related to eigenperturbations that are not normalisable within the Hilbert space $\mathcal{L}$ that defines the discrete spectrum of critical exponents.
\end{itemize}
With this interpretation in mind, the partial fixed point collision that causes the \WGB{} is likely to be a truncation artefact in our system.

In section \autoref{sec:gravmatt}, we studied the shift-symmetric scalar field minimally coupled to gravity. With the understanding that the partial fixed point collision is a truncation artefact, it is important to search for approximation schemes that can avoid such a spurious behaviour. We started by performing a detailed investigation of the fixed point structure obtained via polynomial truncations, extending previous analyses by including higher order terms in the scalar sector. We find that the partial fixed point collision seems to be a general artefact of truncations characterised by an even $\Nmax$. For odd $\Nmax$, the \SGFP{} remains under control for large values of $\Gdimless$. We also introduced a new notion of the \WGB{} relying on the observation that the \SGFP{} receives additional relevant operators in the strong gravity regime. This new notion thus distinguishes the weak from the strong gravity regime not by the presence or absence of a \UV{} completion via the \SGFP{}, but rather by a comparison with the spectrum of the \GFP{}. A \UV{} completion that happens to be in the strong gravity regime is thus less predictive than if it would be in the weak gravity regime.

To get further insight into the nature and stability of the \SGFP{}, we studied two other expansion schemes. Expanding $\kineticfunctiondimless(\kineticoperatordimless)$ in powers of $\Gdimless$, we find fixed point solutions $\kineticfunctiondimlessFP(\kineticoperatordimless)$ that are also polynomial in $\kineticoperatordimless$. Thus, the truncation scheme based on an expansion in $\Gdimless$ does not capture global information of $\kineticfunctiondimlessFP(\kineticoperatordimless)$. Inspired by this result, we investigated a combined expansion scheme where the fixed point solution retains global information on the product $\Gdimless\,\kineticoperatordimless$. This expansion scheme shows a divergence at $\Gdimless\,\kineticoperatordimless=1/16\pi$, indicating a hard limit of the region where the expansion is applicable. This pole is a consequence of the fact that the combined expansion is about an off-shell background. To test the quality of the truncation schemes employed in this work, we performed an analysis of the relative error of our fixed point solutions. The combined expansion seems to be more efficient than the expansions in $\kineticoperatordimless$ and $\Gdimless$.

In conclusion, we find evidence that the pure scalar system only features one stable fixed point, namely the \GFP{}. Coupling the system to gravity, a partial fixed point collision appears in an expansion of the system only to even powers in the kinetic term. However, for odd powers, and also for expansions in different variables, the \SGFP{} remains stable. Hence, we find indications that the \WGB{} as discussed in the literature seems to be a truncation artefact, but there is nevertheless a way to separate the weak from the strong gravity regime.

As a disclaimer, let us point out that our ansatz for the action for the scalar field is not based on a systematic expansion in derivatives or the classical mass dimension. Instead, it consists of a very specific cut in theory space, similar to the $f(R)$ truncation in quantum gravity. As a consequence, at a given order in the expansion, there are several additional tensor structures of lower canonical mass dimension that we neglected in this study, and that are likely more important than higher order terms that we have retained. These might give rise to a stable, non-trivial fixed point in the pure matter sector that eventually could give rise to a \WGB{} induced by a partial fixed point collision of the \SGFP{}.  

The \WGB{} based on such  a partial fixed point collision has also been found for Abelian gauge fields \cite{Christiansen:2017gtg, Eichhorn:2021qet}. It would be interesting to understand if the induced gauge interactions feature a similar structure as the induced scalar interactions discussed here. The universality for different matter fields discovered in \cite{Laporte:2022ziz} might suggest this. However, a careful study of the coupled gravity-photon system is necessary to investigate this further. We plan to report on this elsewhere.

\begin{figure*}
	\includegraphics[width=.8\linewidth]{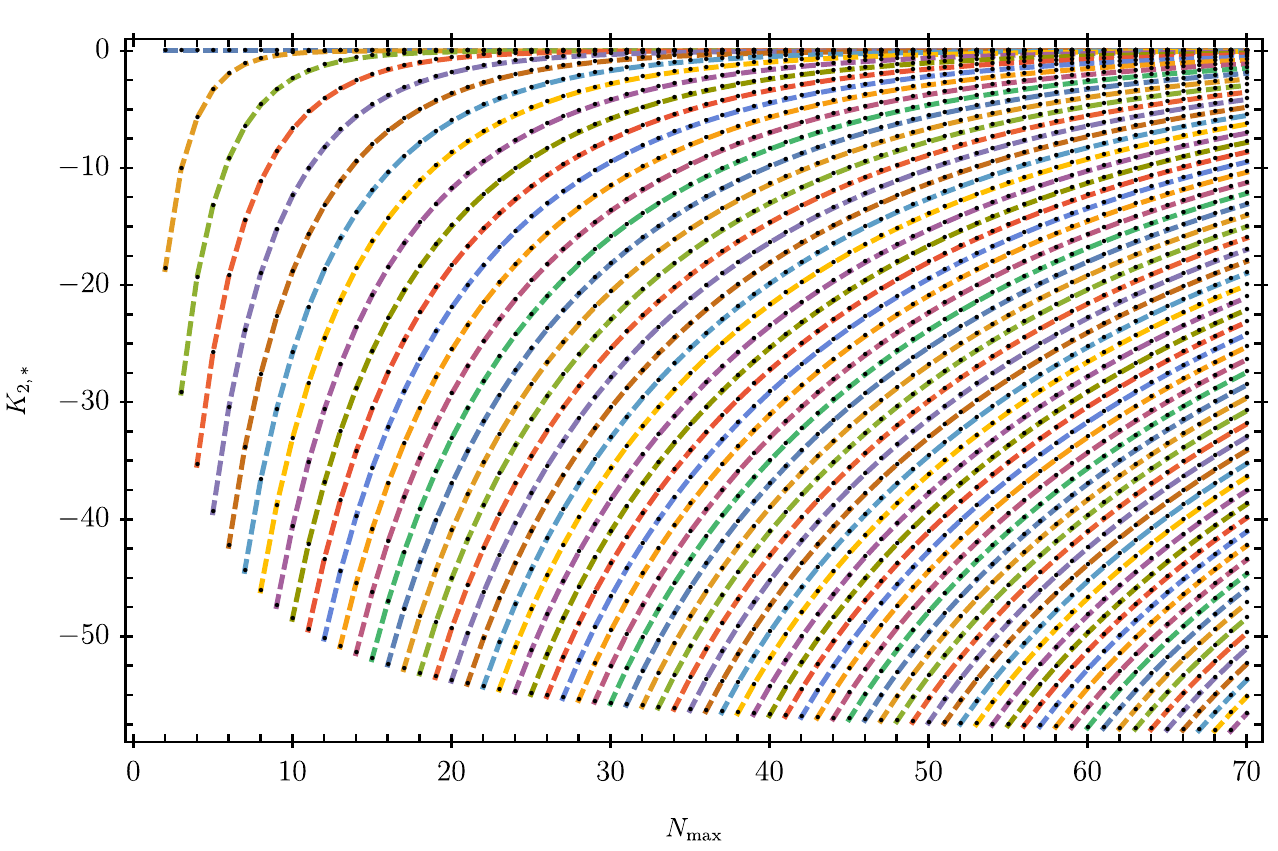}
	\caption{Fixed point structure of the pure scalar system for the exponential regulator as a function of the maximal order of the polynomial expansion $\Nmax$, up to $\Nmax=70$. The black markers indicate the real fixed point values $\KcoupFP{2}$ for a given truncation. The dashed lines indicate the evolution of a given fixed point when increasing \Nmax{}.}
	\label{fig:PMFPs_Exp}
\end{figure*}

\section*{Acknowledgements}

We would like to thank Astrid Eichhorn, Holger Gies, Yannick Kluth, Jan Pawlowski, Ant\^onio Pereira, Manuel Reichert, and Frank Saueressig for interesting discussions, and Astrid Eichhorn and Frank Saueressig for helpful comments on the manuscript. G.\ P.\ B.\ is supported by research grant (29405) from VILLUM fonden. B.\ K.\ and M.\ S.\ acknowledge support by Perimeter Institute for Theoretical Physics. Research at Perimeter Institute is supported in part by the Government of Canada through the Department of Innovation, Science and Economic Development and by the Province of Ontario through the Ministry of Colleges and Universities. B.\ K.\ furthermore acknowledges support from Nordita. Nordita is supported in part by NordForsk. B.\ K.\ would also like to thank CP3-Origins and Southern Denmark University for hospitality during the final stages of this project.

\appendix

\section{Regulator, gauge and parameterisation dependencies}
\label{app:SpuriousDep}

Off-shell quantities like beta functions or fixed point values, but within truncations of the flow equation even physical quantities like critical exponents, depend on unphysical choices like that of the gauge parameters and the regulator. Such dependencies can be interpreted as a quantifier of the robustness of the truncation: results that are qualitatively or even quantitatively independent of unphysical choices are likely to be physical, and not artefacts of the truncation. Conversely, if results depend on these unphysical choices on a qualitative level, one has to critically assess the quality of the truncation. In this appendix, we discuss gauge, regulator and parameterisation dependencies of the gravity-scalar theory studied in the main part.

\subsection{Pure scalar system}\label{app:pmSpuriousDep}

\begin{figure*}
	\includegraphics[width=.8\linewidth]{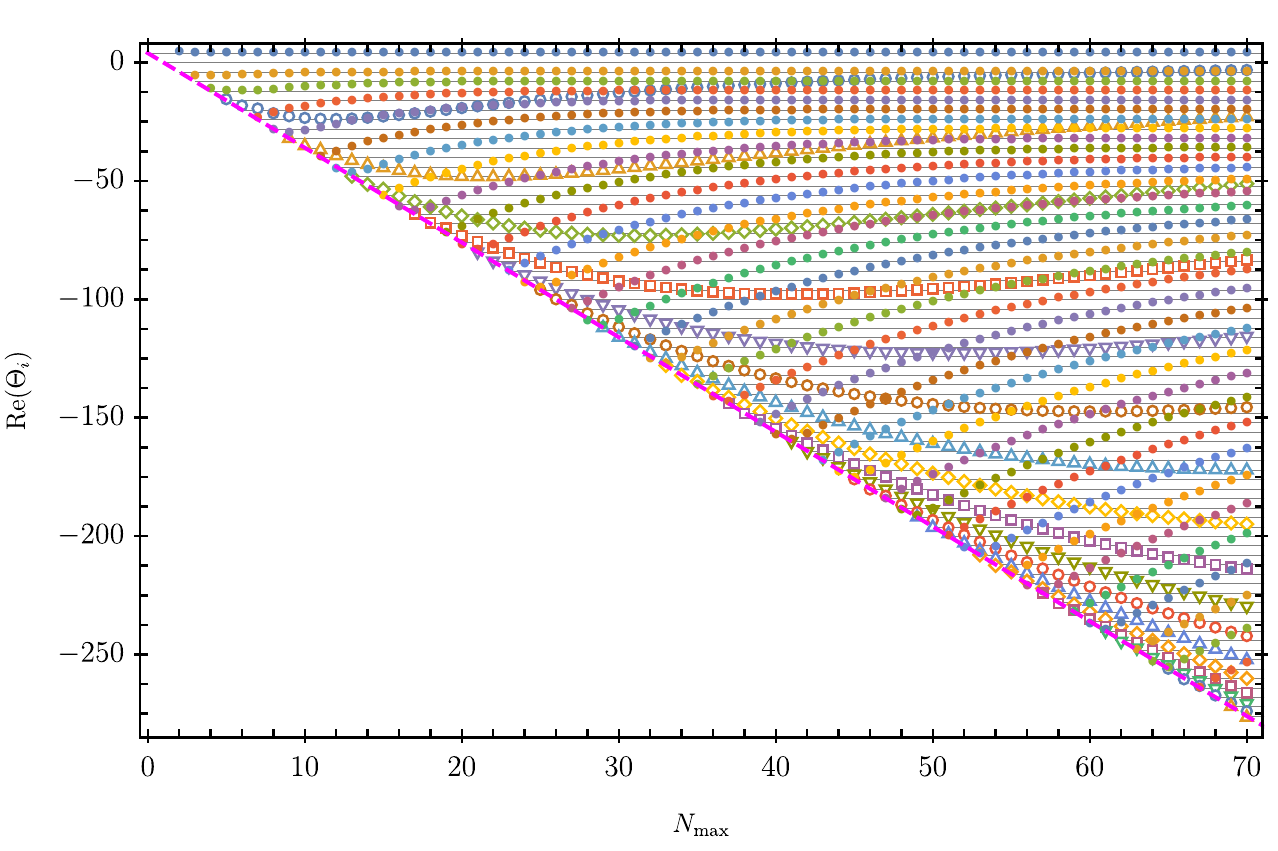}
	\caption{Critical exponents of the first interacting pure scalar fixed point computed with exponential regulator as a function of the truncation order $\Nmax$. The horizontal lines indicate the spacing of $-4$ between critical exponents that is expected at the Gaussian fixed point. Furthermore, the magenta dashed line is given by $y=4-4\Nmax$, and indicates the canonical mass dimension of the canonically most irrelevant coupling added in a given truncation.}
	\label{fig:PMTheta_Exp}
\end{figure*}
\begin{figure}
	\includegraphics[width=\linewidth]{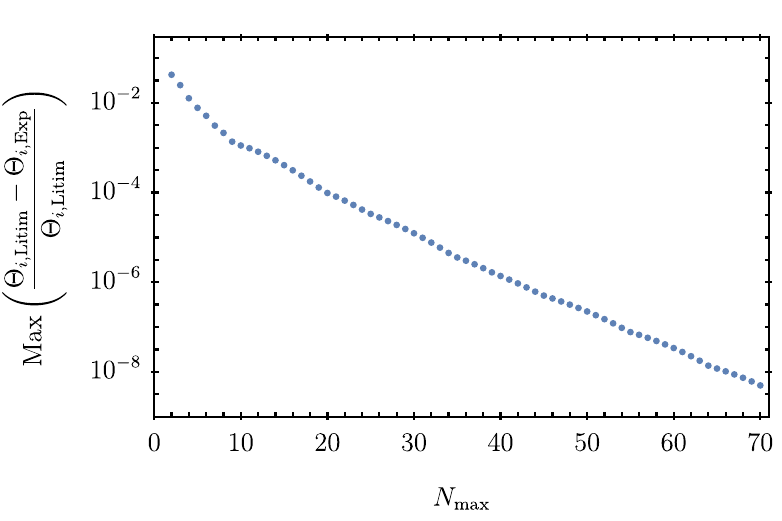}
	\caption{Maximum relative difference of critical exponents of the pure scalar system between the Litim  \eqref{eq:shapeLitim} and the exponential regulator \eqref{eq:shapeExp}. While the critical exponents show a deviation of few percent for $\Nmax=2$, the maximal deviation decreases with $\Nmax$ to $\sim 5 \times 10^{-9}$ at $\Nmax=70$.}
	\label{fig:CritExpDeviation}
\end{figure}

We start with the pure scalar system, where the only spurious dependence is related to the regulator choice. The results presented in \autoref{sec:purematt} were obtained with the Litim regulator \eqref{eq:shapeLitim}. In the following, we supplement our analysis with results obtained with the exponential regulator \eqref{eq:shapeExp}. 

In \autoref{fig:PMFPs_Exp}, we show the values of $\KcoupFP{2}$ for the different fixed point solutions of the pure scalar system. The fixed point structure obtained with the exponential  regulator shares the same qualitative behaviour as the fixed point structure obtained with Litim regulator (c.f. \autoref{fig:PMFPs}). In particular, all fixed point values $\KcoupFP{2}$ show an exponential fall-off as a function of $\Nmax$.

\begin{figure*}[t]
	\includegraphics[width=\linewidth]{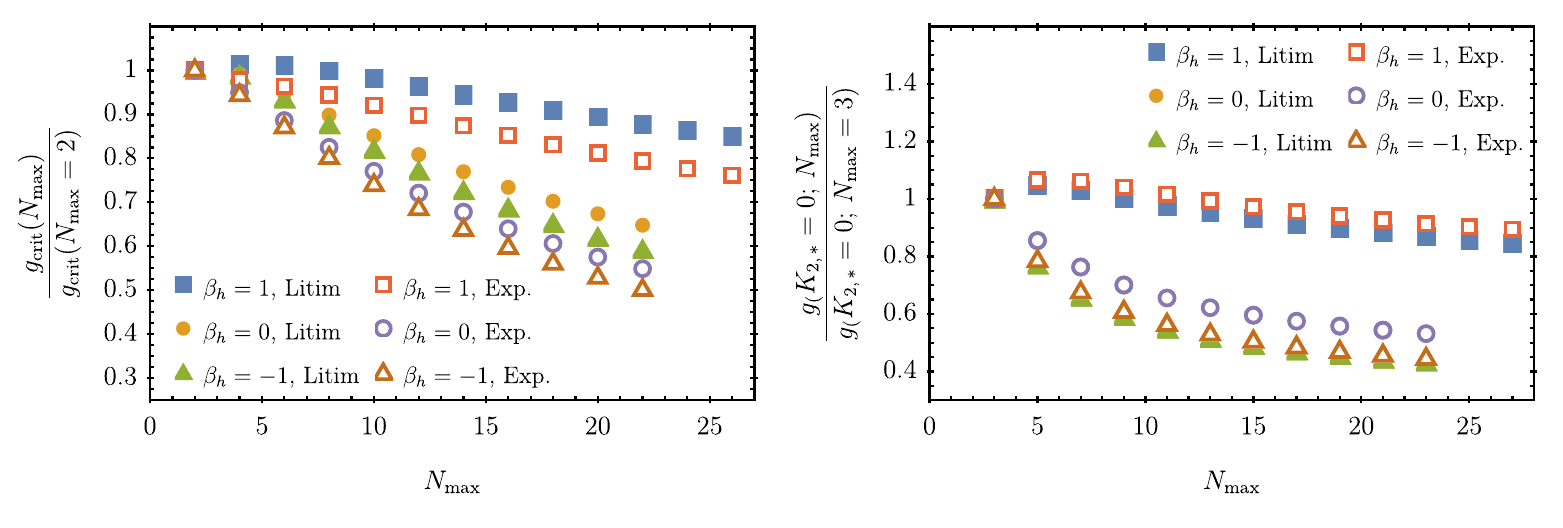}
	\caption{Aspects of gauge and regulator dependences of the \SGFP{}, obtained with the exponential parameterisation. Left panel: we show the (normalised) critical value of the Newton coupling $\Gdimlesscrit$ as a function of $\Nmax$ for different choices of the gauge parameter $\GFbeta$ and the regulator. For all displayed choices, the fixed point collision of the \SGFP{} that gives rise to $\Gdimlesscrit$, is only present for even $\Nmax$. Right panel: we show the value of the Newton coupling where the partial fixed point value $\KcoupFP{2}$ at the \SGFP{} crosses zero, normalised to the $\Nmax=3$ case. We see that the qualitative behaviour agrees for all displayed choices of gauge fixing and regulator. Furthermore, we note that the results obtained with the exponential parameterisation are qualitatively similar to the results obtained with the linear parameterisation (c.f. \autoref{fig:GcritGK2Zero}).}
	\label{fig:GcritGK2Zero_ExpParam}
\end{figure*}

On the quantitative level, the values of $\KcoupFP{2}$ for the Litim regulator are approximately seven times the corresponding values obtained with the exponential  regulator. Such a difference on numerical values is not surprising given that fixed point values are, in general, non-universal quantities. Nevertheless, it is remarkable that the ratio between $\KcoupFP{2}|_{\textmd{Litim }}$ and $\KcoupFP{2}|_{\textmd{exp }}$ is approximately the same for all fixed points and for all values of $\Nmax$. Furthermore, the deviation of the exponential fall-off of the first, second, and eighth interacting fixed point (the ones displayed in \autoref{fig:FPPMExpv2}) between the Litim and the exponential regulator is on the level of $0.2\%$, $0.3\%$, and $0.1\%$, respectively.

The critical exponents associated with the first pure scalar interacting fixed point (cf.~\autoref{fig:PMTheta_Exp}) agree \emph{quantitatively} between both regulators. For the simplest truncation ($\Nmax = 2$), the critical exponents computed with the Litim and the exponential regulator deviate by approximately $4 \%$. The agreement improves significantly for larger \Nmax{}: for each \Nmax{} we compute the relative difference between the critical exponents for both regulators. In \autoref{fig:CritExpDeviation} we show the maximal relative difference over all exponents for each \Nmax{}. It decreases monotonically for increasing \Nmax, down to $5\times 10^{-9}$ for $\Nmax =70$.

Besides the critical exponents, also the eigenvectors agree on a quantitative level. Also with the exponential regulator, we find that: the coupling with the largest overlap with the relevant direction of the first interacting fixed point is the canonically most irrelevant coupling for each \Nmax{}; the same is true for the complex-conjugate pairs of critical exponents; for the irrelevant directions, the largest overlap is with the couplings of the canonical mass dimension closest to the critical exponent. For $\Nmax\geq35$, the maximum relative deviation of a component of any of the eigenvectors does not exceed $0.5\%$ between the eigenvectors obtained using a Litim or an exponential regulator.

In summary, crucial properties of the pure scalar system, like the exponential fall-off of fixed point values, critical exponents and eigenvectors, agree on a qualitative, and for large enough $\Nmax$ even on a quantitative level. This is an indication that these properties of the system are robust and trustworthy, and not induced by the regulator choice.

\subsection{Gravity-scalar system}\label{app:gmSpuriousDep}

When coupling the pure scalar system to gravity, additional spurious dependences are introduced. Besides the choice of the regulator discussed for the pure scalar case, the choice of the gauge parameter $\GFbeta$ as well as the parameterisation of metric fluctuations can influence the results in truncations.

We again consider the Litim and the exponential regulator introduced in \eqref{eq:shapeLitim} and \eqref{eq:shapeExp}, respectively.
Regarding the dependence on the gauge parameter \GFbeta{}, we focus on the choices that are the most common in the Asymptotic Safety literature, namely $\GFbeta \in \{ -1, 0, 1 \}$, and the limit $\GFbeta \to -\infty$. We do not consider choices for the gauge parameter \GFalpha{} other than the Landau limit, $\GFalpha \to 0$. This choice is well-motivated as it corresponds to a fixed point for both gauge parameters \cite{Litim:2002ce, Knorr:2017fus}. 

Concerning parameterisation dependence, besides the linear split \eqref{eq:linsplit} used in the main text, we also explore the exponential parameterisation defined as
\begin{equation}
	g_{\mu\nu} = \bar{g}_{\mu\alpha} \big( e^{h^{\cdot}_{\,\,\cdot}} \big)^\alpha_{\,\,\,\nu} \,.
\end{equation}
In contrast to the linear parameterisation, the exponential parameterisation is an example of a bijective and signature-preserving map between the space of metric fluctuations and the space of Euclidean metrics \cite{Nink:2014yya, Demmel:2015zfa, Knorr:2022mvn}.\footnote{The exponential parameterisation is also particularly convenient in the context of unimodular quantum gravity, since it allows us to express the unimodularity condition $\det (g_{\mu\nu}) = 1$ as a tracelessness condition $h^{\mu}_{\,\,\,\mu} = 0$ \cite{Eichhorn:2013xr, Benedetti:2015zsw, Eichhorn:2015bna, DeBrito:2019gdd, deBrito:2020rwu, deBrito:2020xhy, deBrito:2021pmw}.}

\begin{figure*}
	\includegraphics[width=\linewidth]{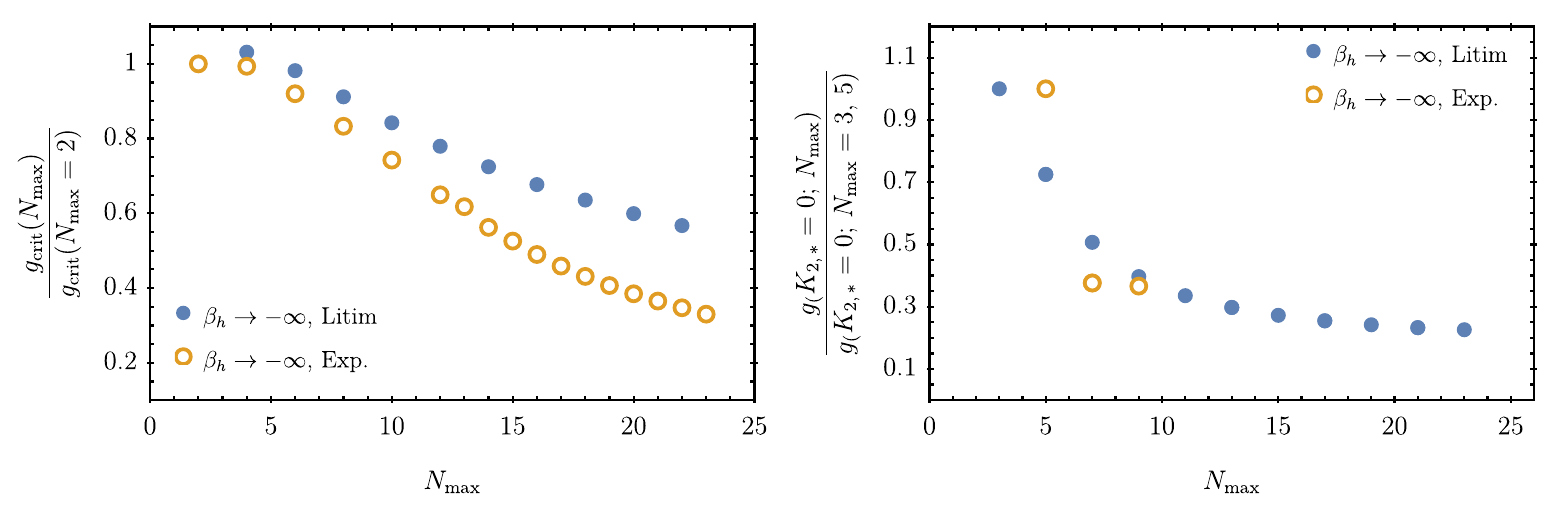}
	\caption{
		Aspects of the \SGFP{} obtained with the exponential parameterisation and the gauge choice $\GFbeta \to -\infty$, with both the Litim and the exponential regulator.
		Left panel: we show the normalised value of $\Gdimlesscrit$ as a function of $\Nmax$.
		Right panel: we show the value of the Newton coupling where the partial fixed point value $\KcoupFP{2}$ at the \SGFP{} crosses zero. We normalise the results with respect to $\Nmax=3$ (Litim regulator) or $\Nmax=5$ (exponential regulator). For the Litim regulator, the results shown in this plot are qualitatively similar to the other gauge choices depicted in \autoref{fig:GcritGK2Zero_ExpParam}. For the exponential regulator, we see a more irregular pattern. In this case, we observe partial fixed point collision for all values of $\Nmax$, except for $\Nmax = 5,7,9$.}
	\label{fig:GcritGK2Zero_betaInf_ExpParam}
\end{figure*}

\begin{figure}
	\includegraphics[width=\linewidth]{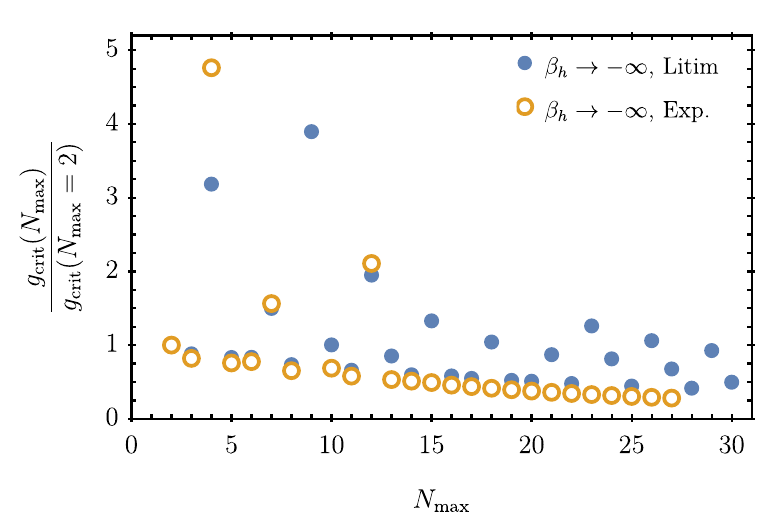}
	\caption{Normalised value of $\Gdimlesscrit$ as a function of $\Nmax$ for the gauge choice $\GFbeta \to -\infty$ and linear metric parameterisation for both the Litim and the exponential regulator. Different from the gauge choices depicted in \autoref{fig:GcritGK2Zero}, here we observe partial fixed point collision for even and odd values of $\Nmax$. Furthermore, the pattern of the partial fixed point collision is much more irregular than what we observed for $\GFbeta \in \{ -1,0,1\}$.}
	\label{fig:Gcrit_betaInf_Lin}
\end{figure}

On a more technical level, together with the exponential parameterisation and in order to reduce the computational complexity, we employ an additional approximation regarding the anomalous dimension $\scalaranomalousdimension$. Specifically, we consider a perturbative treatment for the scalar anomalous dimension $\scalaranomalousdimension$, corresponding to neglecting its contributions due to the regulator insertion $k \partial_k \mathfrak R_k(p^2)$.

In the following, for definiteness we focus on the regulator, gauge, and parameterisation dependences of specific quantities in the polynomial expansion in \kineticoperatordimless{}, see \eqref{eq:polynomial_expansion}. As in \autoref{sec:rhoexp}, we separate our analysis for even and odd values for $\Nmax$. For even $\Nmax$, we focus on the critical value \Gdimlesscrit{} defined by the collision between the \SGFP{} and the first interacting fixed point. For odd $\Nmax$, we focus on $\Gdimless(\KcoupFP{2}=0)$, \ie, the value of \Gdimless{} for which the fixed point value of $\KcoupFP{2}$ (associated with the \SGFP{}) is zero.

For the exponential parameterisation, we show these two quantities for different choices of the regulator and the gauge in \autoref{fig:GcritGK2Zero_ExpParam}. Furthermore, the same quantities, but obtained within a linear parameterisation of metric fluctuations, are shown in \autoref{fig:GcritGK2Zero}. Overall, the qualitative picture for the exponential parameterisation is very similar to that obtained with the linear parameterisation. Except for a small bump for small values of $\Nmax$, our results show that \Gdimlesscrit{} and $\Gdimless(\KcoupFP{2}=0)$ decrease for increasing values of $\Nmax$. Comparing \autoref{fig:GcritGK2Zero} and \autoref{fig:GcritGK2Zero_ExpParam}, for the linear parameterisation the regulator dependence is smaller than the gauge dependence, while for the exponential parameterisation the regulator dependence is dominant over the gauge dependence. Also with an exponential parameterisation the fiducial partial fixed point collision can be circumvented by restricting to odd values of $\Nmax$.

\begin{figure}
	\includegraphics[width=\linewidth]{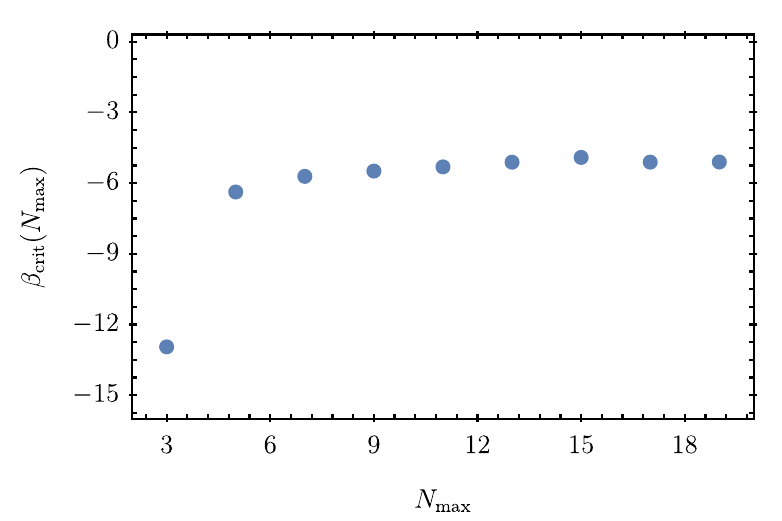}
	\caption{$\beta_\textmd{crit}$ as a function of $\Nmax$. This result correspond to calculations done with the Litim regulator and the linear metric parameterisation. The value of $\beta_\textmd{crit}$ seems to stabilise around a value of $-5$ for large \Nmax{}.}
	\label{fig:BetaCrit}
\end{figure}

Now, we turn to the gauge choice $\GFbeta \to -\infty$, both for the linear and the exponential parameterisation. The results obtained with this gauge choice show some qualitative differences in comparison with the previous cases.

For the exponential parameterisation and the Limit regulator, we obtain results that are qualitatively similar to the cases with $\GFbeta \in \{ -1,0,1 \}$, c.f. \autoref{fig:GcritGK2Zero_ExpParam} and \autoref{fig:GcritGK2Zero_betaInf_ExpParam}. For the exponential regulator, we observe a qualitative difference. In this case, the \SGFP{} features a fixed point collision for most values of $\Nmax$, including odd values. However, for some specific odd values ($\Nmax = 5,7,9$), the \SGFP{} remains real for all values of $\Gdimless$ within the range of investigation.

In \autoref{fig:Gcrit_betaInf_Lin}, we show our results for the linear parameterisation. Here, we see that the \SGFP{} features a partial fixed point collision for all investigated values of $\Nmax$. This is a significant difference with respect to the cases $\GFbeta \in \{ -1,0,1\}$, where we observed partial fixed point collisions involving the \SGFP{} only for even values of $\Nmax$. Furthermore, the pattern for the critical value of \Gdimless{} for $\GFbeta \to -\infty$ is much more irregular than what we observe for the other explored gauge choices. Such an irregular pattern is realised both for the Litim and the exponential regulator. 

To better understand the behaviour for $\GFbeta \to - \infty$, we perform a more detailed analysis of the gauge dependence of the quantities $\Gdimlesscrit{}$ and $\Gdimless(\KcoupFP{2}=0)$. For simplicity, we focus only on results obtained with the Litim regulator and the linear metric parameterisation. We have checked that for odd values of $\Nmax$, there is a critical value $\beta_\textmd{crit}$ that separates two regimes: i) for $\GFbeta > \beta_\textmd{crit}$ we observe a pattern similar to \autoref{fig:GcritGK2Zero}, where there is no partial fixed point collision involving the \SGFP{} for odd $\Nmax$. ii) for $\GFbeta < \beta_\textmd{crit}$, the pattern of partial fixed point collisions becomes more irregular (similar to \autoref{fig:Gcrit_betaInf_Lin}), and we observe collisions involving the \SGFP{} for all values of $\Nmax$. In \autoref{fig:BetaCrit}, we plot $\beta_\textmd{crit}$ as a function of $\Nmax$. As we see, the value of $\beta_\textmd{crit}$ increases for small values of $\Nmax$ and seems to stabilise around a value of about $-5$ for large values of $\Nmax$.

\bibliographystyle{apsrev4-1}
\bibliography{THE_ULTIMATE_BIB.bib}

\end{document}